\pdfoutput=1
\documentclass[11pt]{article}
\usepackage[utf8]{inputenc}
\usepackage[T1]{fontenc}
\usepackage{cite}
\usepackage{hyperref}
\hypersetup{colorlinks=true,linkcolor=blue2,citecolor=red2,urlcolor=green2,pdfencoding=auto,linktocpage}
\usepackage{amsmath,amssymb,amsbsy,amstext,amsthm,simplewick,amsfonts}
\usepackage{mathrsfs}
\usepackage{graphicx}
\usepackage{wrapfig}
\usepackage{upgreek}
\usepackage{bm} 
\usepackage{framed}
\usepackage{bbm}
\usepackage{textcomp}
\usepackage{adjustbox}
\usepackage{makecell}
\usepackage{tcolorbox}
\usepackage{empheq}
\usepackage[normalem]{ulem}
\usepackage{enumitem}
\usepackage{braket} 
\usepackage{array}
\usepackage{dsfont}
\usepackage{physics}
\usepackage{ulem}
\usepackage{xcolor}
\usepackage{caption}
\usepackage{subcaption}
\usepackage{tocloft}
\usepackage{tikz}
\usetikzlibrary{decorations.pathmorphing,decorations.markings}
\usetikzlibrary{arrows.meta}
\usetikzlibrary{positioning}
\usepackage{circuitikz}
\usepackage{booktabs}
\usepackage{pifont}

\usepackage[framemethod=default]{mdframed}

\newmdenv[backgroundcolor=gray!15,%
skipabove=5pt,%
skipbelow=5pt,%
leftmargin=2pt,%
rightmargin=2pt,%
innertopmargin=-6pt,%
innerbottommargin=5pt,%
innerleftmargin=5pt,%
innerrightmargin=5pt,%
splittopskip=0pt,%
splitbottomskip=0pt,%
linewidth=0pt,%
nobreak=true]%
{keyeqn}

\newmdenv[backgroundcolor=gray!15,%
skipabove=5pt,%
skipbelow=5pt,%
leftmargin=2pt,%
rightmargin=2pt,%
innertopmargin=5pt,%
innerbottommargin=5pt,%
innerleftmargin=5pt,%
innerrightmargin=5pt,%
splittopskip=0pt,%
splitbottomskip=0pt,%
linewidth=0pt,%
nobreak=true]%
{keyeqn2}

\graphicspath{{Fig/}}

\definecolor{red2}{RGB}{214, 39, 40}
\definecolor{green2}{RGB}{0,170,0}
\definecolor{blue2}{RGB}{0,100,200}
\definecolor{magenta2}{RGB}{191,64,191}
\definecolor{purple2}{RGB}{112,48,160}
\definecolor{orange2}{RGB}{255,192,0}

\newcommand{\cmark}{\ding{51}}
\newcommand{\xmark}{\ding{55}}

\def\d{\mathrm{d}}

\def\bfk{\mathbf{k}}
\def\bfx{\mathbf{x}}
\def\Res{\mathrm{Res}\,}
\def\Re{\mathrm{Re}\,}
\def\Im{\mathrm{Im}\,}
\def\Poly{{\rm Poly}}

\def\Disc{\mathrm{Disc}\,}
\newcommand{\mdreg}{$m$\&$d$ reg }

\newcommand{\ex}[1]{\left\langle #1 \right\rangle}

\renewcommand{\star}{*}

\numberwithin{equation}{section}

\allowdisplaybreaks[1]
\begin{document}

\begin{titlepage}
	\setcounter{page}{1} \baselineskip=15.5pt 
	\thispagestyle{empty}

     \begin{center}
		{\fontsize{18}{18}\centering {\bf Unitary and Analytic Renormalisation \\[0.2cm] of Cosmological Correlators}} 
	\end{center}
 
	\vskip 18pt
	\begin{center}
		\noindent
		{\fontsize{12}{18}\selectfont Diksha Jain\footnote{\tt dj428@cam.ac.uk}$^{,a}$, Enrico Pajer\footnote{\tt enrico.pajer@gmail.com}$^{,a}$, and Xi Tong  \footnote{\tt xt246@cam.ac.uk}$^{,a}$}
	\end{center}
	
	\begin{center}
		\vskip 12pt
		$^{a}$ \textit{Department of Applied Mathematics and Theoretical Physics, University of Cambridge,\\Wilberforce Road, Cambridge, CB3 0WA, UK} 
	\end{center}

	    \vskip 40pt
		
		\noindent\rule{\textwidth}{0.5pt}
		\noindent \textbf{Abstract} ~~       
    Loop contributions to cosmological correlators and to the associated wavefunction are of key theoretical and phenomenological interest. Here, we investigate and compare different renormalisation schemes proposed in the literature to handle ultraviolet divergences and develop new schemes adapting $\upeta$ regulators to de Sitter spacetime. We focus on one-loop contributions to the quadratic wavefunction coefficient of a shift-symmetric massless scalar in de Sitter spacetime, which is a good toy model of primordial curvature perturbations. We show that different implementations of dimensional regularisation agree with each other and with unitarity and scale invariance in the final renormalised result. Imposing unitarity in the form of the cosmological optical theorem, we define a class of \emph{unitary and analytic $\upeta$ regulators} that agree with dim reg but feature considerable technical and conceptual simplifications. We show that the imaginary part of all one-loop wavefunction coefficients is \textit{universally} fixed in terms of the logarithmic running of the real part, under the assumptions of scale invariance, Bunch-Davies vacuum and light bulk fields. Our work resolves discrepancies in the literature, establishes regulator-independent predictions for the imaginary part at one loop, and provides a practical framework for computing quantum contributions to cosmological correlators.

	
\end{titlepage} 


\newpage
\setcounter{page}{2}
{
	\tableofcontents
}




\newpage

\section{Introduction}\label{IntroSect}
 

The generation of primordial perturbations in our universe is studied within the paradigm of quantum field theory (QFT) in curved spacetime. The main object of study are cosmological correlators, namely quantum expectation values of the product of fields, usually taken at the same time, corresponding to the beginning of the hot big bang. An associated and more primitive object is the field theoretic wavefunction, which can be decomposed in terms of wavefunction coefficients, whose relationship to correlators is somewhat analogous to that of amplitudes to cross sections. Realistic theories include the non-linear dynamics of gravity and are therefore necessarily described by interacting QFTs. Such theories cannot, in general, be solved exactly, but are amenable to a perturbative expansion around free, non-interacting theories. The leading order results, corresponding to the saddle point or ``tree-level'' approximation of the path integral that defines the theory, are so far perfectly suitable to describe observations. This is because of the strong bounds on interactions that have been derived from the non-observations of primordial non-Gaussianity \cite{Planck:2019kim}. These bounds suggest that the very early universe may be described in terms of a weekly coupled effective QFT. Nevertheless, there are several conceptual as well as practical reasons to go beyond this tree-level, saddle-point approximation and study loop contributions.

\paragraph{Motivations} The first reason is that we should understand our theory as well as we possibly can. On the theoretical side, we need to develop a fully consistent prescription of how to compute, at least in principle, any desired observable to any desired precision. This consistency is almost trivial at tree level. However, it becomes subtle at loop order, due to the emergence of divergences associated with loop integrals over momenta. Understanding how to treat these divergences in flat spacetime via renormalisation was one of the major achievement of fundamental physics of the 20th century. Because we live in a curved rather than flat spacetime, it is important to develop a similar understanding of QFTs in curved spacetime. On the observational side, loop contribution lead to small but distinctive observational signatures, which may suggest different experimental strategies and analysis of cosmological data. \\

A second reason to study loop contributions has emerged recently, and to understand it we have to review some interesting results in the study of cosmological correlators. In \cite{Liu:2019fag}, it was noticed by Liu, Tong, Wang, and Xianyu that for massless scalar fields in exact de Sitter spacetime all parity-odd correlation functions vanish when computed at tree level and assuming a standard Bunch-Davies initial state. This result is remarkable because de Sitter space is a good leading-order approximation to many realistic models of inflation, suggesting parity-odd correlators would vanish in the simplest models. The derivation in \cite{Liu:2019fag} relied on the careful examination of the brute-force calculation of parity-odd correlators, using the standard in-in formalism. Later on, this result was further elucidated and generalised by employing general results on the consequences of unitarity for QFTs in cosmological spacetimes. These unitarity constraints, collectively known as the Cosmological Optical Theorem (COT) \cite{Goodhew:2020hob,Melville:2021lst,Goodhew:2021oqg}, provide an infinite set of relations for perturbative contributions to wavefunction coefficients, from which correlators can be computed with the well known Born rule (see also \cite{Cespedes:2020xqq,Ghosh:2024aqd,Stefanyszyn:2024msm,Goodhew:2024eup} for extensions of these rules). The COT relies on the unitarity of time evolution and the assumption of a Bunch-Davies initial state. Using the COT, one can show that party-odd correlators with any number of scalar fields must vanish at tree level in exact de Sitter \cite{Cabass:2022rhr} (see \cite{Stefanyszyn:2023qov,Thavanesan:2025kyc} for further generalisations). Many different possibilities have been considered in the literature \cite{Cabass:2022rhr,Jazayeri:2023kji,Creque-Sarbinowski:2023wmb,Stefanyszyn:2025yhq} to violate the assumptions of these theorems and hence generate non-zero parity-odd correlators, the simplest of which is the four-point function or trispectrum. Many of these signals have been searched for in the data \cite{Philcox:2021hbm,Cahn:2021ltp,Philcox:2022hkh,Philcox:2023ypl,Hou:2022wfj,Cabass:2022oap,Slepian:2025kbb}. A particularly interesting possibility is to go beyond the tree-level approximation and consider loop contributions. In \cite{Lee:2023jby}, it was indeed shown that at one-loop order the parity-odd trispectrum is not zero and remarkably it takes a surprisingly simple form, as compared to the much more complicated parity-even trispectrum at one loop. This observation highlights a unique opportunity to study an observable, namely, the parity-odd trispectrum, where the classical tree-level contribution vanishes, and the intrinsically quantum contribution is leading and possibly large,  depending on the model. This has been one of the main reasons of our interest in loops.\\

\paragraph{State of the art}  The standard Feynman rules allow one to straightforwardly write down integral expressions for cosmological correlators or wavefunction coefficients to any desired order, including any number of loops. The difficulty is that loop integrals typically lead to divergences that need to be first regulated and then renormalised. In this work, we are exclusively interested in \textit{ultraviolet (UV) divergences}, which come from the region of loop integration where some of the momenta of the propagators run off to infinity. Other divergences could emerge in the infrared, but we do not study them here as they do not occur in the large class of models that we consider, which is outlined in \ref{ModelSection}. Furthermore, there is a rich literature that studies late-time divergences associated with the behaviour of time integrals towards the future conformal boundary of de Sitter spacetime 
(see \cite{Cespedes:2023aal} and references therein). These late-time divergences are also absent in the class of models that we study here, which involve a shift-symmetric scalar field. Because all interactions involve derivatives, which redshift with time, all interactions turn off at the future conformal boundary. Loops in de Sitter have been studied for a long time, and there is already a rich literature \cite{Tsamis:1996qq,Weinberg:2005vy,Weinberg:2006ac,Senatore:2009cf,Marolf:2010zp,Marolf:2010nz,Pimentel:2012tw,Premkumar:2021mlz,Melville:2021lst,Green:2020txs,Cohen:2020php,DiPietro:2023inn,Heckelbacher:2022hbq,Chowdhury:2023khl,Lee:2023jby,Green:2024fsz,Huenupi:2024ztu,Huenupi:2024ksc,Palma:2025oux,Benincasa:2024ptf,Braglia:2025cee,Nowinski:2025cvw,Bhowmick:2024kld,Bhowmick:2025mxh, Chowdhury:2023arc,Chowdhury:2024snc} on this subject. One of our motivation for this project is the apparent tension among different regularisation schemes present in the literature. To elucidate this tension, we give here a very brief and admittedly \textit{biased} historical account.\\

In \cite{Weinberg:2005vy}, Weinberg popularised the so called in-in formalism, which had been first developed for very different reasons half a century ago by Schwinger \cite{Schwinger:1960qe} and Keldysh \cite{Keldysh:1964ud}. The main application of this formalism was the calculation of the one-loop contribution to the power spectrum. It was found that this contribution depends logarithmically on the external momentum.  If correct, this result would potentially lead to observable effects in the measured scalar spectral tilt. Key to this result was a particular proposal for how \textit{dimensional regularisation} should be extended from flat to curved spacetime. Subsequently, in \cite{Senatore:2009cf}, Senatore and Zaldarriaga criticised this result and this formulation of dimensional regularisation. One of their main points is that, on physical grounds, the correct result should not contain any logarithmic contributions because every mode experiences the same history and hence scale invariance should be preserved also at the quantum level. Furthermore, they showed that a different and more natural implementation of dimension regularisation succeeds in correctly reproducing this physical expectation. The main observation is the following. In flat spacetime, the mode functions take the form $e^{i\omega t}$ irrespectively of the number of dimensions. In curved spacetime they have a complicated dependence on the number $d$ of spacetime dimensions. For example, in de Sitter spacetime, the dimension appears in the mode function through the index of a Hankel function. Since the spirit of dimension regularisation is to extend the \textit{full} theory to $d$ spatial dimensions, consistency requires that the mode functions as well as the spacetime metric are similarly extended. This fact is inconsequential in flat spacetime but in curved spacetime it leads to very different results. We will refer to the procedure of \cite{Senatore:2009cf} simply as dimensional regularisation or \textit{dim reg} for short, while we refer the procedure in \cite{Weinberg:2005vy} as \textit{partial dim reg}, because there the mode functions were not continued to $d$ dimensions. A shortcoming of dim reg in curved spacetime is that one has to work with special functions appearing inside integrals, which makes the calculation unwieldy. A modification of dim reg was proposed by Melville and Pajer in \cite{Melville:2021lst}. There it was observed that if the mass of the field is also analytically continued together with the spacetime dimension, it is possible to ensure that the mode functions remain unchanged in the regularisation process (a similar strategy was adopted for conformally-coupled fields \cite{Green:2020txs} and in conformal field theories \cite{Bzowski:2013sza}). We refer to this procedure as \textit{mass and dimensional regularisation}, or \mdreg for short. The advantage of this procedure is that for massless and conformally-coupled fields, calculations involve only exponential and polynomial functions so that integrals are much easier to perform.\\

A first confusing aspect of the state of affairs of the literature is that it naively looks like the above three versions of dim reg give different results. This is in contrast with the expectation that all consistent regularisation schemes should give the same physical results, albeit presenting very different technical challenges.\footnote{There are certainly ``bad'' regulators (such as sharp cutoffs for gauge theories) that violate physical principles and give inconsistent results. The useful question is, of course, how to sharpen the boundary between good regulators and bad ones, which is also one of our main motivations.} A second aspect is that no one has yet studied how the choice of regularisation schemes affects the parity-odd part of correlators, or equivalently, the imaginary part of the associated wavefunction coefficient. Again, the naive applications of the above regularisations seem to lead to contradictory results for the imaginary part of wave function coefficients. This casts doubt on the overall consistency of the theory. This motivated us to perform a systematic study to understand how different renormalisation procedures work. \\

A final motivation for our work is that it may well be that some renormalisation scheme other than a version of dim reg turns out to be particularly convenient in cosmology, especially considering that a lot of computational effort has been poured into the phenomenology of loops in the early universe in recent years \cite{Chen:2016nrs,Wang:2021qez,Xianyu:2022jwk,Qin:2023bjk,Qin:2024gtr,Green:2020txs,Green:2024fsz,Bodas:2025wuk,Kristiano:2023scm,Firouzjahi:2023ahg,Beneke:2023wmt,Fumagalli:2023zzl,Choudhury:2023rks,Firouzjahi:2023aum,Frolovsky:2025qre,Inomata:2025bqw,Fang:2025vhi,Inomata:2025pqa,Braglia:2025qrb,Ema:2025ftj,Wang:2025qfh,Maru:2021ezc,Niu:2022fki,Fujita:2023inz,Reinhard:2024evr,Garcia-Saenz:2025jis}. As mentioned above, partial dim reg is at best confusing because the free and interaction parts of the theory live in different dimension, leaving the possibility to include counterterms in $3$ or $d$ dimension or both. Dim reg is surely well-defined and gives consistent results in $d$ dimensions, but the appearance of special functions makes it technically very challenging. Finally, \mdreg is just as well-defined as dim reg and removes the technical difficulties of the calculation. However, for massless fields with a shift symmetry, which are a good toy model of the curvature perturbations we measure in cosmology, the technical simplification of \mdreg comes at the cost of breaking the shift symmetry during renormalisation. We have not found this to be a problem per se, but we anticipate it may make certain general properties of the theory, such as soft theorems, less manifest. In light of this, we chose to explore new renormalisation schemes. Here we have extensively discussed how to define and utilize \textit{$\upeta$ regularisation} in cosmological spacetimes, by generalising the corresponding procedure in flat spacetime recently developed by Padilla and Smith in \cite{Padilla:2024cvk,Padilla:2024mkm}. We found that this scheme offers many advantages and will likely be useful for further study of loop contributions to cosmological correlators. 

Before we dive into the summary of our results, we note that dim-reg variants and $\upeta$ regularisation are by no means the only ways to tame loops in de Sitter. Alternative regularisation schemes include, for instance, employing the Mellin-Barnes transform \cite{Premkumar:2021mlz,Green:2024fsz,Qin:2024gtr} and hybrid \mdreg via weight-shifting operators \cite{Xianyu:2022jwk,Zhang:2025nzd}. A complete analysis is beyond the scope of this paper.\\

 
\setcounter{footnote}{0}

\paragraph{Summary of results} For the convenience of the busy modern reader, we summarise our main results below.

\begin{itemize}
    \item \textbf{The renormalised wavefunction:} We focus on theories of a single massless scalar field in exact de Sitter space, working in flat (a.k.a. Poincar\'e) slicing. The existence of a shift symmetry implies that UV divergences appear exclusively in the calculation of the one-loop wavefunction and not in the second step of averaging the wavefunction to obtain correlators. Focussing on the cubic interaction $\dot \phi^3$ as a working example, we compute the renormalised quadratic wavefunction coefficient $\psi_2$ at one loop. We perform the calculation in dim reg, mass and dimensional regularisation (\mdreg) and with $\upeta$ regulators. In Appendix \ref{partialDimRegAppendix}, we also discuss partial dim reg.
    In all cases, we find the same final renormalised late-time result
    \begin{align}
	   \psi_2 = \psi_2^{\text{tree}}+\widehat{\psi}_2^{\rm 1L}= \frac{1}{H^2}\left(\frac{i k^2}{\eta_0}-k^3\right)+ \frac{1}{15} \frac{\lambda^2 H^2}{(4\pi)^2}k^3\left(   \frac{i \pi}{2} + \log\frac{\mu}{H}\right)~.
    \end{align}
    where the first term corresponds to the well-known tree-level result and the last term is the renormalised one-loop result including counterterms. Here $\mu$ is an arbitrary constant. In agreement with \cite{Senatore:2009cf} we find that the final result does not contain terms proportional to $\log k$. To the best of our knowledge, this is the first time the full renormalised $\psi_2$ including the imaginary part was computed at one loop. 
    
    \item \textbf{Unitary $\upeta$ regulators:} We have developed for cosmological spacetimes a new class of regulators known as $\upeta$ regulators by generalising the seminal work for flat spacetime in \cite{Padilla:2024cvk,Padilla:2024mkm}. For the one-loop quadratic wavefunction, the regularisation consists in adding a window function $\upeta$ inside the loop integral. Since this function is independent of time this can be done before or after performing the time integrals over the position of the vertices. Schematically the loop contribution then takes the form
    \begin{align}
	\psi_2 \sim \int_k^\infty \frac{\d q_+}{q_+} \frac{\Poly_{m+n}(q/M)}{\Poly_{m}(q/M)}\upeta\left(\frac{q_+/k}{\Lambda/H}\right)~.
    \end{align}
    The argument of $\upeta$ is chosen to be proportional to $q_+/k$ such that the regulator respects scale invariance. Assuming $\upeta(0)=1$ and $\upeta(+\infty)=0$ regulates the integral and the regulator can be removed in the limit $\Lambda \to \infty$. We show that different choices of $\upeta$ give different results for $\Im \psi_2$, for which no unitary counterterms exist. Moreover, generic choices of $\upeta$ violate unitarity in the form of the Cosmological Optical Theorem (COT) \cite{Melville:2021lst}. We then argue that the only consistent $\upeta$ regulators are those respecting the following additional properties: for any $x \in \mathbb{C}^+$ 
    \begin{align}
        \text{Convergence: }&& \lim_{|x|\to \infty}\upeta(x)&=0\,, \\ \text{Hermitian analyticity: }&& \upeta(x)&=[\upeta(-x^*)]^*\,.
    \end{align}
    which we dub \textit{unitary and analytic} $\upeta$ regulators. We prove that all unitary and analytic $\upeta$ regulators respect the COT and give the same value for $\Im \psi_2$.
    
    \item \textbf{Regulator comparison:} While dim reg, \mdreg and the unitary and analytic $\upeta$ regulators all give the same final renormalised $\psi_2$ at one loop, they differ in the intermediate step of the calculation. In Table \ref{regulatorBenchmarkTable} we summarise this for the regulators we have considered. There are a few notable facts. In dim reg the loop contribution violates 3-dimensional scale invariance, while respecting $d$-dimensional scale invariance by construction. This violation, indicated by \xmark~in Table \ref{regulatorBenchmarkTable}, manifests itself in the presence of a $\log k$ in the one-loop contribution, which is precisely cancelled by a counterterm, a fact which is indicated by \xmark $\to$ \cmark. This appears to be in agreement with \cite{Senatore:2009cf,Bhowmick:2024kld,Braglia:2025cee}. An analogous phenomenon takes place in \mdreg where the intermediate result contains a $\log \eta$ contribution, which is cancelled by counterterms.
    
    \item \textbf{Universality of the imaginary part:} We find that under the constraint of unitarity and analyticity, all renormalised one-loop wavefunction coefficients yield the \textit{same} imaginary part relative to the logarithmic running of their real part,
    \begin{align}
        \left(\mu\frac{\partial}{\partial\mu}-\frac{2}{\pi}\Im\right) \widehat{\psi}_n^{\rm 1L}=0~,
    \end{align}
    assuming scale invariance and the Bunch-Davies vacuum. This relation holds for arbitrary diagram topology with any scale-invariant and IR-convergent interaction vertices and any number of bulk fields whose masses lie in the complementary series representation of the de Sitter group. It does not assume de Sitter boost invariance and actually incorporates cases with non-trivial sound speed and chemical potential. The universality of the emergent imaginary part fixes parity-odd correlators and points to an underlying connection to the renormalisation group flow of QFTs in de Sitter.
\end{itemize}

\begin{table}[h]
	\centering
	\begin{tabular}{c|cccc}
		\toprule[1.1pt] 
		  Regularisation schemes \quad &\quad Scale inv.  \quad & \quad Im. part \quad & \quad Time-indep.\quad & \quad Diff inv.\\ 
		\midrule[1pt]
          Dim reg \quad &\quad \xmark $\,\to$ \cmark  \quad &\quad $0\to +\pi/2$ \quad &\quad \cmark \quad & \quad \cmark\\[1ex]
          \mdreg \quad &\quad \cmark  \quad &\quad $+\pi/2$\quad &\quad \xmark $\,\to$ \cmark\quad & \quad \cmark\\[1ex]
		 $\upeta$ reg (unitary \& analytic)\quad &\quad \cmark  \quad &\quad $+\pi/2$\quad &\quad \cmark\quad & \quad \cmark\\[1ex]
         \midrule[1pt]
		 $\upeta$ reg (real)\quad &\quad \cmark  \quad &\quad 0\quad &\quad \cmark\quad & \quad \cmark\\[1ex]
         $\upeta$ reg (complex)\quad &\quad \cmark  \quad &\quad Arbitrary\quad &\quad \cmark\quad & \quad \cmark\\[1ex]
         Partial dim reg (scheme A)\quad &\quad \xmark  \quad &\quad 0\quad &\quad \cmark\quad & \quad \xmark\\[1ex]
         Partial dim reg (scheme B) \quad &\quad \xmark$\,\to$ \cmark  \quad &\quad $0\to +\pi/2$ \quad &\quad \cmark\quad & \quad \xmark\\[1ex]
		\bottomrule[1.1pt]
	\end{tabular}
	\caption{A benchmark of the regularisation schemes when applied to the two-point wavefunction coefficient at one-loop order. $a\to b$ means $a$ before renormalisation and $b$ after renormalisation. Entries without an arrow indicates that the property remains unchanged after renormalisation. \textit{Upper rows}: Examples of unitary and analytic regulators. They land on the same scale-invariant and time-independent result with a common imaginary part $+\pi/2$ relative to the coefficient of logarithmic divergences, consistent with the cosmological optical theorem. \textit{Lower rows}: Examples of regulators that \textit{do not} respect unitarity or analyticity. Apparently, these regulators give inconsistent results that strongly depend on the regulator choice while also violating some of the fundamental principles.}\label{regulatorBenchmarkTable}
\end{table}


\paragraph{Notations and conventions}
Throughout this paper, we work with the $(-, +, +, +)$ metric sign convention, and use the following conventions for Fourier transforms
\begin{align}
    f(\bfx)=\int \frac{\d^3k}{(2\pi)^3} e^{i\bfk\cdot\bfx}f(\bfk)\equiv \int_{\mathbf{k}} e^{i\bfk\cdot\bfx}f(\bfk) \,,
\end{align}
where bold letters represent three-dimensional spatial vectors. We shall work in the Poincaré patch of de Sitter space with the metric
\begin{align}
	\d s^2=a^2(\eta)(-\d\eta^2+\d\mathbf{x}^2)~,\quad a(\eta)=-\frac{1}{H\eta}~.
\end{align}
The cosmological wavefunction evaluated at a late-time boundary $\eta=\eta_0$ is parametrised by the exponentiation of a power series of boundary fields multiplied by the corresponding wavefunction coefficients,
\begin{align} 
	\label{WFU}
	\Psi[\bar\phi] = \exp\left[+\sum_{n=2}^{\infty}\frac{1}{n!} \int_{\mathbf{k}_{1} \cdots \mathbf{k}_{n}} \psi_{n}(\{ k \}, \{ \mathbf{k} \}) (2 \pi)^3 \delta^3 \bigg(\sum_{{\sf a}=1}^n {\mathbf{k}}_{\sf a} \bigg)\bar\phi(\mathbf{k}_{1}) \cdots \bar\phi(\mathbf{k}_{n}) \right] \,\,,
\end{align}
where $\bar\phi\equiv\phi(\eta_0)$ denotes the boundary value of the field. The bulk-boundary and bulk-bulk propagators in the wavefunction formalism are given by
\begin{subequations}
    \begin{align}
	K(\eta,k)&=\frac{\phi^+(\eta,k)}{\phi^+(\eta_{0},k)}~,\\
	G(\eta_1,\eta_2;k)&=P(k)\left[\Big(K^*(\eta_1,k)K(\eta_2,k)\theta(\eta_1-\eta_2)+(\eta_1\leftrightarrow\eta_2)\Big)-K(\eta_1,k)K(\eta_2,k)\right]~,\label{GdefIntro}
    \end{align}
\end{subequations}\label{propDefs}
with $P(k)=| \phi^+(\eta_0,k)|^2$ denoting the free-theory power spectrum. We adopt the amplitude Feynman rules by including a factor of $i$ for every vertex (our bulk-bulk propagator therefore differs from that in \cite{Goodhew:2020hob,Goodhew:2021oqg,Melville:2021lst} by a factor of $i$). When performing Wick rotations, we denote the Euclidean conformal time by $\chi>0$, which is related to the Wick-rotated Lorentzian conformal time via
\begin{align} \label{WickRotation}
	\eta=e^{i\pi/2}\chi~.
\end{align}
Very often we will be working with complexified momenta that live in different parts of the complex plane,
\begin{subequations}
    \begin{align}
        \text{Upper half:}&\quad \mathbb{C}^+\equiv \{a+i b: b>0\}~,\\
        \text{Lower half:}&\quad \mathbb{C}^-\equiv \{a+i b: b<0\}~.
    \end{align}
\end{subequations}
The discontinuity operation is defined by
\begin{align}
    \nonumber &\underset{k_1\cdots k_j}{\Disc} \left[f(k_1,\cdots, k_n;\{p\};\{\mathbf{k}\})\right] \\
    &\equiv f(k_1,\cdots,k_j,.,k_n;\{p\};\{\mathbf{k}\})- \Big[f(k_1,\cdots,k_j,-k_{j+1}^*,...,-k_n^*;\{p\};\{-\mathbf{k}\})\Big]^*~,~ k_{\sf a}\in \mathbb{C}^-~.\label{DiscDef}
\end{align}
We use the following definition for the Mellin transform:
\begin{align}
    C_s[f]\equiv \int_{0}^{\infty} \d x\, x^{s-1} f(x)~.
\end{align}


\section{Prelude: setting the model and questions}\label{ModelSection}

In this section, we begin by introducing our toy model for the loop analysis and then we specify the relevant observables and the questions we shall be interested in. We consider the Effective Field Theory (EFT) of a single massless scalar field $\phi$ with unit sound speed living in $(3+1)$-dimensional de Sitter space. To single out the physics of UV divergences, we demand the IR convergence of the interactions. This is most easily achieved by imposing a shift symmetry $\phi\to \phi+C$ and thus requiring the scalar to be derivatively coupled,
\begin{equation}\label{toy}
   S = \int \d^4 x ~\sqrt{-g}\left[-\frac{1}{2} g^{\mu\nu}\partial_\mu \phi \partial_\nu \phi + \mathcal{L}_{\rm int}(\partial\phi,\partial^2\phi,\cdots)\right]~.
\end{equation}
We will first perform a case study on the two-point wavefunction coefficient $\psi_2$ and move to general lessons later. Henceforth, we single out the simplest shift-symmetric cubic coupling,
\begin{align}
    \mathcal{L}_{\rm int}(\partial\phi,\partial^2\phi,\cdots)=\frac{\lambda}{3!} (-n^\mu\partial_\mu\phi)^3+\cdots~,\label{interactionLagrangianIn3d}
\end{align}
and compute $\psi_2$ up to order $\mathcal{O}(\lambda^2)$. Here $n_\mu = a(\eta)\delta_\mu^0$ is a future directed normal vector satisfying $n^\mu n_\mu = -1$. 


\paragraph{Quantum vs classical loops.}

In the Schrödinger picture of quantum field theory, the state of the field at a late time $\eta_0\to 0$ is characterised by the Bunch-Davies cosmological wavefunction,
\begin{align}
    \Psi[\bar{\phi}]=\int_{\phi(-\infty)=0}^{\phi(\eta_0)=\bar{\phi}} [\mathcal{D}\phi] \, e^{i S[\phi]}~.\label{WFUDefinSect2}
\end{align}
Correlators of observables are then derived from the (un-normalised) cosmological wavefunction via the Born rule,
\begin{align}
    \ex{O( \bar{\phi}, \bar{\pi} )}=\frac{\int\mathcal{D} {\bar\phi} \,\Psi^*[\bar\phi] ~ O\left( \bar\phi, -i\frac{\delta}{\delta{\bar\phi}}\right)\, \Psi[\bar\phi]}{\int\mathcal{D}{\bar\phi}\, \Psi^*[\bar\phi] \Psi[\bar\phi]}~,\label{BornRule}
\end{align}
where $\bar{\pi}$  denotes the canonical momentum conjugate of the field $\phi$ evaluated at the boundary $\eta=\eta_0$. The wavefunction is often parametrised by the exponential of a power series of boundary fields. In momentum space, this reads (see also \eqref{WFU})
\begin{align}
    \nonumber\log \Psi&=+\frac{1}{2!}\int_{\mathbf{k}}\psi_2(k)\bar{\phi}(\mathbf{k}) \bar{\phi}(-\mathbf{k})\\
    &\quad+\frac{1}{3!}\int_{\mathbf{k}_1,\mathbf{k}_2,\mathbf{k}_3}\psi_3(k_1,k_2,k_3)\,(2\pi)^3 \delta^3(\mathbf{k}_1+\mathbf{k}_2+\mathbf{k}_3)\bar{\phi}(\mathbf{k}_1)\bar{\phi}(\mathbf{k}_2)\bar{\phi}(\mathbf{k}_3)+\cdots~,
\end{align}
where the wavefunction coefficients $\psi_n(\{k\},\{\mathbf{k}\})$ can be computed order-by-order in perturbation theory using diagrammatics.\footnote{In this paper, all wavefunction coefficients are in momentum space unless otherwise stated.} For instance, one can organise the diagrammatic expansion in the number of loops,
\begin{align}
    \psi_n=\psi_n^{\rm 0L}+\widehat{\psi}_n^{\rm 1L}+\widehat{\psi}_n^{\rm 2L}+\cdots~,
\end{align}
where $\widehat{\cdot}$ denotes the renormalised wavefunction coefficients at each loop order after adding the corresponding counterterms. We stress that these loop contributions are inherently quantum as they come from performing the quantum path integral in \eqref{WFUDefinSect2}, which pushes the evolution history away from the classical saddle at $\delta S=0$. Consequently, these loops are known as \textit{quantum loops}. At two-point level, it is well-known that the free theory of a massless scalar yields \cite{Guven:1987bx,Maldacena:2002vr}
\begin{align}
    \psi_2^{\rm 0L}=\frac{ik^2}{H^2\eta_0(1-ik\eta_0)}=-\frac{1}{H^2}\left(k^3-\frac{i k^2}{\eta_0}+\mathcal{O}(\eta_0)\right)
\end{align}
in the late-time limit $\eta_0\to 0$, giving rise to a scale-invariant scalar power spectrum at tree level, 
\begin{align}
    \ex{\phi(\mathbf{k})\phi(-\mathbf{k})}_{\rm 0L}'=\frac{1}{-2\Re \psi_2^{\rm 0L}(k)}=\frac{H^2}{2k^3}~.\label{powerspectrumAt0L}
\end{align}
At higher orders, the power spectrum receives contribution from both quantum loops (e.g. $\widehat{\psi}_n^{\rm 1L}$) and loops involving the contraction of tree-level wavefunction coefficients\footnote{Here we have dropped a term proportional to $\Re \psi_3(k,k,0)$, since this vanishes identically for the class of theories we will consider (see \cite{Pueyo:2024twm} around Eq. 3.21 for a brief discussion).},
\begin{align}
    \nonumber\ex{\phi(\mathbf{k})\phi(-\mathbf{k})}_{\rm 1L}'&=\frac{1}{(2\Re \psi_2^{\rm 0L}(k))^2}\Bigg[2\Re \widehat{\psi}_2^{\rm 1L}(k)\\
    &\quad -\frac{1}{2}\int_{\mathbf{q}}\frac{2\Re \psi_4^{\rm 0L}(k,q,q,k)}{2\Re \psi_2^{\rm 0L}(q)}+\frac{1}{2}\int_{\mathbf{q}_1,\mathbf{q}_2}\frac{[2\Re \psi_3^{\rm 0L}(k,q_1,q_2)]^2}{2\Re \psi_2^{\rm 0L}(q_1)\,2\Re \psi_2^{\rm 0L}(q_2)}\Bigg]~.\label{powerspectrumAt1L}
\end{align}
The loops encountered in averaging the wavefunction over the boundary configurations $\phi(\bfx)$ to compute $\phi$ correlators are purely classical in nature and are hence known as \textit{classical loops}. These are completely analogous to the averages over the noise, as encountered in open dissipative theories or in solving stochastic differential equations. These classical loops are very familiar to cosmologists as they are the main object of study in perturbative approaches to the clustering of dark matter, as for example in standard perturbation theory \cite{Bernardeau:2001qr} or the effective field theory of large scale structures \cite{Baumann:2010tm,Carrasco:2012cv}. The renormalisation of these classical loops has been the object of study for a long time, including using dim reg, see e.g. \cite{Scoccimarro:1995if,Pajer:2013jj}. In contrast, the loops encountered in the calculation of the wavefunction coefficients are all quantum loops and are associated with contributions to the path integral beyond the saddle approximation. In general, both classical and quantum loops may display ultraviolet divergences, which in turn need to be regulated and removed by counterterms as we will do in this paper. For a recent discussion of classical and quantum loops in the context of late-time divergences in de Sitter see \cite{Cespedes:2023aal}.

In this work, we have chosen to work with the model \eqref{toy}, which enjoys a shift symmetry. Because of this shift symmetry, one can check that the classical loops from averaging over $\phi(\bfx)$ turn out to be UV-convergent. To see this, note that for the loop integrals above the integrand is proportional to $[\Re \psi_3^{\rm 0L}(k,q_1,q_2)]^2$ or $\Re \psi_4^{\rm 0L}(k,q,q,k)$. The UV-limit of the integral corresponds to $q,q_{1,2}\to \infty$. Thank to the scale covariance of $\psi_n$ we can trade this limit for the limit $k\to 0$, which corresponds to one or two external legs being very soft. The shift symmetry ensures that in this limit $\psi_3 \sim (k/q)^2$ and $\psi_4\sim k^4/q $ (see the explicit result in \eqref{psi3psi4}). Power counting then shows that the integral is convergent in the UV. In summary, \textit{only the quantum loops in the wavefunction require regularisation and renormalisation}. The renormalised wavefunction derived in this paper therefore can be used to directly compute the renormalised correlators, using the standard formulae, without the need of any additional regularisation or renormalisation. We postpone to future work the interesting study of theories in which an additional renormalisation step is required to go from the wavefunction to correlators. 

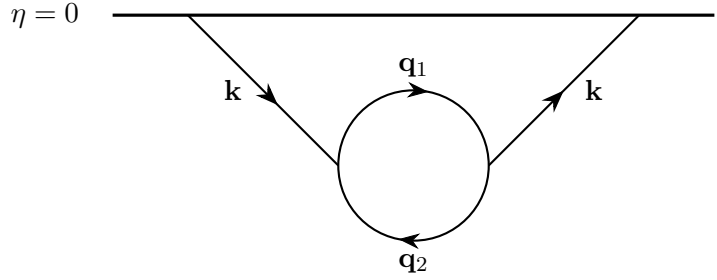
\begin{figure}[ht]
  \centering
  \begin{tikzpicture}[scale=1]

    \draw[thick] (0,0) circle (1cm);

    \draw[postaction={decorate},
          decoration={markings, mark=at position 0.6 with {\arrow[scale=1.3]{Stealth}}},thick]
          (-3,2)-- (-1,0); 
    \node at (-2.4,1.0) {\(\mathbf{k}\)}; 

    \draw[postaction={decorate},
          decoration={markings, mark=at position 0.5 with {\arrow[scale=1.3]{Stealth}}},thick]
          (1,0) -- (3,2); 
    \node at (2.4,1.0) {\(\mathbf{k}\)}; 

    \draw[-{Stealth[length=8pt, width=6pt]}] (0,1) -- (0.2,0.98);    
    \draw[-{Stealth[length=8pt, width=6pt]}] (0,-1) -- (-0.2,-0.98); 

    \node at (0,1.3) {\(\mathbf{q}_1\)};  
    \node at (0,-1.3) {\(\mathbf{q}_2\)}; 

    \draw[very thick] (-4,2) -- (4,2);

    \node at (-4.9,2.0) {\(\eta = 0\)};

  \end{tikzpicture}
  \caption{The quantum one-loop diagram that contributes to the two-point wavefunction coefficient $\psi_2^{\rm 1L}$.}
  \label{figloop}
\end{figure}

\paragraph{The imaginary can be real.}
Another notable fact in \eqref{powerspectrumAt0L} and \eqref{powerspectrumAt1L} is that only the real part of $\psi_n$ contributes to the scalar power spectrum. However, this does not necessarily imply that the imaginary part of all wavefunction coefficients are unobservable and thereby unphysical. There are at least three cases in which the imaginary part of the wavefunction coefficients is important or even directly observable:
\begin{itemize}
    \item {\it Parity-odd field correlators}. When parity violation is present in the scalar EFT, the parity-odd component of field correlators is purely imaginary in momentum space, and is directly related to the imaginary part of wavefunction coefficients \cite{Liu:2019fag,Cabass:2022rhr,Stefanyszyn:2023qov},
    \begin{align}
        \ex{\phi^n}_{\rm PO}\supset \frac{2i\,\Im (\psi_n)_{\rm PO}}{(-2\Re \psi_2)^n}~.\label{POCorrelatorHasIm}
    \end{align}
    A measurement of the parity-odd correlators, such as for example in \cite{Philcox:2021hbm,Cahn:2021ltp,Philcox:2022hkh,Philcox:2023ypl,Hou:2022wfj,Cabass:2022oap,Slepian:2025kbb}, is thus a direct measurement of the imaginary part of wavefunction coefficients.
    
    \item {\it Conjugate momentum correlators}. The conjugate momentum $\pi$ is, in principle, an observable equally physical as the field value $\phi$. And their mixed correlators typically involve the imaginary part of the wavefunction coefficients e.g. \cite{Guven:1987bx, Cespedes:2020xqq}
    \begin{align}
        \ex{\phi\pi}\supset \frac{\Im \psi_2}{-2\Re \psi_2}~.\label{phipiCrossCorrelator}
    \end{align}
    Such conjugate momentum correlators, despite being challenging to measure in reality (due to the exponential decay of $\dot\phi$ outside the horizon), should be treated as physical observables nevertheless. In addition, these conjugate momentum correlators contribute to the time gradient of the power spectrum via \cite{Melville:2021lst}
    \begin{align}
    	\lim_{\eta \to 0} \frac{1}{3!}\partial_\eta^3 P(k,\eta)=-\frac{H^4}{3k^3}\Im \psi_2^{\rm 1L}~.
    \end{align}
    
    \item {\it Decoherence rates}. The imaginary part of wavefunction coefficients also contributes to the decoherence of perturbations outside the horizon. In the perturbative regime, the decoherence rate can be schematically characterised by \cite{Nelson:2016kjm,Pueyo:2024twm,Burgess:2024eng}
    \begin{align}
        |\gamma-1|\sim-\log D_2 \supset   \int \frac{\left|\Re\psi_3\right|^2+\left|\Im\psi_3\right|^2}{(-2\Re \psi_2)^3}~.
    \end{align}
    where $\gamma$, $D_2$ are the purity and the off-diagonal element of the one-mode reduced density matrix.
\end{itemize}
Interestingly, by scale invariance, Bunch-Davies vacuum and unitarity, one can show that all tree-level wavefunction coefficients are purely real \cite{Cabass:2022rhr,Stefanyszyn:2023qov,Goodhew:2024eup}. Combining with the previous observation on quantum loops, we thus conclude that \textit{the imaginary part of wavefunction coefficients are physically observable, and can only come from UV-divergent quantum loops in our EFT model}.\footnote{The fact that UV divergences can give rise to a finite imaginary parity-odd correlator was found previously in \cite{Lee:2023jby}.} This then leads to the following questions: 
\begin{enumerate}
    \item How is the imaginary part of wavefunction coefficients determined from the regularisation and renormalisation of quantum loops?
    \item Do different regularisation schemes give consistent results after renormalisation?
    \item If they do not agree in general, are there constraints that uniquely determine the imaginary part from fundamental principles?
\end{enumerate}
To address the above questions, we will investigate the two-point wavefunction coefficient $\psi_2$ up to one-loop order i.e. $\mathcal{O}(\lambda^2)$. The quantum loop contribution comes from the Feynman diagram in Figure \ref{figloop}. Notice that we are not including diagrams with a tadpole attached. This is allowed because such diagrams involve a propagator of zero momentum. When attached to a $\dot\phi^3$ interaction this vanishes. For curvature perturbations in the presence of dynamical gravity and for more general non-derivative interactions these tadpole terms have been studied in \cite{Senatore:2009cf,Pimentel:2012tw,Braglia:2025cee}. There it was noticed that non-linearly realised symmetries imply that tadpole counterterms are accompanied by associated quadratic counterterms, which in turn are responsible for crucial cancellations of divergences and time dependence. 

In the following sections and in Appendix \ref{partialDimRegAppendix}, we will apply to this diagram three variants of dimensional regularisation and then propose a new infinite family of $\upeta$ regulators in de Sitter. Our main target will be their resulting imaginary parts.

\section{Dimensional regularisation}\label{dimRegSection}

Dimensional regularisation (dim reg) is perhaps the most familiar scheme used in modern quantum field theory. However, unlike Minkowski space, the analytic continuation of space dimensions of de Sitter space is subtle because the geometry depends non-trivially on the space dimensionality $d$.\footnote{For instance, the Ricci scalar in $(d+1)$-dimensional de Sitter space is $R=d(d+1) H^2$.} In addition, the dynamics of fields living in de Sitter also depends on $d$ through the mode functions and interactions. This then allows for multiple variants of dim reg in de Sitter, each corresponding to a different deformation of a massless scalar field in $(3+1)$-dimensional de Sitter. In this section, we discuss the scheme used by Senatore and Zaldarriaga in \cite{Senatore:2009cf}. We refer to this scheme simply as dimensional regularisation, or \textit{dim reg} for short, since it is the closest in spirit to the procedure in Minkowski \cite{Bollini:1972ui,tHooft:1972tcz}.

In dim reg, one analytically continues the space dimension to $d=3-\epsilon$ while holding the Lagrangian mass of the scalar field fixed at $m^2=0$. The mode function then takes the following form
\begin{align}
    \phi^+(\eta,k)=-i\frac{\sqrt{\pi}}{2} H^{-1/2} (-H \eta)^{d/2} H_{d/2}^{(2)}(-k\eta)~.
\end{align}
where $H_\nu^{(2}(z)$ denotes the Hankel function of the second kind. The bulk-boundary propagator reads
\begin{align}\label{btbmd}
	K(\eta,k)=\frac{\phi^+(\eta,k)}{\phi^+(\eta_0,k)}\approx \frac{-i\pi}{2^{d/2}\Gamma(d/2)} (-k\eta)^{d/2} H_{d/2}^{(2)}(-k\eta)~.
\end{align}
The bulk-bulk propagator is given by
\begin{align}\label{butobumd}
	G(\eta_1,\eta_2;q)=\frac{\pi H^{d-1}}{2}(-\eta_1)^{d/2} (-\eta_2)^{d/2}\left[J_{d/2}(-q\eta_1) H_{d/2}^{(2)}(-q\eta_2)\theta(\eta_1-\eta_2)+(\eta_1\leftrightarrow\eta_2)\right]~.
\end{align}
where $J_{\nu}(z)$ is the Bessel-$J$ function. Crucially, both propagators become independent of the boundary time in the late-time limit $\eta_0\to 0$. 

The interaction vertex \eqref{interactionLagrangianIn3d} becomes
\begin{align}
	S_{\rm  int}\supset 
	\int \frac{\d\eta\, \d^d x}{(-H\eta)^{d-2}} \,\frac{\lambda_d}{3!}\,[\partial_\eta \phi(\eta,\mathbf{x})]^3  ~,\quad \lambda_d\equiv \mu_0^{(3-d)/2}\lambda~,
\end{align}
where we have inserted a factor $\mu_0^{(3-d)/2}$ as usually done in dim reg to keep the mass dimension of the coupling fixed i.e. $\lambda\sim [\rm mass]^{-2}$. This arbitrary energy scale $\mu_0>0$ is unphysical and will eventually be removed after renormalisation.

In the rest of this section, we compute and renormalise the one-loop contribution to $\psi_2$ in dim reg and then comment on some of the properties of this regularisation scheme. 


\subsection{Loop contributions}\label{masslessDimRegLoopSubsection}
The one-loop contribution to $\psi_2$ arising from the diagram in Figure \ref{figloop} is given by
\begin{align}
	\nonumber\psi_2^{\rm 1L}=-\frac{\lambda_d^2}{2 H^{2d-4}}\int\frac{\d^{d} q_1}{(2\pi)^d}&\int_{-\infty}^{0} \frac{\d\eta_1}{(-\eta_1)^{d-2}} \frac{\d\eta_2}{(-\eta_2)^{d-2}} \\
	\times & \partial_{\eta_1} K(\eta_1,k) \partial_{\eta_2} K(\eta_2,k) \partial_{\eta_1} \partial_{\eta_2} G(\eta_1,\eta_2; q_1)\, \partial_{\eta_1} \partial_{\eta_2} G(\eta_1,\eta_2; q_2)~.\label{toSenatoreZalda}
\end{align}
Before proceeding two comments are in order. First, we stress that we continued \textit{all} propagators to $d$ dimensions, including those corresponding to external lines. This is in contrast with what was done in \cite{Senatore:2009cf}, where the external propagators remained fixed to 3 dimensions. As discussed there, the two choices give the same result as long as one consistently uses the same external propagators in the calculation of the counterterms. For us, this means we will use $d$-dimensional propagators for the counterterms, which instead were computed in 3 dimensions in \cite{Senatore:2009cf}. While the final, renormalised results is the same, the intermediate steps are different. We find a $\log k$ contribution from the loop, which is then cancelled by a corresponding $\log k$ contribution in the counterterm. In contrast, this $\log k$ is  absent from both the loop and the counterterms in the prescription of \cite{Senatore:2009cf}. The presence of a $\log k$ term in dim reg, where everything is continued to $d$ dimensions precisely to respect scale invariance, may be confusing and so we briefly elaborate on this point. In 3 dimensions, one may define scale invariance as the requirement that $F_3(k)\equiv \psi_2(k)/k^3$ is invariant under rescaling, $F_3(\alpha k)=F_3(k)$. Conversely, in $d=3-\epsilon$ dimension, one would require that $F_d(k)=\psi_2(k)/k^d$ is invariant, $F_d(\alpha k)=F_d(k)$. In the absence of divergences, these two conditions would agree with each other as $\epsilon\to0$. However, since $\psi_2\sim 1/\epsilon$, in the limit $\epsilon\to 0$ one finds that $F_3$ and $F_d$ differ by a finite term proportional to $\log k$. Dim reg guarantees by construction that the regulated loop contribution is scale invariant in $d$ dimensions, but not in $3$ dimensions. In the end, as  we show in Sec. \ref{masslessDimRegRenormSubsection} and further discuss around \eqref{furtherscaleinv}, these $\log k$ are exactly cancelled by counterterms, leaving a final renormalised $\widehat\psi_2$ that is scale invariant in 3 dimensions. The intermediate presence of $\log k$ terms was also noticed recently in \cite{Bhowmick:2024kld,Braglia:2025cee}.

Second, we note that we can already extract some useful information by inspecting the phase and the $k$-scaling of the integrand. To make manifest the phase information, we perform a Wick rotation $\eta_i = e^{i\pi/2}\chi_i$ to obtain
\begin{align}\label{psi2step}
	\nonumber\psi_2^{\rm 1L}= \frac{\lambda_d^2 \,e^{i\pi(d-2)}}{2 H^{2d-4}}\int&\frac{\d^{d} q_1}{(2\pi)^d} \int^{\infty}_{0} \frac{\d\chi_1 \d\chi_2}{(\chi_1 \chi_2)^{d-2}}\\
	\times & \partial_{\chi_1} K(i\chi_1,k) \partial_{\chi_2} K(i\chi_2,k) \partial_{\chi_1} \partial_{\chi_2} G(i\chi_1,i\chi_2; q_1)\, \partial_{\chi_1} \partial_{\chi_2} G(i\chi_1, i\chi_2; q_2).
\end{align}
Under the Wick rotation, the bulk-boundary and bulk-bulk propagators analytically continue to
\begin{align}\label{analybb}
 	K(i\chi,k) & = \frac{1}{2^{d/2-1}\Gamma(d/2)} (k\chi)^{d/2}  K_{d/2}(k\chi)~\,,  \\
 	G(i\chi_1,i\chi_2;q) &= i e^{-\frac{i\pi d}{2}} H^{d-1}(\chi_1\chi_2)^{d/2} \left[I_{d/2}(q\chi_1) K_{d/2}(q\chi_2)\theta(\chi_2-\chi_1)+(\chi_1\leftrightarrow\chi_2)\right]~\,,
\end{align}
where $I_{\nu}(z)$ and $K_{\nu}(z)$ are the modified Bessel functions of the first and second kind, respectively and we used 
\begin{align}
	J_{\nu}(-i z) &= e^{-\frac{\nu \pi i}{2}}I_\nu(z)\,, &
	H_{\nu}^{(2)}(-i z) &= - \frac{2}{\pi i}e^{\frac{\nu \pi i}{2}} K_\nu(z)~.  
\end{align}
Therefore, we note the Wick-rotated propagators satisfy the following phase rules,
\begin{subequations}\label{argdimreg}
    \begin{align}
	\arg [K(i\chi,k)] &= 0~,  \\
	\arg [G(i\chi_1,i\chi_2;q)] &=  -\frac{\pi (d-3)}{2}~.
    \end{align}
\end{subequations}
Since $I_{\nu}(z)$ and $K_{\nu}(z)$ are real and positive for real and positive arguments, we find that in dim reg the 1-loop wavefunction coefficient satisfies 
\begin{align}
	\arg [\psi_2^{\rm 1L}]= 0~.
\end{align}
The $k$-dependence of $\psi_2^{\rm 1L}$ can be made explicit by rescaling the variables to $\chi_i=\tilde{\chi}_i/k$ and $q_i=\tilde{q}_i k$. Noting
\begin{align}
	K\left(\frac{i \tilde{\chi}}{k},k\right) &=  \frac{1}{2^{d/2-1}\Gamma(d/2)} \tilde{\chi}^{d/2}  K_{d/2}(\tilde{\chi})~,  \\
	G\left(\frac{i\tilde{\chi}_1}{k},\frac{i\tilde{\chi}_2}{k};\tilde{q} k\right) &= i e^{-\frac{i\pi d}{2}} k^{-d} H^{d-1}(\tilde{\chi}_1 \tilde{\chi}_2)^{d/2} \left[I_{d/2}(\tilde q \tilde{\chi}_1) K_{d/2}(\tilde q \tilde{\chi}_2)\theta(\tilde{\chi}_2-\tilde{\chi}_1)+(\tilde{\chi}_1\leftrightarrow \tilde{\chi}_2)\right]~,
\end{align}
 we find that $\psi_2^{\rm 1L}$
 scales as $k^d$. This scaling is consistent with expectations and the result does not depend on the boundary time $\eta_0$. Consequently, we expect that after expanding around $d=3 - \epsilon$, the result takes the following form:
\begin{align}
	\psi_2^{\rm 1L}&=\mu_0^{3-d} k^d \left(\frac{f_0}{3-d} + \text{finite \& real const.}\right)=f_0 k^3\left(\frac{1}{\epsilon}-\log \frac{k}{\mu_0}+\text{real \& finite const.} \right)~,
\end{align}
where $f_0$ is a finite and real constant to be computed. Notice that the $\log k$ piece inevitably appears as required by scale invariance in $d$ spatial dimensions. As mentioned above, this differs from \cite{Senatore:2009cf} due to the usage of different bulk-boundary propagators. This technical difference will be disappear in the final renormalised result.

To compute the coefficient $f_0$, we note that it suffices to approximate the phase- and $k$-scaling-stripped integrand up to the leading order in $\epsilon$ and track only the divergent terms,
\begin{align}
	\nonumber\psi_2^{\rm 1L}&=\frac{\lambda_d^2 H^2 \,k^d}{2}\int\frac{\d^{d} \tilde{q}_1}{(2\pi)^d} \Bigg[e^{i\pi(d-2)}k^{2d} H^{2-2d}\int^{\infty}_{0} \frac{\d\tilde{\chi}_1 \d\tilde{\chi}_2}{(\tilde{\chi}_1 \tilde{\chi}_2)^{d-2}}  \partial_{\tilde{\chi}_1} K\left(\frac{i\tilde{\chi}_1}{k},k\right) \partial_{\tilde{\chi}_2} K\left(\frac{i\tilde{\chi}_2}{k},k\right) \\
	&\qquad\qquad\qquad\qquad\qquad\quad\times\partial_{\tilde{\chi}_1} \partial_{\tilde{\chi}_2} G\left(\frac{i\tilde{\chi}_1}{k},\frac{i\tilde{\chi}_2}{k};\tilde{q}_1 k\right)\, \partial_{\tilde{\chi}_1} \partial_{\tilde{\chi}_2} G\left(\frac{i\tilde{\chi}_1}{k},\frac{i\tilde{\chi}_2}{k};\tilde{q}_2 k\right)\Bigg]\label{psi2dimRegf0computationStep1}\\
	&=\frac{\lambda_d^2 H^2\, k^d}{8}\int\frac{\d^{d} \tilde{q}_1}{(2\pi)^d} \tilde{q}_1 \tilde{q}_2 \int_{0}^\infty \d\tilde\chi_1 \d\tilde\chi_2\, \tilde\chi_1^2 \tilde\chi_2^2 \Bigg[e^{-(1+\tilde q_1+\tilde q_2)(\tilde \chi_1+\tilde \chi_2)}\nonumber\\
	&\qquad\qquad\qquad\qquad\qquad\quad+e^{-(\tilde\chi_1+\tilde\chi_2)}e^{(\tilde q_1+\tilde q_2)(\tilde\chi_1-\tilde\chi_2)}\left(1-e^{-2 \tilde q_1\tilde \chi_1}-e^{-2 \tilde q_2 \tilde \chi_1}\right)\theta(\tilde \chi_2-\tilde \chi_1) \nonumber\\
	\nonumber&\qquad\qquad\qquad\qquad\qquad\quad+e^{-(\tilde\chi_1+\tilde\chi_2)}e^{(\tilde q_1+\tilde q_2)(\tilde\chi_2-\tilde\chi_1)}\left(1-e^{-2 \tilde q_1\tilde\chi_2}-e^{-2 \tilde q_2 \tilde\chi_2}\right)\theta(\tilde\chi_1-\tilde\chi_2) \Bigg]\\
	&\qquad\qquad\quad+ \text{finite \& real}~,\label{psi2dimRegf0computationStep2}
\end{align}
where in the second step \eqref{psi2dimRegf0computationStep2}, we have isolated the leading piece by setting $d=3$ explicitly in the square bracket of \eqref{psi2dimRegf0computationStep1}. The higher-order corrections are finite and real. Finishing the integrals over $\tilde{\chi}_1$ and $\tilde{\chi}_2$, we obtain
\begin{align}
    \nonumber \psi_2^{\rm 1L}=&\,\frac{\lambda_d^2 H^2\, k^d}{16} \int\frac{\d^{d} \tilde{q}_1}{(2\pi)^d} \Bigg[\tilde q_1 \tilde q_2\frac{3(\tilde q_1+\tilde q_2)^2+9(\tilde q_1+\tilde q_2)+8}{(1+\tilde q_1+\tilde q_2)^3}\\
    &\quad +\frac{1}{(1+\tilde q_1+\tilde q_2)^3} \left(\frac{\Poly_{4}}{(1+\tilde q_1)^5}+\frac{\Poly_{4}}{(1+\tilde q_2)^5}\right)
    +\frac{\Poly_{2}}{(1+\tilde q_1+\tilde q_2)^6}+ \text{finite \& real}\Bigg]~,\label{OriginalPsi2IntegraldimReg}
\end{align}
where $\Poly_{n}(\tilde q_1,\tilde q_2)$ denotes a polynomial of degree $n$. It is easy to check that the first row in the above expression diverges as $\tilde q_1^4$ near $d=3$, which is expected since the interaction \eqref{interactionLagrangianIn3d} is a non-renormalisable operator of dimension 6. Terms in the second row, on the other hand, are convergent and give a finite and real constant. Also note that the divergent piece entirely comes from the Feynman part of the loop integral, resonating with our previous comment on the UV convergence of classical loops. Using \eqref{genintans} with $n=2$ and 
\begin{equation}
    g(\tilde q_+,1) = \frac{3 \tilde q_+^2+9 \tilde q_++8}{(1+\tilde q_+)^3}~,
\end{equation}
and expanding the result around $d=3$, we obtain
\begin{align}
	f_0=\frac{1}{15}\frac{\lambda^2 H^2}{(4\pi)^2}~,\label{dimRegf0}
\end{align}
and
\begin{align}
	\psi_2^{\rm 1L}&=\frac{1}{15}\frac{\lambda^2 H^2}{(4\pi)^2}k^3\left(\frac{1}{\epsilon}-\log \frac{k}{\mu_0} +\text{finite \& real const.\,}\right)~.\label{masslessdimregresult}
\end{align}
We see that the one-loop contribution in the partial dim reg scheme gives a \textit{real and scale-dependent} wavefunction coefficient. 

Before tackling renormalisation, a final comment is in order. In principle, one should renormalise $\psi_2$ at all times, not just at $\eta =0$. Indeed this was done both in \cite{Senatore:2009cf} and \cite{Braglia:2025cee}. Renormalisation at finite $\eta$ is interesting because the structure of UV divergences is slightly different and, as we will see, the degeneracy of counterterms encountered at $\eta=0$ is lifted at $\eta < 0$. Moreover, \cite{Pimentel:2012tw,Braglia:2025cee} showed that renormalisation at finite time displays a beautiful interplay between tadpole contributions and non-linearly realised symmetries. In this work however, we will only concern ourselves with renormalisation at $\eta =0$. This allows us to derive and present more transparently one of our main results, namely the universal relation between logarithmic divergences and $\Im \psi_n$ at $\eta=0$ (see \eqref{RG=Im}). We hope to present a finite-time discussion of renormalisation in a future publication.


\subsection{Renormalisation}\label{masslessDimRegRenormSubsection}

In this subsection, we will renormalise the action given in \eqref{toy} to remove the divergent piece of $\psi_2^{\rm 1L}$. Consider adding quadratic counterterms in $(d+1)$-dimensional de Sitter i.e.
\begin{equation}\label{ctac}
    S_{\rm ct} =  \int  \d\eta\,\d^d x \sqrt{-g} \, \mathcal{L}_{\rm ct}~.
\end{equation}
Note that the counterterms are computed in $d$ dimensions, which is the same prescription as for all propagators appearing in our loop calculation. This makes our dim reg prescription simple and straightforward. As we will see shortly, this consistency is essential to ensure the cancellation of the $\log(k)$ terms we encountered in the loop calculation. A different ``mixed prescription'' is also possible \cite{Senatore:2009cf}, as discussed in detail below \eqref{toSenatoreZalda}. \\

The most general IR-convergent counterterm Lagrangian consistent with shift symmetry, scale invariance, rotational and translational invariance is given by
\begin{equation}\label{ct1}
     \mathcal{L}_{\rm ct} =\sum_{2l+p+q\geq 3}  {c}_{l,p,q} ~(-H\eta\partial_\eta)^{p}\phi~ (g^{ij}\partial_i\partial_j)^{l}(-H\eta\partial_\eta)^{q}\phi~.
\end{equation}
The wavefunction gets the following contribution from the counterterms\footnote{One could in principle integrate the time derivatives by part to relate terms with the same value of $p+q$, but we see no need to do this here.}
\begin{align}
	\psi_2^{\rm ct}=i\sum_{2l+p+q\geq 3} {c}_{l,p,q} k^{2l}\int_{-\infty}^{0}\d \eta [a(\eta)]^{d+1-2l}(-H\eta\partial_\eta)^{p} K(\eta,k)(-H\eta\partial_\eta)^{q} K(\eta,k)~.
\end{align}
Now we can play the same trick as in the previous subsection and extract the phase and $k$-scaling information of the counterterm contribution. After Wick rotation and rescaling the momenta,\footnote{Note that the counterterm Lagrangian is IR-finite, hence the Wick rotation does not give rise to any additional terms.} we arrive at
\begin{equation}
    \psi_2^{\rm ct} = i^{d-3} H^{3-d} k^d \sum_{2l+p+q\geq 3} \,c_{l,p,q} \tilde{I}_{l,p,q}(d)~,
\end{equation}

where 
\begin{equation}
	\tilde{I}_{l,p,q}(d) \equiv  (-1)^{l+p+q} H^{p+q+2l-4} \int_{0}^{\infty}\d \tilde\chi \,\tilde\chi^{2l-d-1} (\tilde\chi\partial_{\tilde\chi})^{p} K\left(\frac{i \tilde\chi}{k},k\right) (\tilde\chi\partial_{\tilde\chi})^{q} K\left(\frac{i \tilde\chi}{k},k\right)\label{masslessDimRegCountertermIlpqd}
\end{equation}
is a finite and real constant. To cancel the divergence in the one-loop contribution $\psi_2^{\rm 1L}$, we choose $c_{l,p,q}$ to satisfy 
\begin{equation}
 	\sum_{2l+p+q\geq 3} c_{l,p,q} \tilde{I}_{l,p,q}(d) = -f_0\left( \frac{1}{\epsilon}+ \log\frac{\mu_0}{\mu} +\text{finite \& real const.\,}\right)  \,,
\end{equation}
with $f_0$ given in \eqref{dimRegf0} and the finite \& real constant chosen to match that in \eqref{masslessdimregresult}. The parameter $\mu>0$ here represents an arbitrary energy scale to be fixed by renormalisation conditions. Because we focus on the renormalisation of $\psi_2$ at $\eta=0$, we see that all counterterms are equally suitable, since their kinematic dependence $k^d$ is fixed by $d$-dimensional scale invariance. Had we insisted on renormalising at finite $\eta <0$, we would have needed to consider specific counterterms, as discussed in \cite{Senatore:2009cf,Braglia:2025cee}. This technical aspect is inconsequential for the derivation of one of our main results in Sec. \ref{imaginaryuniversal}, namely the one-loop universality of $\Im \psi_2$ at $\eta=0$.

Finally in $d = 3-\epsilon$ dimensions, $\psi_2^{\rm ct}$ is given by 
\begin{equation}
    \psi_2^{\rm ct} =  -\frac{1}{15} \frac{\lambda^2 H^2}{(4\pi)^2} k^3 \left(\frac{1}{\epsilon}-\frac{i \pi}{2}-\log \frac{k}{H}+ \log\frac{\mu_0}{\mu} +\text{finite \& real const.\,}\right) ~.
\end{equation}
Adding the counterterm contribution to \eqref{masslessdimregresult}, we find that the renormalised one-loop wavefunction coefficient in dim reg is
\begin{keyeqn}
	\begin{align}\label{drfresult}
	   \widehat{\psi}_2^{\rm 1L}= \psi_2^{\rm 1L} + \psi_2^{\rm ct} = \frac{1}{15} \frac{\lambda^2 H^2}{(4\pi)^2}k^3\left(   \frac{i \pi}{2} + \log\frac{\mu}{H}\right)~.
	\end{align}
\end{keyeqn}
where $\mu$ is an arbitrary real and positive constant to be fixed by renormalisation conditions. 

We can use the above result to compute the correlator at one-loop. Using \eqref{powerspectrumAt1L} and \eqref{psi3psi4}, we obtain
\begin{equation}
   \ex{\phi(\mathbf{k})\phi(-\mathbf{k})}_{\rm 1L}'=  \frac{\lambda ^2 H^6 }{15 (32 \pi )^2 k^3} \left(32 \log \left(\frac{\mu }{H}\right)-10 \pi ^2+\frac{2131}{30}\right)~.
\end{equation}


\subsection{Properties: unitarity, scale invariance and reality}
Let us analyse some properties of the one-loop result obtained after dim reg.
\begin{itemize}
    \item \textbf{Scale invariance}: The action \eqref{toy} is invariant under the de Sitter dilation isometry $(\eta,\mathbf{x})\to (\alpha^{-1}\eta,\alpha^{-1} \mathbf{x})$ in any number of spacetime dimensions. As a result, we expect the wavefunction \( \Psi[\bar\phi] \) to be invariant under dilation. In momentum space at the boundary $\eta=\eta_0$, this isometry corresponds to the scaling transformation $\mathbf{k}\to \alpha \mathbf{k}$ when the boundary time $\eta_0$ is not involved. Since $\phi(\eta,x)$ and therefore $\Psi[\phi(\eta_0,x)]$ is scale invariant and becomes independent of time in the limit $\eta_0\to 0$, it follows from \eqref{WFU} that the wavefunction coefficients in \( d \) spatial dimensions must scale as  
	\begin{equation}\label{furtherscaleinv}
    	\psi_n\left(\alpha k\right) \rightarrow \alpha^d \psi_n(k)~.
	\end{equation}
	However, the unrenormalised one-loop result \( \psi_2^{\rm 1L} \), given in \eqref{masslessdimregresult}, explicitly violates the scaling symmetry in \( 3+1 \) dimensions due to the presence of a \( \log k \) term. This is understandable as the scale invariance in $d+1$ dimensions is different from that in 3+1 dimensions. Fortunately, this logarithmic term is precisely cancelled by the contribution from the counterterms, ensuring that the fully renormalised result \( \widehat\psi_2^{\rm 1L} \) respects the scaling symmetry. 
    \item \textbf{Unitarity}: The Cosmological Optical Theorem (COT) \cite{Goodhew:2020hob, Melville:2021lst} is a manifestation of unitarity, analyticity and the choice of the Bunch-Davies initial state in any Friedmann–Lema\^{i}tre–\\Robertson–Walker (FLRW) spacetime and in particular in de Sitter space. Below, we check if the full one-loop result obtained after dim reg obeys the COT or not. Let us first compute the discontinuity of the unrenormalised loop result,
    \begin{align}\label{discdr}
            \begin{split}
             i\Disc [i\psi_2^{\rm 1L}] &= \frac{1}{15}\frac{\lambda^2 H^2}{(4\pi)^2} k^3\left[\log k - \left(\log (e^{-i \pi} k)\right)^*\right]\\
            &= \frac{1}{15}\frac{\lambda^2 H^2}{(4\pi)^2}k^3 \times (-i \pi)~,
            \end{split}
    \end{align}
    which matches the right-hand side prediction of the COT (see Appendix \ref{COTApplicationToOurModelSubAppendix} for more details). Here the discontinuity comes from a scale-dependent $\log k$ piece. Then we turn to the fully renormalised loop result and take the discontinuity,
    \begin{align}\label{discdr}
            \begin{split}
                 i\Disc [i\widehat\psi_2^{\rm 1L}] &= -\frac{1}{15}\frac{\lambda^2 H^2}{(4\pi)^2} k^3\left[\frac{i\pi}{2} - \left(\frac{i\pi}{2}\right)^*\right]\\
                &= \frac{1}{15}\frac{\lambda^2 H^2}{(4\pi)^2}k^3 \times (-i \pi)~,
            \end{split}
    \end{align}
    again matching the right-hand side of the COT. Interestingly, after renormalisation, the discontinuity comes from the imaginary part as opposed to any unwanted $\log k$ piece. In summary, the result of dim reg indeed satisfies the COT and is thereby unitary and analytic. 

	\item \textbf{Reality}: Despite being real at tree-level \cite{Stefanyszyn:2023qov}, the wavefunction coefficient spontaneously grows an imaginary part after renormalisation at one-loop level (as previously found in \cite{Lee:2023jby}). To be more precise, even though the unrenormalised one-loop result in dim reg was purely real, from \eqref{drfresult}, we see that the renormalised one-loop contribution acquires an imaginary part from the counterterm,
	\begin{align}
		\Im[\psi_2^{\rm 0L}]&=\Im[\psi_2^{\rm 1L}]=0\quad \& \quad \Im[\psi_2^{\text{ct}}]\neq 0  \qquad \Rightarrow \qquad \Im[\widehat\psi_2^{\rm 1L}]\neq 0\,.
	\end{align}
	Since we have mentioned that only quantum loops are relevant for renormalisation in our model, the spontaneous emergence of an imaginary part can be thought of as an  \textit{intrinsic quantum effect}. As mentioned in Section \ref{ModelSection}, this imaginary part is irrelevant for the two-point correlator of $\phi$, since only the real part of $\psi_2$ contributes to it. On the other hand, the imaginary part of $\psi_2$ does have an effect both on the two-point cross correlator $\ex{\phi\pi}$ (see e.g. \eqref{phipiCrossCorrelator}) and, relatedly, on the time dependence of the power spectrum of $\phi$ for finite $\eta < 0$. In particular, in \cite{Melville:2021lst} it was found that
	\begin{align}\label{Impsi2}
    	\lim_{\eta \to 0} \frac{1}{3!}\partial_\eta^3 P(k,\eta)=-\frac{H^4}{3k^3}\Im [\widehat\psi_2^{\rm 1L}(k)]\,.
	\end{align}
	Moreover, at higher points, the imaginary part of wavefunction coefficients contributes to the parity-odd component of field correlators (see e.g. \eqref{POCorrelatorHasIm}), just like the real part of wavefunction coefficients contributes to parity-even correlators, making the above result particularly important when generalised to higher order correlators. 
\end{itemize}


\section{Mass-dimensional regularisation}\label{MassDimRegSect}

Although dimensional regularisation (dim reg) introduced in the previous section provides the most natural extension to de Sitter space, it is difficult to implement in practice due to the complicated form of the propagators. To address this computational challenge, an alternative regularisation scheme was proposed in \cite{Melville:2021lst}, which we call \textit{mass-dimensional regularisation (\mdreg)}.\footnote{Notice that a similar scheme was also proposed by \cite{Bzowski:2013sza} in the context of conformal field theory. Detailed loop calculations using this regulator were performed in \cite{Beneke:2023wmt}.} In this scheme, in addition to the space dimensions, one simultaneously continues the mass of the scalar field such that the mode function and the propagators stay simple. Consider the mode function of a scalar field with mass $m<3H/2$ in $d$-spatial dimensions,
\begin{align}
    \phi^+(\eta,k)=-i\frac{\sqrt{\pi}}{2} H^{-1/2} (-H \eta)^{d/2}H_{\nu}^{(2)}(-k\eta)~,\quad \nu\equiv \sqrt{\frac{d^2}{4}-\frac{m^2}{H^2}}~.
\end{align}
In \mdreg, we deform the mass so that $\nu$ remains fixed at the value $3/2$ \cite{Melville:2021lst}, i.e.\footnote{Note that another viable choice for the external fields is the conformal limit $\nu=1/2$ \cite{Green:2020txs} which is then mapped back to the massless case via weight-shifting. This is the strategy adopted, for instance, in \cite{Xianyu:2022jwk,Zhang:2025nzd}.}
\begin{align}
	\nu= \sqrt{\frac{d^2}{4}-\frac{m^2(d)}{H^2}}\equiv \frac{3}{2}~,\quad \text{or}\quad m^2(d)= H^2\left(\frac{d^2 -9}{4}\right)~.
\end{align}
With this choice, the mode functions and propagators revert to simpler forms. The mode function becomes
\begin{align}
	\phi^+(\eta,k)=-i\frac{\sqrt{\pi}}{2}H^{-1/2} (-H\eta)^{d/2}H_{3/2}^{(2)}(-k\eta)= H^{-1/2} (-H\eta)^{d/2}(-k\eta)^{-3/2}(1-ik\eta)e^{ik\eta}~,
\end{align}
and in the $\eta_0\to 0$ limit, the propagators take the following form
\begin{subequations}\label{massdimregprop}
    \begin{align}
	K(\eta,k)&=\frac{\phi^+(\eta,k)}{\phi^+(\eta_0,k)}\approx\left(\frac{-\eta}{-\eta_0}\right)^{(d-3)/2}(1-ik\eta)e^{ik\eta}~,\\
	\nonumber G(\eta_1,\eta_2;q)&\approx\frac{i H^{d-1}}{q^3} (-\eta_1)^{(d-3)/2}(-\eta_2)^{(d-3)/2}\\
        &\quad\times\Big\{\left[q \eta_1 \cos(q\eta_1)-\sin (q\eta_1)\right] (1-i q\eta_2)e^{iq\eta_2}\theta(\eta_1-\eta_2)+(\eta_1\leftrightarrow\eta_2)\Big\}~.
    \end{align}
\end{subequations}
Comparing with the $(3+1)$-dimensional propagators (see \eqref{MasslessBtobPropIn3dim} and \eqref{MasslessBtoBPropIn3dim}), we see that the \mdreg propagators have extra factors of $(-\eta)^{(d-3)/2}$.

The interaction vertex, on the other hand, remains the same as in dim reg,
\begin{align}
	S_{\rm  int}\supset 
	\int \frac{\d\eta\, \d^d x}{(-H\eta)^{d-2}} \,\frac{\lambda_d}{3!}\,[\partial_\eta \phi(\eta,\mathbf{x})]^3  ~,\quad \lambda_d\equiv \mu_0^{(3-d)/2}\lambda~,
\end{align}
with $\mu_0>0$ an arbitrary scale to balance the mass dimension of the coupling $\lambda$.


\subsection{Loop contributions}
Let us now compute the one-loop contribution to $\psi_2$ arising from the diagram in Figure \ref{figloop} using \mdreg,
\begin{align}
    \nonumber\psi_2^{\rm 1L}=-\frac{\lambda_d^2}{2 H^{2d-4}}\int\frac{\d^{d} q_1}{(2\pi)^d} &\int_{-\infty}^{0} \frac{\d\eta_1}{(-\eta_1)^{d-2}} \frac{\d\eta_2}{(-\eta_2)^{d-2}}\\
    \times &\partial_{\eta_1} K(\eta_1,k) \partial_{\eta_2} K(\eta_2,k) \partial_{\eta_1} \partial_{\eta_2} G(\eta_1,\eta_2; q_1)\, \partial_{\eta_1} \partial_{\eta_2} G(\eta_1,\eta_2; q_2)~.
\end{align}
To exact the phase information, we perform a Wick rotation $\eta_i \rightarrow e^{i\pi/2}\chi_i$ to obtain
\begin{align}\label{psi2step}
    \nonumber\psi_2^{\rm 1L}= \frac{\lambda_d^2\, e^{i\pi(d-2)}}{2 H^{2d-4}} &\int\frac{\d^{d} q_1}{(2\pi)^d} \int^{\infty}_{0} \frac{\d\chi_1 \d\chi_2}{(\chi_1 \chi_2)^{d-2}}\\
    \times & \partial_{\chi_1} K(i\chi_1,k) \partial_{\chi_2} K(i\chi_2,k) \partial_{\chi_1} \partial_{\chi_2} G(i\chi_1,i\chi_2; q_1)\, \partial_{\chi_1} \partial_{\chi_2} G(i\chi_1, i\chi_2; q_2).
\end{align}
Using \eqref{massdimregprop}, we see that under Wick rotation, the bulk-boundary and bulk-bulk propagators pick up the following phases,
\begin{subequations}\label{argumassdimregprop}
    \begin{align}
	\arg[K(i\chi,k)]&= -\frac{\pi}{2}\left(\frac{d-3}{2}\right)~,\\
	\arg[G(i\chi_1,i\chi_2;q)]&= -\frac{ \pi}{2}(d-3)~,
    \end{align}
\end{subequations}
up to $\pm \pi$ phases. Using these properties, we obtain
\begin{align}
    \arg \psi_2^{\rm 1L}=-\frac{\pi(d-3)}{2}~.
\end{align}
To extract the $k$-scaling information, we rescale the variables by $\chi_i=\tilde{\chi}_i/k$ and $q_i=\tilde{q}_i k$, under which the propagators changes as
\begin{subequations}
    \begin{align}
	K\left(\frac{i\tilde \chi}{k},k\right)&=e^{-i\pi(d-3)/4}\left(-k\eta_0\right)^{-(d-3)/2}\, \tilde{\chi}^{(d-3)/2} (1+\tilde\chi)e^{-\tilde{\chi}}~,\\
	\nonumber G\left(\frac{i\tilde \chi_1}{k},\frac{i\tilde \chi_2}{k};\tilde{q} k\right)&= e^{-i\pi(d-3)/2}\frac{H^{d-1}}{k^d} (\tilde\chi_1 \tilde \chi_2)^{(d-3)/2}\\
    &\times\Big\{[\tilde q \tilde \chi_1 \cosh (\tilde q \tilde \chi_1)-\sinh (\tilde q\tilde \chi_1)] (1+\tilde q\tilde\chi_2)e^{-\tilde q\tilde\chi_2}\theta(\tilde\chi_2-\tilde\chi_1)+(\tilde\chi_1\leftrightarrow\tilde\chi_2)\Big\}~.
    \end{align}
\end{subequations}
Counting the power of $k$, we find $\psi_2^{\rm 1L}$ scales as $k^3$. However, notice that the \mdreg wavefunction explicitly depends on the boundary time $\eta_0$ as well. Stripping all $d$-dependent powers from the integrand, we expect
\begin{align}
    \nonumber\psi_2^{\rm 1L}&=e^{-i\pi(d-3)/2}\mu_0^{3-d}  (-\eta_0)^{3-d} k^3\left(\frac{f_0}{3-d} + \text{finite \& real const.}\right)\\
    &=f_0 k^3\left(\frac{1}{\epsilon}+\frac{i\pi}{2}+\log (-\mu_0\eta_0)+\text{real \& finite const.} \right)~,
\end{align}
Analogous to the situation in Section \ref{masslessDimRegLoopSubsection}, the constant $f_0$ can also be computed by expanding the phase-, $k$-scaling and $\eta_0$-scaling stripped integrand at leading order in $\epsilon$. We find the same result\footnote{Logarithmic divergences in Minkowski quantum field theories are generally believed to be universal as they represent the characteristics of the renormalisation group flow \cite{Schwartz:2014sze}. Yet it remains remarkable that even in de Sitter, such universality still persists. This is so far confirmed for two regulator choices only, but later in Section \ref{EtaRegSect}, we will reinforce this universality with a much broader class of regulators.} as in dim reg i.e. \eqref{dimRegf0}. In summary, we have
\begin{align}
    \psi_2^{\rm 1L}&=\frac{1}{15}\frac{\lambda^2 H^2}{(4\pi)^2}k^3\left(\frac{1}{\epsilon}+\frac{i\pi}{2}+\log (-\mu_0\eta_0)+\text{real \& finite const.} \right)~.\label{mdra1}
\end{align}
Interestingly, \mdreg predicts a \textit{finite imaginary part} already at the level of the unrenormalised one-loop wavefunction. Instead of a scale-dependent $\log k$ term, \mdreg offers a time-dependent $\log(-\eta_0)$ term in the form of secular growth.


\subsection{Renormalisation}

The one-loop contribution obtained above is UV-divergent, and it can be cancelled again by adding higher derivative counterterms as in Section \ref{masslessDimRegRenormSubsection}. Using the counterterm Lagrangian given in \eqref{ct1}, we find that the counterterm contribution is given by
\begin{align}
	\psi_2^{\rm ct}=i\sum_{2l+p+q\geq 3}c_{{l,p,q}}\,k^{2l}\int_{-\infty}^{\eta_0}\d \eta \,[a(\eta)]^{d+1-2l}(-H\eta\partial_\eta)^{p} K(\eta,k)(-H\eta\partial_\eta)^{q} K(\eta,k)~.
\end{align}
Recall that the bulk-boundary propagator reads
\begin{align}
	K(\eta,k)=\left(\frac{-\eta}{-\eta_0}\right)^{\frac{d-3}{2}}\frac{(1-ik\eta)e^{ik\eta}}{(1-ik\eta_0)e^{ik\eta_0}}~.
\end{align}
In the late-time limit $\eta,\eta_0\to 0$,
\begin{align}
	(-H\eta\partial_\eta)^{p}K(\eta,k)\sim \left[\left(-\frac{\epsilon}{2}\right)^p+k^2\eta^2+\cdots\right] \left(\frac{-\eta}{-\eta_0}\right)^{-\epsilon/2}~,
\end{align}
therefore, for $\epsilon\to 0^-$, the time integral is always IR-convergent. However, we note that there are overall factors of $(-\eta_0)^{\epsilon/2}$ coming from the bulk-boundary propagators. These factors do not converge when $\eta_0\to 0$. Rather, they become secularly growing $\log(-\eta_0)$ terms in the small-$\epsilon$ expansion. We will see that these $\log(-\eta_0)$ terms serve to cancel the time dependence in $\psi_2^{\rm 1L}$. We then extract the phase, the $k$- and $\eta_0$-scaling information of the counterterm contribution by Wick-rotating and rescaling the variables,
\begin{equation}
    \psi_2^{\rm ct} = H^{3-d} (-\eta_0)^{3-d} k^3 \sum_{2l+p+q\geq 3} \,c_{l,p,q} \tilde{I}_{l,p,q}(d)~,
\end{equation}
where 
\begin{equation}
	\tilde{I}_{l,p,q}(d) \equiv  (-1)^{l+p+q} H^{p+q+2l-4} \int_{0}^{\infty}\d \tilde\chi \,\tilde\chi^{2l-d-1} (\tilde\chi\partial_{\tilde\chi})^{p} [(1+\tilde\chi)e^{-\tilde\chi}]\, (\tilde\chi\partial_{\tilde\chi})^{q} [(1+\tilde\chi)e^{-\tilde\chi}]
\end{equation}
is a finite and real constant. Same as in dim reg, to cancel the loop divergence, we choose $c_{l,p,q}$ to satisfy 
\begin{equation}
 	\sum_{2l+p+q\geq 3} c_{l,p,q} \tilde{I}_{l,p,q}(d) = -f_0\left( \frac{1}{\epsilon}+ \log\frac{\mu_0}{\mu} +\text{finite \& real const.\,}\right) ~
\end{equation}
where $\mu>0$ is an arbitrary constant and the finite \& real const. is chosen to match that in \eqref{mdra1}. Expanding around $d=3-\epsilon$, we have
\begin{equation}
    \psi_2^{\rm ct} =  -\frac{1}{15} \frac{\lambda^2 H^2}{(4\pi)^2} k^3 \left(\frac{1}{\epsilon}+\log (-H\eta_0)+ \log\frac{\mu_0}{\mu} +\text{finite \& real const.\,}\right) ~.
\end{equation}
We see that the $\log(-\eta_0)$ term magically appears and serves to cancel those in $\psi_2^{\rm 1L}$ together with the $1/\epsilon$ divergence. Combining the counterterm contribution with the one-loop contribution \eqref{mdra1}, we arrive at the renormalised one-loop wavefunction coefficient in \mdreg:
\begin{keyeqn}
	\begin{align}\label{mdrfresult}
	   \widehat{\psi}_2^{\rm 1L}= \psi_2^{\rm 1L} + \psi_2^{\rm ct} = \frac{1}{15} \frac{\lambda^2 H^2}{(4\pi)^2}k^3\left(   \frac{i \pi}{2} + \log\frac{\mu}{H}\right)~.
	\end{align}
\end{keyeqn}
with $\mu$ an arbitrary real and positive constant to be fixed by renormalisation conditions. We note that this result agrees with the one previously derived using dim reg.


\subsection{Properties: unitarity, scale invariance and reality}

Even though the full renormalised one-loop result agrees for dim reg and \mdreg schemes, the properties of the individual components—the loop contribution and the counterterm—differ between schemes. Below, we analyse some features of the result obtained using \mdreg:

\begin{itemize}

    \item \textbf{Scale invariance}:  
    To check scale invariance, we first note that both the unrenormalised one-loop contribution $\psi_2^{\rm 1L}$ and the counterterm contribution $\psi_2^{\rm ct}$ depend explicitly on the boundary time $\eta_0$. Under de Sitter dilation isometry, both the boundary time and comoving momenta are rescaled i.e. $(\eta_0,k)\to (\alpha^{-1}\eta_0, \alpha k)$, whereas under boundary scale transformation, only momenta are rescaled i.e. $k\to \alpha k$. In cases where the wavefunction does not depend on time, these two transformations are identical. Thus we conclude that the renormalised wavefunction coefficient is both scale-invariant and dilation-invariant (by not depending on time),
    \begin{align}
        \widehat\psi_2^{\rm 1L}(\alpha k)=\alpha^3 \widehat\psi_2^{\rm 1L}(k)~.
    \end{align}
    By contrast, the unrenormalised one-loop wavefunction coefficient \( \psi_2^{\rm 1L} \) and the counterterm contribution $\psi_2^{\rm ct}$ are time-dependent. They are indeed scale-invariant in $3+1$ dimensions thanks to the absence of $\log k$ terms,
    \begin{equation}
        \psi_2^{\rm 1L}(\eta_0,\alpha k) = \alpha^3 \psi_2^{\rm 1L}(\eta_0,k)\quad ,\quad \psi_2^{\rm ct}(\eta_0,\alpha k) = \alpha^3 \psi_2^{\rm ct}(\eta_0,k)~.
    \end{equation}
    Yet they are not dilation-invariant in $3+1$ dimensions (due to the $\log (-\eta_0)$ term)\footnote{Of course, they are still dilation-invariant in $d+1$ dimensions, namely $\psi_2^{\rm 1L}(\alpha^{-1}\eta_0,\alpha k) = \alpha^d \psi_2^{\rm 1L}(\eta_0,k)$ and $ \psi_2^{\rm ct}(\alpha^{-1}\eta_0,\alpha k) = \alpha^d \psi_2^{\rm ct}(\eta_0,k)$.},
    \begin{equation}
        \psi_2^{\rm 1L}(\alpha^{-1}\eta_0,\alpha k) \neq \alpha^3 \psi_2^{\rm 1L}(\eta_0,k)\quad ,\quad \psi_2^{\rm ct}(\alpha^{-1}\eta_0,\alpha k) \neq \alpha^3 \psi_2^{\rm ct}(\eta_0,k)~.
    \end{equation}

    \item \textbf{Unitarity}:  
    To check unitarity, we first compute the $\Disc$ of the one-loop contribution as well as that of the fully renormalised result,
    \begin{align}
            \begin{split}
                i\Disc [i\widehat\psi_2^{\rm 1L}]= i\Disc [i\psi_2^{\rm 1L}]= &= -\frac{1}{15}\frac{\lambda^2 H^2}{(4\pi)^2} k^3\left[\frac{i\pi}{2} - \left(\frac{i\pi}{2}\right)^*\right]\\
                &= \frac{1}{15}\frac{\lambda^2 H^2}{(4\pi)^2}k^3 \times (-i \pi)~,
            \end{split}
    \end{align}
    both matching the right-hand side of the cosmological optical theorem (COT) (see Appendix \ref{COTApplicationToOurModelSubAppendix}) thanks to the imaginary part.\footnote{We thank Ayngaran Thavanesan for pointing out a missing minus sign in the previous computation of the left-hand side of the COT in \mdreg.} We thus conclude that \mdreg is fully compatible with unitarity and analyticity.

    \item \textbf{Reality}:  
    Unlike in dim reg, the one-loop contribution in \mdreg (see \eqref{mdra1}) already contains the \textit{correct imaginary part}, which persists in the final renormalised result:
    \begin{align}
        \Im[\psi_2^{\rm 0L}] = \Im[\psi_2^{\rm ct}] = 0\,, \qquad \Im[\psi_2^{\rm 1L}] \neq 0 \quad \Rightarrow \quad \Im[\widehat{\psi}_2^{\rm 1L}] \neq 0~.
    \end{align}
    Additionally, the value of the imaginary part of $\widehat{\psi}_2^{\rm 1L}$ agrees for the two variants of dim reg so far. While the discussion so far has focused on the two-point wavefunction coefficient, and is limited to variants of dim reg, in Section \ref{imaginaryuniversal} we will show that this is a general phenomenon: the imaginary part of the one-loop renormalised \( n \)-point wavefunction coefficients $\Im \widehat\psi_n^{\rm 1L}$ is universal under the constraint from unitarity and analyticity. Since the imaginary part contributes to the parity-odd component of the correlators, its universality suggests a deeper understanding of parity violation in de Sitter space. We will discuss this further in Section \ref{imaginaryuniversal}.

\end{itemize}


\section{$\upeta$ regularisation}\label{EtaRegSect}

While analytically continuing the number of space dimensions appears as the most natural way to regularise loop divergences, it is less straightforward in de Sitter than in Minkowski since the background geometry depends non-trivially on dimensionality. As we have seen in the previous two sections, the mode functions can become complicated and the time integrals are challenging to evaluate. Regularisation is therefore more involved. To overcome this difficulty, in this section, we introduce to de Sitter EFTs a new regularisation scheme named $\upeta$ regularisation, and examine what $\upeta$ regulators are compatible with unitarity and analyticity. In particular, we show that constraints from unitarity and analyticity pin down a \textit{unique} answer for the imaginary part of the renormalised one-loop wavefunction that agrees with dim reg and \mdreg.

$\upeta$ regularisation was first introduced in \cite{Padilla:2024mkm} as a gauge-invariant regulator for Minkowski quantum field theories. Inspired by the study of smoothed asymptotics and divergent power series in mathematics, $\upeta$ regularisation formally implements a window function $\upeta(x)$ that cuts off the divergent tail of loop integrals. The UV divergences are then exposed as the Mellin transform of the regulator function, and one can make general arguments without assuming any explicit form for the regulator function $\upeta(x)$. Consider a general Minkowski loop integral of the form,
\begin{align}
	\mathcal{I}(M)=\int_0^\infty \frac{\d q}{q} \frac{\Poly_{m+n}(q/M)}{\Poly_{m}(q/M)}~,\quad m,n\in \mathbb{N}~,
\end{align}
where $M$ denotes a physical energy scale of the theory. This integral is UV-divergent for any non-negative integer $n$. To regulate the divergence, one inserts a regulator function in the integrand,
\begin{align}
	\mathcal{I}_\upeta(M,\Lambda)=\int_0^\infty \frac{\d q}{q} \frac{\Poly_{m+n}(q/M)}{\Poly_{m}(q/M)}\upeta\left(\frac{q}{\Lambda}\right)~,
\end{align}
where $\Lambda>0$ denotes an arbitrary energy cutoff that will eventually drop out after renormalisation. The regulator function is smooth and satisfies
\begin{align}
	\upeta(0)=1~,\quad \upeta(+\infty)=0~.
\end{align}
In addition, $\upeta(x)$ is required to fall off sufficiently fast such that the integral is convergent at large $x=q/\Lambda$.\footnote{Sometimes $\upeta(x)$ is chosen to be a ``Schwartz function'' that falls off faster than any polynomial in the large-$x$ limit, thereby regularising all UV divergences in quantum field theories.} To examine the large-$\Lambda$ asymptotics in $\mathcal{I}_\upeta$, we can expand it as
\begin{align}
	M^{-n}\mathcal{I}_\upeta(M,\Lambda)= f_n C_n[\upeta]\left(\frac{\Lambda}{M}\right)^n+\cdots+ f_1 C_1[\upeta] \left(\frac{\Lambda}{M}\right)^1+f_0\left(\log\frac{\Lambda}{M}+\gamma[\upeta]+g\right)+\mathcal{O}\left(\frac{M}{\Lambda}\right)~,\label{etaRegGeneralForm}
\end{align}
where $f_i$'s and $g$ are functions of kinematics that depend solely on the form of the polynomials in the integrand and
\begin{align}
	C_s[\upeta]\equiv \int_{0}^{\infty} \d x\, x^{s-1} \upeta(x)
\end{align}
is the Mellin transform of the regulator and
\begin{align}\label{gammaeta1}
	\gamma[\upeta]=-\int_{0}^{\infty}\d x \log x\, \upeta'(x)~.
\end{align}
We see that all the power-law divergent terms depend on the regulator choice, whereas the logarithmically divergent term appears universal. In Minkowski quantum field theories, such logarithmic divergences bear a special meaning as they characterise the renormalisation group running of couplings and anomalous dimensions. In \cite{Padilla:2024mkm}, it has been shown that the logarithmic divergence obtained in $\upeta$ regularisation agrees with that obtained from other regularisation schemes such as dim reg. The finite term $\gamma[\upeta]$, however, does depend on the regulator. In particular, a rescaling $\upeta(x)\to \upeta(\lambda x)$ gives a constant shift
\begin{align}
	\gamma[\upeta(\lambda x)]=\gamma[\upeta(x)]-\log \lambda~.\label{finShiftAfterEtaRescaling}
\end{align}
Note also that this is degenerate with a rescaling of the cutoff scale $\Lambda\to \lambda^{-1} \Lambda$. Upon renormalisation, all divergent terms in \eqref{etaRegGeneralForm} are removed via a suitable choice of counterterms, leaving a finite result that is fixed by the renormalisation conditions. To summarise, the advantage of $\upeta$ regularisation is that it exposes the divergence structure of loop integrals without explicitly referring to the form of the regulator $\upeta(x)$, while also avoiding subtleties of going to non-integer dimensions.\footnote{It was shown later in \cite{Padilla:2024cvk} that for Minkowski quantum field theories, dim reg also falls into a subcategory of a generalised version of $\upeta$ regularisation.}


\subsection{Loop contributions}

Now we make the first application of $\upeta$ regularisation to loops in de Sitter EFTs. We shall again focus on the one-loop bubble diagram for the two-point wavefunction (see e.g. \eqref{OriginalPsi2IntegraldimReg}),
\begin{align}
    \psi_2^{\rm 1L}=\frac{\lambda^2 H^2}{16} k^3 \int\frac{\d^3 q_1}{(2\pi)^3}q_1 q_2\frac{3(q_1+q_2)^2+9k(q_1+q_2)+8k^2}{k^4 (k+q_1+q_2)^3}+\text{finite and real}~.
\end{align}
Since the finite and real piece is not the focus of our interest, we will henceforth neglect it. Using the symmetric two-body phase space formula given in \eqref{s2bdy}, we can change variables to $q_\pm\equiv q_1\pm q_2$ and finish the integral over $q_-$ to obtain
\begin{align}
    \psi_2^{\rm 1L}=\frac{\lambda^2 H^2}{16} \frac{\pi}{8(2\pi)^3 k}\int_{k}^\infty \d q_+ \left(q_+^4 -\frac{2}{3} k^2 q_+^2+ \frac{1}{5}k^4\right)\frac{3 q_+^2+9 k q_++8k^2}{(k+q_+)^3}~.
\end{align}
To regulate the UV divergences, we now insert the window function,
\begin{align}
    \psi_2^{\rm 1L}=\frac{\lambda^2 H^2}{16} \frac{\pi}{8(2\pi)^3 k}\int_{k}^\infty \d q_+ \left(q_+^4 -\frac{2}{3} k^2 q_+^2+ \frac{1}{5}k^4\right)\frac{3 q_+^2+9 k q_++8k^2}{(k+q_+)^3}\upeta\left(\frac{q_+/k}{\Lambda/H}\right)~.\label{psi2OneLoopEtaRegAfterInsertion}
\end{align}
Notice that with the benefit of hindsight, we have chosen a regulator function that implicitly refers to the external momentum $k$ such that $x=(q_+/k)/(\Lambda/H)$ is both scale-invariant and dimensionless. This will ensure the scale invariance of the final result. One can now expand the integrand at large $x$ and obtain (see Appendix \ref{FiniteRemainderAppendix} for more details)
\begin{align}
    \nonumber\psi_2^{\rm 1L}&=\frac{\lambda^2 H^2}{15(4\pi)^2} k^3 \int_{H/\Lambda}^\infty \d x \left[\frac{45}{64} x^3\left(\frac{\Lambda}{H}\right)^4 - \frac{45}{64} x \left(\frac{\Lambda}{H}\right)^2 +\frac{1}{x}+\mathcal{O}\left(\frac{H}{\Lambda}\right)\right]\upeta\left(x\right)\\
    &=\frac{\lambda^2 H^2}{15(4\pi)^2} k^3 \left[\frac{45}{64} C_4[\upeta]\left(\frac{\Lambda}{H}\right)^4 - \frac{45}{64} C_2[\upeta] \left(\frac{\Lambda}{H}\right)^2 +\left(\log\frac{\Lambda}{H}+\gamma[\upeta]-\frac{59}{256}-\log 2\right)+\mathcal{O}\left(\frac{H}{\Lambda}\right)\right]~,\label{psi2oneloopEtaRegResult}
\end{align}
where in the last step, we have used $\int_{H/\Lambda}^{\infty}=\int_{0}^{\infty}-\int_0^{H/\Lambda}$ to put the result into the standard form \eqref{etaRegGeneralForm}. All $\mathcal{O}(H/\Lambda)$ terms here have been omitted. As expected, the quartic and quadratic divergences are regulator-dependent but the logarithmic divergence is universal. The finite piece is also regulator-dependent. In Table \ref{etaRegExampleResultsTable}, we explicitly choose several different regulator functions and exhibit their results for the one-loop wavefunction $\psi_2^{\rm 1L}$.

\begin{table}[h!]
	\centering
	\begin{tabular}{c|ccc}
		\toprule[1.1pt]
		  $\upeta(x)$ \quad &\quad $C_4[\upeta]$ \quad &\quad $C_2[\upeta]$ \quad&\quad $\gamma[\upeta]$\\
		\midrule[1pt]
		 $e^{-x}$ \quad &\quad $6$ \quad &\quad $1$  \quad &\quad $-\gamma_E$\\ [1ex]
		 $e^{-x^2/2}$ \quad &\quad $2$ \quad &\quad $1$ \quad &\quad $\frac{-\gamma_E+\log 2}{2}$\\ [1ex]
		 $\theta(1-x)$ \quad &\quad $\frac{1}{4}$ \quad &\quad $\frac{1}{2}$ \quad &\quad $0$\\ [0.8ex]
		 $\left(1+x\right)^{-6}$ \quad &\quad $\frac{1}{20}$ \quad &\quad $\frac{1}{20}$  \quad &\quad $-\frac{137}{60}$\\ [1ex]
		 $\left(1+x^2\right)^{-3}$ \quad &\quad $\frac{1}{4}$ \quad &\quad $\frac{1}{4}$ \quad &\quad $-\frac{3}{4}$\\ [1ex]
		 \midrule[1pt]
         $\frac{1}{2}\left(3 e^{-x}-e^{-x^2/6}\right)$ \quad &\quad $0$ \quad &\quad $0$ \quad &\quad $-\frac{5\gamma_E+\log 6}{4}$\\ [1ex]
         $\frac{1}{4}\left[5 (1+x)^{-6} -\left(1+x^2\right)^{-3}\right]$ \quad &\quad $0$ \quad &\quad $0$ \quad &\quad $-\frac{8}{3}$\\ [1ex]
         \midrule[1pt]
		 $\frac{1}{2}\left(3 e^{-e^{i\theta}x}-e^{-e^{2i\theta}x^2/6}\right)$~,\quad$|\Re \theta|<\frac{\pi}{4}$ \quad &\quad $0$ \quad &\quad $0$ \quad &\quad $-\frac{5\gamma_E+\log 6}{4}-i\theta$\\ [1ex]
		 $\frac{1}{4}\left[5 (1+e^{i\theta} x)^{-6} -\left(1+e^{2i\theta}x^2\right)^{-3}\right]$~,\quad $\theta\in \mathbb{C}$\quad &\quad $0$ \quad &\quad $0$ \quad &\quad $-\frac{8}{3}-i\theta$\\ [1ex]
		\bottomrule[1.1pt]
	\end{tabular}
	\caption{Different explicit choices of the $\upeta$ regulator and their corresponding results. The \textit{upper} rows correspond to choices of real function $\upeta(x)$ that no doubt produce real results. The \textit{middle} rows correspond to the so-called enhanced regulators for which the power-law divergences are engineered to vanish. The \textit{lower} rows correspond to examples of complex enhanced regulators for which one obtains an seemingly arbitrary imaginary part for $\psi_2^{\rm 1L}$ and hence the renormalised wavefunction coefficient $\widehat{\psi}_2^{\rm 1L}$. Note that in Section \ref{UnitaryAnalyticEtaRegSubsection} we give a more stringent criterion on the selection of $\upeta(x)$ based on unitarity and analyticity.}\label{etaRegExampleResultsTable}
\end{table}

\paragraph{Discussion} Before moving on to renormalisation, we make a few important remarks: \label{EtaRegOneLoopSubsectionDiscussions}
\begin{itemize}
    \item \textbf{Scale invariance:} The regularised one-loop wavefunction coefficient \eqref{psi2oneloopEtaRegResult} is covariant under the $(3+1)$-dimensional scale transformations, i.e. $\psi_2^{\rm 1L}(\alpha k)=\alpha^3 \psi_2^{\rm 1L}(k)$. This is a consequence of the choice of the general form of the regulator function, $\upeta=\upeta(x)$, with $x\propto q_+/k$. In Minkowski quantum field theories, the regulator function $\upeta(q/\Lambda)$ serves as a low-pass filter for physical loop momentum $q$ lower than the cutoff scale $\Lambda$. In de Sitter, however, the loop momentum merely characterises comoving scales. Had we put a cutoff directly on a comoving scale $q_+$, scale invariance would have been broken, and the result would no longer be covariant under scaling. Instead, we chose to cut off the \textit{scale separation} between the loop modes and external modes, namely $q/k<\Lambda/H$. Since the external modes typically carry a physical momentum scale of order $k_{\rm p}\sim H$, our choice constrains the physical loop momentum by $q_{\rm p}<k_{\rm p}\Lambda/H\sim \Lambda$, consistently with the Minkowski picture.
    
    \item \textbf{Power-law divergences:} The power-law divergences in \eqref{psi2oneloopEtaRegResult} depend on the regulator choice. In fact, given a specific loop integral, one can tune the regulator function $\upeta(x)$ such that its Mellin transform vanishes at finite integer values, $C_n[\upeta]=0$, $n=1,\cdots,N$. In practice, this can be achieved by taking linear superpositions of different regulator functions. We have shown two such examples in the middle rows of Table \ref{etaRegExampleResultsTable}. Such $\upeta$ regulators are known as \textit{enhanced} regulators, and are essential for preserving gauge invariance in Minkowski quantum field theories \cite{Padilla:2024mkm}.
    
    \item \textbf{Reality:} Obviously, any real regulator function that obeys $\upeta^*(x)=\upeta(x)$ inevitably leads to a real loop contribution, $(\psi_2^{\rm 1L})^*=\psi_2^{\rm 1L}$. This is because the integrand \eqref{psi2OneLoopEtaRegAfterInsertion} is purely real and convergent after the regulator insertion. However, despite being less natural, there is no clear obstruction to allow complex regulator functions. The only potential issue is that divergences may become complex and require a complex counterterm. Unitarity of the bare Lagrangian thus demands that any sensible complex regulator function must give real divergences. This leaves two possibilities:
    \begin{itemize}
        \item[(a)] $\upeta(x)=\rho(-i x)$ with $\rho^*=\rho$. In this case, since $C_{2m}[\rho(-ix)]=(-1)^m C_{2m}[\rho(x)]\in \mathbb{R}$, $\upeta(x)$ gives purely real power-law divergences if the divergences are always of an even order, which is indeed the case in \eqref{psi2oneloopEtaRegResult}. This possibility will be important later when we consider the constraint from unitarity and analyticity.
        \item[(b)] $\upeta$ is an enhanced regulator and $C_n[\upeta]=0$ for any positive integer $n$. In this case, no power-law divergences are present from the beginning. We have shown two such examples in the lower rows of Table \ref{etaRegExampleResultsTable}, where the imaginary part of $\psi_2^{\rm 1L}$ seems completely \textit{arbitrary}.
    \end{itemize}
\end{itemize}
We will now turn to consider counterterms. 


\subsection{Renormalisation}

Renormalisation in $\upeta$ regularisation is much simpler than in schemes involving dimensionality changes, as the field lives in $d=3$ spatial dimensions where the computation is almost trivial. We again consider the counterterm Lagrangian \eqref{ct1} with $d=3$. The time integral can be trivially finished to give
\begin{align}
    \psi_2^{\rm ct}=\delta_{\rm ct}k^3~, 
\end{align}
as expected from scale invariance and IR convergence. Here the coefficient
\begin{align}
    \delta_{\rm ct}=\sum_{2l+p+q\geq 3} \,c_{l,p,q} \tilde{I}_{l,p,q}(3)~,
\end{align}
is purely real as dictated by the unitarity of the counterterm Lagrangian. Here the expression of $\tilde{I}_{l,p,q}(3)$ is given in \eqref{masslessDimRegCountertermIlpqd} (setting $d=3$). To cancel the divergences in $\psi_2^{\rm 1L}$, we set
\begin{align}
    \delta_{\rm ct}=-\frac{\lambda^2 H^2}{15(4\pi)^2} \left[\frac{45}{64} C_4[\upeta]\left(\frac{\Lambda}{H}\right)^4 - \frac{45}{64} C_2[\upeta] \left(\frac{\Lambda}{H}\right)^2 +\left(\log\frac{\Lambda}{\mu_0}-\frac{59}{256}-\log 2\right)\right]~,\label{psi2countertermEtaRegChoice}
\end{align}
where $\mu_0>0$ is an arbitrary reference energy scale to be fixed by the renormalisation condition on the scalar power spectrum. Crucially, unitarity demands the reality of each terms in \eqref{psi2countertermEtaRegChoice}. Combining \eqref{psi2oneloopEtaRegResult} and \eqref{psi2countertermEtaRegChoice}, we obtain the renormalised wavefunction coefficient at order $\lambda^2$,
\begin{align}
    \widehat{\psi}_2^{\rm 1L}=\psi_2^{\rm 1L}+\psi_2^{\rm ct}=\frac{\lambda^2 H^2}{15(4\pi)^2} k^3\left(\gamma[\upeta]+\log\frac{\mu_0}{H}\right)~,\label{psi2oneloopEtaRegResultRenormedGeneral}
\end{align}
with $\mu_0$ being an arbitrary real and positive constant to be fixed by the renormalisation condition on the power spectrum. Notice that $\gamma[\upeta]$ can, in principle, have an imaginary part that is protected from any adjustment of $\mu_0$ in the real counterterm.



\subsection{An ambiguous imaginary part?}

As discussed above, $\psi_2$ is not necessarily real: Taking the imaginary part, we have
\begin{align}
    \Im \widehat{\psi}_2^{\rm 1L}=\frac{\lambda^2 H^2}{15(4\pi)^2} k^3\,\Im \gamma[\upeta]~.
\end{align}
Unfortunately, this equation does not make sense since the left-hand side is a physical quantity, even observable according to \eqref{Impsi2}, whereas the right-hand side is a mathematical quantity that is intrinsically hypothetical. As discussed in Section \ref{EtaRegOneLoopSubsectionDiscussions}, the imaginary part of $\gamma[\upeta]$ is arbitrary. For all real regulator functions, we have $\Im \widehat{\psi}_2^{\rm 1L}=\Im \gamma[\upeta]=0$. For complex regulators, however, even if there is a constraint on the reality of the counterterm $\delta_{\rm ct}$, the finite remainder $\gamma[\upeta]$ remains unconstrained and can apparently take any value for its imaginary part. As an example, the complex enhanced regulators in the lower rows of Table \ref{etaRegExampleResultsTable} give $\Im \widehat{\psi}_2^{\rm 1L}\propto \theta$, with $\theta$ being arbitrary.

This apparent ambiguity of the imaginary part of the renormalised wavefunction coefficient is a serious problem. In the computation of loop diagrams in Minkowski quantum field theories, the removal of infinities is always accompanied by the freedom to add arbitrary constants to the finite part of loop diagrams that is consistent with redefining the counterterms. There the freedom is fulfilled by the renormalisation conditions, after which there is no further undetermined degrees of freedom. Here, we do observe such freedom in the parameter $\mu_0$: any real part of the finite remainder $\Re \gamma[\upeta]$ can be removed by a redefinition $\mu_0\to \mu_0 e^{-\Re \gamma[\upeta]}$. Yet due to the constraint from unitarity of the counterterm contributions, $\mu_0$ cannot shift in the imaginary direction. Therefore, there is no continuously tunable parameter in the counterterm Lagrangian that is capable of affecting this arbitrary imaginary part. This mathematical artefact is thus directly brought upon the physical world. 

There can be three ways out of this dilemma in increasing order of radicality:
\begin{enumerate}
    \item The first possibility is that the imaginary part of wavefunction coefficients is not arbitrary and that there exists \textit{constraints from fundamental principles} on the choice of regulators that force the imaginary part to take a specific value. We have already observed that dim reg and \mdreg schemes indeed give the same imaginary part after renormalisation. If we can show this for consistent $\upeta$ regulators, this option would constitute as a solution of the problem and the imaginary part would be an unambiguous \textit{prediction} of fundamental principles.
    \item A less elegant resolution is to give up unitarity and allow for complex counterterms, where one now has an additional continuous parameter to renormalise the imaginary part of a wavefunction coefficient. This option suggests that any such massless EFTs in de Sitter would be unavoidably non-unitary at loop level. One might then attempt to make sense of the theory as an \textit{open} EFT \cite{Salcedo:2024smn,Salcedo:2024nex,Salcedo:2025ezu}, in which case some couplings are real while some are purely imaginary. This approach would require working with a density matrix from the very beginning. 
    \item The perhaps most radical response is to give up predictability and admit that the imaginary part of the wavefunction coefficients is not computable within a field theory framework. It is simply a \textit{supplementary} parameter given in addition to the Lagrangian when defining a quantum field theory. Ideally, in a truly UV complete theory such as string theory, there is a preferred regulator of UV divergences that spits out a unique answer. Yet from the perspective of a low-energy EFT, this answer is somehow concealed. This radical perspective also suggests that any observation of the imaginary part would directly probe the UV completion of field theories in de Sitter.
\end{enumerate}

Depending on which assumptions one insists when defining a theory, all three options could be acceptable. However, in the following discussions, we show that the simplest solution, namely the first possibility, can indeed be realised at least at one-loop level. Therefore, in favour of simplicity, we shall advocate that there are consistency conditions for choosing regulators and the imaginary part of wavefunction coefficients is an unambiguous prediction of unitarity and analyticity.


\subsection{Constraint from the cosmological optical theorem}

Specifically speaking, we shall leverage unitarity and analyticity on the $\upeta$ regulators. In perturbation theory, one useful manifestation of unitarity and analyticity is the cosmological optical theorem (COT). As reviewed in Appendix \ref{COTApplicationToOurModelSubAppendix} (see \eqref{oneloopBubbleCOT}), at order $\mathcal{O}(\lambda^2)$ COT implies
\begin{align}
	\nonumber i \Disc [i\widehat\psi_2(k)]=&\int_{\mathbf{q}_1} P(q_1)\, (-i)\underset{q_1}{\Disc} \left[i\widehat\psi_4(k,q_1,q_1,k)\right]\\
	&+\frac{1}{2}\int_{\mathbf{q}_1} P(q_1)P(q_2)\, i\,\underset{q_1,q_2}{\Disc} \left[i\widehat\psi_3(k,q_1,q_2)\right]\times i\,\underset{q_1,q_2}{\Disc}  \left[i\widehat\psi_3(q_1,q_2,k)\right]~.
\end{align}
where the $\Disc$ operation is defined in \eqref{DiscDef}. Note that since $\Disc[i\psi_2^{\rm ct}]=0$\footnote{As shown in Appendix \ref{COTApplicationToOurModelSubAppendix}, the $\Disc$ of all tree-level diagrams vanish.}, the counterterm does not contribute on both sides. The right-hand side is given by (see Appendix \ref{COTApplicationToOurModelSubAppendix})
\begin{align}
    \text{right-hand side}=\frac{1}{15}\frac{\lambda^2H^2}{(4\pi)^2}k^3\times (-i\pi)~.
\end{align}
The left-hand side reads
\begin{align}
    \text{left-hand side}=-\left(\widehat\psi_2^{\rm 1L}(k)+\left[\widehat\psi_2^{\rm 1L}(-k^*)\right]^*\right)=\frac{1}{15}\frac{\lambda^2H^2}{(4\pi)^2}k^3\times \left(-2i\Im \gamma[\upeta]\right)~.
\end{align}
Therefore, to comply with the cosmological optical theorem, the imaginary part of the finite remainder of $\upeta$ regularisation must be fixed as
\begin{align}
    \Im \gamma[\upeta]=+\frac{\pi}{2}~,
\end{align}
and therefore
\begin{keyeqn}
    \begin{align}
        \widehat{\psi}_2^{\rm 1L}=\frac{1}{15}\frac{\lambda^2 H^2}{(4\pi)^2} k^3\left(\frac{i\pi}{2}+\log\frac{\mu}{H}\right)~,\label{psi2oneloopEtaRegResultRenormedFixedByCOT}
    \end{align}
\end{keyeqn}
where $\mu=\mu_0 e^{\Re \gamma[\upeta]}$ is an arbitrary reference energy scale to be fixed by renormalisation condition on the power spectrum. Note that \eqref{psi2oneloopEtaRegResultRenormedFixedByCOT} agrees with the results from dim reg \eqref{drfresult} and \mdreg \eqref{mdrfresult} schemes. In other words, from the requirement of unitarity and analyticity, the cosmological optical theorem fixes a unique imaginary part of wavefunction coefficients. In doing so, it also poses a selection rule on compatible $\upeta$ regulators.


\subsection{Unitary and analytic $\upeta$ regulators}\label{UnitaryAnalyticEtaRegSubsection}

To further investigate how unitarity and analyticity constrain the $\upeta$ regulators, we note that the cosmological optical theorem is derived diagram-by-diagram from the propagator identities,
\begin{align}
    \Disc [K(\eta,k)]&=0~,\\
    \Disc_p[i G(\eta_1,\eta_2,p)]&=-i P(p)\Disc_p K(\eta_1,p) \times \Disc_p K(\eta_2,p)~,
\end{align}
as well as the fact that the interaction vertices satisfy Hermitian analyticity,
\begin{align}
    V\left(\eta\partial_\eta,i\mathbf{k} \eta \right)=\left[V\left(\eta\partial_\eta,i(-\mathbf{k}) \eta \right)\right]^*~.
\end{align}
Using these identities, one can show that the cosmological optical theorem is always satisfied at the level of the integrand. However, to commute the $\Disc$ operation out of the regularised loop integrals and establish an equation for the wavefunction coefficients, the loop integrals must schematically satisfy
\begin{align}
    \Disc_{q} \left[~\int_{\mathbf{q}}~\right]_{\text{regularised}}=0~,
\end{align}
which, in our case, reduces to
\begin{align}
    \Disc_{q_+} \left[~\int\d q_+ \,\upeta\left(\frac{q_+/k}{\Lambda/H}\right)~\right]=0~.
\end{align}
Since the integral itself commutes with the $\Disc$, we are left with the Hermitian analyticity of the regulator function,
\begin{align}
    \upeta\left(\frac{q_+/k}{\Lambda/H}\right)=\left[\upeta\left(\frac{q_+/(-k^*)}{\Lambda/H}\right)\right]^*~,\quad k\in\mathbb{C}^-~.\label{HermitianAnalyticityOfEtaRegFun}
\end{align}
Note that in the physical domain, the external momentum $k$ is equipped with a small imaginary part i.e. $k=|k|-i\epsilon$ such that the bulk time integrals are convergent within a Bunch-Davies vacuum. In addition, to give a well-defined analytic continuation of the regularised wavefunction to $-k^*$, we require that the regulator has to function properly (i.e. the loop integral must converge) along \textit{any} path joining $k$ with $-k^*$ in the lower-half complex $k$-plane. To see how this requirement rules out some of the regulators presented in Table \ref{etaRegExampleResultsTable}, we take $\upeta_1(x)=e^{-x}$ as an example. After formally continuing the integrand of \eqref{psi2OneLoopEtaRegAfterInsertion} from $k$ to $-k^*$, we arrive at
\begin{align}
    \nonumber\psi_{2,\upeta_1}^{\rm 1L}(-k^*)&=\frac{\lambda^2 H^2}{16} \frac{\pi}{8(2\pi)^3 (-k^*)}\int_{-k^*}^\infty \d q_+ \\
    &\left(q_+^4 -\frac{2}{3} (-k^*)^2 q_+^2+ \frac{1}{5} (-k^*)^4\right)\frac{3 q_+^2+9 (-k^*) q_+ +8(-k^*)^2}{(-k^*+q_+)^3} \exp\left[-\frac{q_+/(-k^*)}{\Lambda/H}\right]~.
\end{align}
Since $\Re k>0$, instead of being regularised, this integral now exponentially \textit{diverges}. Therefore, the analyticity property is not preserved. It is straightforward to show that convergence is maintained only in the right-half complex $k$-plane with $\Re k>0$. To fix the convergence issue while satisfying Hermitian analyticity, we are motivated to rotate the regulator by $90^\circ$ and set $\upeta_2(x)=e^{i x}$. The regularised integral thus becomes
\begin{align}
    \psi_{2,\upeta_2}^{\rm 1L}(k)&=\frac{\lambda^2 H^2}{16} \frac{\pi}{8(2\pi)^3 k}\int_{k}^\infty \d q_+ \left(q_+^4 -\frac{2}{3} k^2 q_+^2+ \frac{1}{5}k^4\right)\frac{3 q_+^2+9 k q_+ +8k^2}{(k+q_+)^3} \exp\left[+i\,\frac{q_+/k}{\Lambda/H}\right]~.
\end{align}
This integral converges in the physical region because the external energy $k=|k|-i\epsilon$ has a small negative imaginary part. Furthermore, for arbitrary complex $k\in\mathbb{C}^-$, the integral is convergent. Consequently, the continuation to $-k^*$ can be smoothly performed,
\begin{align}
    \nonumber\psi_{2,\upeta_2}^{\rm 1L}(-k^*)&=\frac{\lambda^2 H^2}{16} \frac{\pi}{8(2\pi)^3 (-k^*)}\int_{-k^*}^\infty \d q_+ \\
    &\left(q_+^4 -\frac{2}{3} (-k^*)^2 q_+^2+ \frac{1}{5} (-k^*)^4\right)\frac{3 q_+^2+9 (-k^*) q_+ +8(-k^*)^2}{(-k^*+q_+)^3} \exp\left[+i\,\frac{q_+/(-k^*)}{\Lambda/H}\right]~.
\end{align}
In addition, Hermitian analyticity is satisfied,
\begin{align}
	\exp\left[+i\,\frac{q_+/k}{\Lambda/H}\right]=\exp\left[+i\,\frac{q_+/(-k^*)}{\Lambda/H}\right]^*~.
\end{align}
The resulting regularised wavefunction then follows from the property \eqref{finShiftAfterEtaRescaling} with $\lambda=-i$,
\begin{align}
    \psi_{2,\upeta_2}^{\rm 1L}&=\frac{\lambda^2 H^2}{15(4\pi)^2} k^3 \left[\frac{45}{64} \times 6\left(i\frac{\Lambda}{H}\right)^4 - \frac{45}{64} \times  \left(i\frac{\Lambda}{H}\right)^2 +\left(\log\frac{\Lambda}{H}-\gamma_E+\frac{i\pi}{2}-\frac{59}{256}-\log 2\right)\right]~,
\end{align}
where we indeed arrive at the imaginary part consistent with the cosmological optical theorem. Importantly, the power-law divergences stay real due to the even order, and so only unitary counterterms are needed for their removal, as expected from our discussions in Section \ref{EtaRegOneLoopSubsectionDiscussions}. Similarly, one can show that $\upeta_3(x)=(1+x)^{-6}$ does not work as the loop integral runs into a singularity after continuation. To fix the issue, one simply performs another $90^\circ$ rotation and use $\upeta_4(x)=(1-ix)^{-6}$, which gives the same imaginary part while maintaining the reality of power-law divergences. \footnote{Note also that $\upeta_5(x)=(1+ix)^{-6}$ does not work because the integrand is then not analytic for $k\in \mathbb{C}^-$, and the analytic continuation is ambiguous.}

In fact, we can translate the constraints directly into statements about the regulator function $\upeta(x)$ and prove that unitary and analytic $\upeta$ regulators always give the same imaginary part for the one-loop wavefunction. We shall define the following class of regulators:
\begin{keyeqn2}
    A regulator $\upeta$ is unitary and analytic if, for any $x\in \mathbb{C}^+$, it satisfies
    \begin{enumerate}\label{UnitaryAnalyticEtaRegProperties}
    \item Normalisation: $\upeta(0)=1$~,
    \item Convergence: $\lim_{|x|\to\infty}\upeta(x)= 0$~,
    \item Hermitian analyticity: $\upeta(x)=\left[\upeta(-x^*)\right]^*$ and $\upeta(x)$ being analytic~.
\end{enumerate}
\end{keyeqn2}
In summary, an $\upeta$ regulator is compatible with the cosmological optical theorem if and only if $\upeta$ is unitarity and analytic. 

\paragraph{Universality of unitary and analytic $\upeta$ regulators:} Now we can state our main observation of this section: \textit{all unitary and analytic $\upeta$ regulators give the same value for $\Im \gamma[\upeta]$ and hence for $\Im \psi_2^{\rm 1L}$}. 

To see this one can proceed as follow. The first (and third) condition implies that the regulator function admits a Taylor expansion at the origin,
\begin{align}
    \upeta(x)=1+c_1 x+c_2 x^2+\cdots~.
\end{align}
The third condition then gives
\begin{align}
    [\upeta(-x^*)]^*=1-c_1^* x+c_2^* x^2+\cdots=\upeta(x)~,
\end{align}
meaning that all odd (even) power coefficients must be purely imaginary (real),
\begin{align}
    &c_1^*=-c_1~,\quad c_3^*=-c_3~,\quad \cdots~,\\
    &c_2^*=+c_2~,\quad c_4^*=+c_4~,\quad \cdots~.
\end{align}
Alternatively, it must be a real analytic function of the combination $-ix$,
\begin{align}
    \upeta(x)=\rho(-ix)~,\quad \rho^*=\rho~.\label{etaRegAnsatz}
\end{align}
Thus the finite remainder is
\begin{align}
    \gamma[\upeta]=\gamma[\rho]+\log i=\gamma[\rho]+\frac{i\pi}{2}~.
\end{align}
Note that the reason why we have picked $-ix$ instead of $ix$ in \eqref{etaRegAnsatz} is because of the second criterion on convergence, which ensures\footnote{Since $x=q_+/k$ with $k\in \mathbb{C}^-$ and $q_+>0$, the dimensionless ratio $x$ should lie in the upper-half complex plane i.e. $x\in \mathbb{C}^+$. In particular, it can take the value $x=iR$, $R>0$.}
\begin{align}
    \lim_{|x|\to\infty}\upeta(x)=\lim_{R\to+\infty}\rho(-i\times iR)= \lim_{R\to\infty}\rho(R)=0~,\quad x=iR~,~R>0
\end{align}
 is well-defined. Therefore, we have proven that any $\upeta$ regularisation scheme satisfying the above criteria always generates the same imaginary part
\begin{keyeqn}
    \begin{align}
        \Im \gamma[\upeta]=+\frac{\pi}{2}\label{UniversalityOfImGamma}
    \end{align}
\end{keyeqn}
for one-loop wavefunction coefficients. Since spacetime dimensionality is untouched in this scheme, the counterterms are purely real, $\Im \psi_2^{\rm ct}=0$, leaving the imaginary part unchanged. Therefore, we obtain a \textit{universal} imaginary part $+i\pi/2$ for any unitary and analytic $\upeta$ regulators, consistent with both the dimensional regulators and the prediction of cosmological optical theorem.


\section{One-loop universality of the imaginary part}\label{imaginaryuniversal}

In this section, we greatly generalise the result found in the last section, and show that all renormalised one-loop wavefunction coefficients $\widehat\psi_n^{\rm 1L}$ share a universal imaginary part $+i \pi /2$ relative to their logarithmic divergences, provided that the regularisation schemes preserve unitarity and analyticity. More precisely, for any $n$-point wavefunction coefficient $\widehat\psi_n^{\rm 1L}$ at one loop one finds
\begin{align}
    \widehat\psi_n^{\rm 1L}(\{k\},\{s\},\{\mathbf{k}\})=f_0(\{k\},\{s\},\{\mathbf{k}\})\left( \frac{i\pi}{2}+\log\frac{\mu}{H}+g(\{k\},\{s\},\{\mathbf{k}\})\right)~,
\end{align}
where $f_0$ and $g$ are real functions of kinematics obtained from evaluating the loop integrand (see below) and $\mu$ is an arbitrary scale where a renormalisation condition is imposed. Here $\{k\}$ ($\{s\}$) collectively denote the external (internal) energies and $\{\mathbf{k}\}$ represent all momentum vectors.\\


\begin{figure}[h]
\centering
\begin{tikzpicture}[scale=2,
  arrow/.style={
    thick,
    postaction={decorate},
    decoration={
      markings,
      mark=at position 0.5 with {\arrow[scale=1.5]{stealth}}
    }
  }
]
  \def\R{0.5}
  \def\Ytop{1.1}
  \def\Ydots{0.9}
  \def\Ylabel{1.26}
  \def\Yq{-0.7}

  \draw[thick] (0,0) circle (\R);

  \draw[arrow] (300:\R) arc (300:230:\R);

  \node at (0,\Yq) {\(\mathbf{q}_1\)};

  \coordinate (VL) at (180:\R);
  \draw[arrow] (VL) -- (-1.2,\Ytop);
  \draw[arrow] (VL) -- (-0.7,\Ytop);
  \node at (-0.85,\Ydots) {\(\cdots\)};

  \coordinate (VM1) at (105:\R);
  \draw[thick] (VM1) -- (-0.2,\Ytop);
  \node at (-0.42,\Ydots) {\(\cdots\)};

  \coordinate (VM2) at (75:\R);
  \draw[thick] (VM2) -- (0.2,\Ytop);
  \node at (0.42,\Ydots) {\(\cdots\)};

  \coordinate (VR) at (0:\R);
  \draw[arrow] (VR) -- (0.7,\Ytop);
  \draw[arrow] (VR) -- (1.2,\Ytop);
  \node at (0.85,\Ydots) {\(\cdots\)};
  \node at (0,\Ydots) {\(\cdots\)};

  \node at (-0.8,-0.1) {\(v = 1\)};
  \node at (0.8,-0.1) {\(v = V\)};

  \draw[very thick] (-1.6,\Ytop) -- (1.6,\Ytop);

  \node at (-1.3,\Ylabel) {\(\mathbf{k}_1\)};
  \node at (-0.7,\Ylabel) {\(\mathbf{k}_{n_1}\)};
  \node at (0.7,\Ylabel) {\(\mathbf{k}_{n - n_V + 1}\)};
  \node at (1.3,\Ylabel) {\(\mathbf{k}_n\)};

\end{tikzpicture}
\caption{A general one-loop diagram contribution to the $n$-point wavefunction coefficient $\psi_n^{\rm 1L}$. Here we have explicitly spelled out the first and last vertices and the loop momentum $\mathbf{q}_1$.}\label{generalOneloopFeynmanDiagram}
\end{figure}
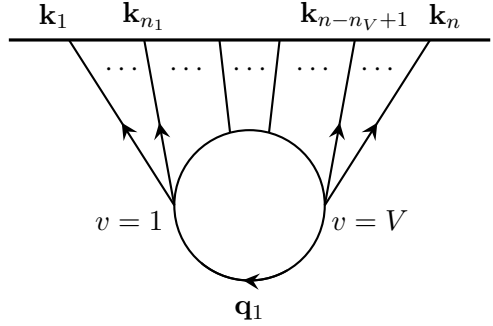


\subsection{Universality from $\upeta$ regulators}

The one-loop universality of the imaginary part is most transparent using the family of unitary and analytic $\upeta$ regulators. In $d=3$ spatial dimensions, consider a one-loop diagram with $V$ vertices, $n=n_1+\cdots +n_V$ external lines and $I$ internal lines (see Figure \ref{generalOneloopFeynmanDiagram}). On the $v$-th vertex, the interaction Lagrangian is characterised by a derivative operator $D_v(\eta_v \partial_{\eta_v},\eta_v\{\partial_i\})$ that is invariant under scale transformation $\alpha: (\eta,x^i)\mapsto (\alpha^{-1}\eta,\alpha^{-1} x^i)$. In momentum space, the regularised wavefunction schematically reads
\begin{align}
	\nonumber&\psi_n^{\rm 1L}(\{k\},\{s\},\{\mathbf{k}\})\\
    =&\int_{\mathbf{q}_1}\upeta\left(\frac{q_+/k_T}{\Lambda/H}\right) \int_{-\infty}^0\left[\,\prod_{v=1}^V   \d\eta_v \, a^4(\eta_v)\,i\lambda_v\, D_{v}(\eta_v \partial_{\eta_v},-i\eta_v\{\mathbf{k}\})\right]\left[\,\prod_{e=1}^n K_e(\eta_e,k_e)\right] \left[\,\prod_{{\rm i}=1}^I G_{\rm i}(\eta_{{\rm i}},\eta_{{\rm i}}';q_{\rm i})\right]~,\label{oneloopPsiGeneralEtaReg}
\end{align}
where $\eta_e\equiv \eta_{v(e)}$ denotes the conformal time for vertex attached to the $e$-th external line and similarly for $\eta_{{\rm i}},\eta_{{\rm i}}'$ and internal lines. Here $q_+\equiv q_1+\cdots +q_I$ denotes the sum of internal energies. We then perform a Wick rotation $\eta=e^{i\pi/2}\chi$ to obtain
\begin{align}
	\nonumber&\psi_n^{\rm 1L}(\{k\},\{s\},\{\mathbf{k}\})=\int_{\mathbf{q}_1} \upeta\left(\frac{q_+/k_T}{\Lambda/H}\right)\\
    &\times\int_0^{\infty}\left[\,\prod_{v=1}^V   \d\chi_v \, a^4(-\chi_v)\lambda_v\, D_{v}(\chi_v \partial_{\chi_v},\chi_v\{\mathbf{k}\})\right]\left[\,\prod_{e=1}^n K_e(i\chi_e,k_e)\right] \left[\,\prod_{{\rm i}=1}^I G_{\rm i}(i\chi_{{\rm i}},i\chi_{{\rm i}}';q_{\rm i})\right]~.\label{oneloopPsiGeneralEtaRegWickRotated}
\end{align}
 Using the property that in de Sitter with $d=3$ spatial dimensions, both the bulk-boundary and bulk-bulk propagators are purely real after Wick rotation (see \eqref{argnaivedimreg}),
\begin{subequations}
    \begin{align}
        \arg [K_e(i\chi_e,k_e)]&=0~,\\
        \arg [G_{\rm i}(i\chi_{{\rm i}},i\chi_{{\rm i}}';q_{\rm i})]&=0~,\label{BkToBkPropRealityIn3d}
    \end{align}
\end{subequations}
we deduce that the loop integrand is a real function of kinematics and that
\begin{align}
    \psi_n^{\rm 1L}(\{k\},\{s\},\{\mathbf{k}\})&=\int_{q_+^{\rm min}}^{\infty} \frac{\d q_+}{q_+} \upeta\left(\frac{q_+/k_T}{\Lambda/H}\right)\times f(\{k\},\{q\},\{\mathbf{k}\})~,\quad f^*=f~.\label{GeneralOneLoopIntegrals}
\end{align}
Here we have finished all integrals along compact dimensions in the phase space and $q_+^{\rm min}= q_+^{\rm min}(\{\mathbf{k}\})$ denotes the threshold for the non-compact radial direction $q_+$. Trading the loop integral with Mellin transforms $C_m[\upeta]$ of the regulator, we have
\begin{align}\label{above2}
    \nonumber \psi_n^{\rm 1L}(\{k\},\{s\},\{\mathbf{k}\})=&\sum_{m >0} f_m(\{k\},\{s\},\{\mathbf{k}\}) \,C_m[\upeta]\left(\frac{\Lambda}{H}\right)^m\\
    &+f_0(\{k\},\{s\},\{\mathbf{k}\})\left(\log \frac{\Lambda}{H}+\gamma[\upeta]+g(\{k\},\{s\},\{\mathbf{k}\})\right)+\mathcal{O}\left(\frac{H}{\Lambda}\right)~,
\end{align}
where $f_m$ are real functions of kinematics derivable from the loop integrand via the large-$q_+$ expansion, 
\begin{align}
     f=\sum_{m\neq 0}  \left(\frac{q_+}{k_T}\right)^m f_m + f_0 ~,\label{generalOneloopLargeqExpansion}
\end{align}
and $g(\{k\},\{s\},\{\mathbf{k}\})$ is a real dimensionless function of kinematics that follows from evaluating the finite piece of the integral, which therefore does not depend on the regulator choice.\footnote{In the two-point case, $g=-\frac{59}{256}-\log 2$ is a pure number (see \eqref{psi2oneloopEtaRegResult}). In more general cases, $g$ can acquire non-trivial kinematic dependence. See Appendix \ref{FiniteRemainderAppendix} for the derivation of the general form of the finite remainder.} The $\gamma[\upeta]$ appearing in \eqref{above2} was defined in \eqref{gammaeta1}, which we report here
\begin{align}
	\gamma[\upeta]=-\int_{0}^{\infty}\d x \log x\, \upeta'(x)~.
\end{align}
In particular, we stress that $\gamma[\upeta]$ depends exclusively on the choice of regulator $\upeta$, and not on any details of the diagram, such as the number of vertices $V$, the number of external legs $n$, or the interaction vertices. 

In $\upeta$ regularisation, the counterterms are always purely real and they serve to remove the real divergences in \eqref{above2}, leaving
\begin{align}
    \widehat\psi_n^{\rm 1L}(\{k\},\{s\},\{\mathbf{k}\})=f_0(\{k\},\{s\},\{\mathbf{k}\})\left(i\, \Im \gamma[\upeta]+\log\frac{\mu}{H}+g(\{k\},\{s\},\{\mathbf{k}\})\right)~,\label{PsiOneloopStructure}
\end{align}
where $\mu>0$ is an arbitrary reference energy scale to be fixed by renormalisation conditions. As stated in Section \ref{UnitaryAnalyticEtaRegSubsection}, a unitary and analytic regulator always gives $\Im \gamma[\upeta]=+\pi/2$, henceforth we conclude that the imaginary part of all renormalised $n$-point one-loop wavefunction coefficients is universally fixed by
\begin{keyeqn}
    \begin{align}
        \Im \widehat\psi_n^{\rm 1L}=f_0(\{k\},\{s\},\{\mathbf{k}\})\times \frac{\pi}{2}~,\label{ImPsiUniversality}
    \end{align}
\end{keyeqn}
In other words, the coefficient of the imaginary part relative to the (real) logarithmic divergence $f_0$ is always $\pi/2$. Since the logarithmic divergence often characterises the renormalisation group running of the theory, one can formally trade $f_0$ with the running with respect to the sliding scale $\mu$ and write,
\begin{keyeqn}
    \begin{align}
        \left(\mu\frac{\partial}{\partial \mu}-\frac{2}{\pi}\,\Im\right)\widehat\psi_n^{\rm 1L}=0~.\label{RG=Im}
    \end{align}
\end{keyeqn}
This form of the universality equation hints at a deep connection between the spontaneous emergence of imaginary parts and the renormalisation group (RG) running of the wavefunction coefficients. However, unlike in Minkowski quantum field theories, the energy scale $\mu$ is not directly associated with a tunable apparatus or observed kinematics. Therefore it remains unclear what \textit{RG running} physically stands for in such IR-convergent de Sitter EFTs.\footnote{One possible way to understand RG running in de Sitter is to slightly deform the geometry. This can be achieved by fiat, i.e. placing the theory in a different geometry, as for example one does when discussing the conformal anomaly. Alternatively, one could introduce an adiabatic time dependence in the Hubble parameter $H(t)$. This softly breaks scale invariance, but as long as the Hubble changes adiabatically in time, one can locally approximate the spacetime by an exact de Sitter and simply dial the Hubble parameter in the scale-invariant results.} A systematic discussion is beyond the scope of this paper and we leave it to future works.


\paragraph{Generalisation to bulk fields with mass and spin} The proof of the one-loop universality of the imaginary part is so far based on an interacting theory of scalar fields. Naturally, we allow for any $n$ external massless states, any IR-convergent interactions (including those breaking the de Sitter boosts), any one-loop topologies, and any function\footnote{We remind that there are \textit{uncountably} many such regulator functions.} $\upeta(x)$ satisfying the three conditions in \eqref{UnitaryAnalyticEtaRegProperties}. However, the validity of \eqref{RG=Im} goes much beyond such massless scalar EFTs. In fact, it is easily generalisable to include massive spinning fields in the bulk. Consider a massive spin-$S$ field within the \textit{cosmological condensed matter} construction in $d=3$ spatial dimensions \cite{Bordin:2018pca,Stefanyszyn:2023qov},
\begin{align} \label{SigmaFreeAction}
	S_0[\sigma]=\frac{1}{2 S!}\int \d\eta \d^3x a(\eta)^{2}\Bigg[&\sigma'^2_{i_1\cdots i_S}-c^2 (\partial_j\sigma_{i_1\cdots i_S})^2-\delta c^2(\partial_j\sigma_{j i_2\cdots i_S})^2 \nonumber \\ &-a(\eta)^2m^2\sigma^2_{i_1\cdots i_S}
	-2Sa(\eta)\kappa\epsilon_{ijk}\sigma_{i l_2\cdots l_S}\partial_j \sigma_{k l_2\cdots l_S}\Bigg]~,
\end{align}
where we have introduced all possible boost-breaking terms up to two derivatives. Decomposing the field in helicity modes,
\begin{align}
	\sigma_{i_{1} \cdots i_{S}}(\eta, \bfx) = \sum_{h = -S}^{S}\int_{\mathbf{k}} \sigma_{h}(\eta, k) a_\sigma^{(h)}(\mathbf{k}) {\rm{e}}^{(h)}_{i_{1} \cdots i_{S}}(\bfk) e^{i \bfk \cdot \bfx}+\rm {h.c.}~,
\end{align}
where $a_\sigma^{(h)}(\mathbf{k})$ denotes the annihilation operator in the canonical formalism. The mode function can be solved in terms of the Whittaker-$W$ function,
\begin{align} \label{CCMmodefunction}
	\sigma_{h}(\eta, k) =  e^{-\pi\tilde{\kappa}_h/2}  \frac{- H \eta}{\sqrt{2 c_{h} k}}  W_{i \tilde{\kappa}_h, \nu}(2 i c_{h} k \eta)~, \quad \nu = \sqrt{\frac{9}{4} - \frac{m^2}{H^2}}~,
\end{align}
where $c^2_{h} = c^2 + \frac{S^2-h^2}{S(2S-1)} \delta c^2$ is the helicity-dependent sound speed and $\tilde{\kappa}_h = \frac{h \kappa}{c_{h} H}$ is the helical chemical potential. It is straightforward to show that the bulk-bulk propagator in the helical basis,
\begin{align}
	G_{\sigma}^{(h)}(\eta_{1}, \eta_{2};k) &=[\sigma_{h}(\eta_{1}, k)\sigma^{\star}_{h}(\eta_{2},k) \theta(\eta_{1}-\eta_{2})  + (\eta_{1} \leftrightarrow \eta_{2})] - \frac{\sigma_{h}(\eta_{0},k)}{\sigma^{\star}_{h}(\eta_{0},k)}\sigma^{\star}_{h}(\eta_{1},k)\sigma^{\star}_{h}(\eta_{2},k)~, \label{bulk-bulkSolution}
\end{align}
satisfies the following property under Wick rotation \cite{Stefanyszyn:2023qov},
\begin{align}
	\left[G_{\sigma}^{(h)}(i\chi_1, i\chi_2; k) \right]^{\star} = G_{\sigma}^{(-h)}(i\chi_1, i\chi_2; k) ~,\label{helicalRealityForLightFields}
\end{align}
subjected to the condition that the field mass is light i.e. $\nu\in\mathbb{R}$ is in the complementary series, and that the boundary time smoothly tends to zero i.e. $\eta_0\to 0$\footnote{For heavy bulk fields that live in the principle series i.e. $\nu=i\mu$, the helical reality relation \eqref{helicalRealityForLightFields} is no longer true, as the last term in \eqref{bulk-bulkSolution} does not admit a well-defined $\eta_0\to 0$ limit. However, it can be shown that there is always a connected piece in the bulk-bulk propagator that does satisfy reality, together with a leftover piece that is factorised, complex and $\eta_0$-dependent \cite{Stefanyszyn:2023qov}. As a result, under such circumstances, the universality relation must be extended with additional $\eta_0$-dependent contributions from the factorised pieces.}. Consequently, the bulk-bulk propagator in the indexed basis
\begin{align}
        G_{i_{1} \cdots i_{S} \,j_{1} \cdots j_{S}}(\eta_1, \eta_2; \bfk) = \sum_{h=-S}^{S} G_{\sigma}^{(h)}(\eta_1, \eta_2;k)  {\rm{e}}^{(h)}_{i_{1} \cdots i_{S}}(\bfk){\rm{e}}^{(h)}_{j_{1} \cdots j_{S}}(-\bfk)
\end{align}
is purely real after Wick rotation,
\begin{align}
        \arg [G_{i_1\cdots i_S \,j_1\cdots j_S}(i\chi_{{1}},i\chi_{{2}};\mathbf{k})]&=0~,
\end{align}
in alignment with the massless case \eqref{BkToBkPropRealityIn3d}. Therefore, the derivation above remains the same and our previous results remain unchanged. This concludes the generalisation of the universality relation \eqref{RG=Im} to theories with arbitrary light bulk fields with integer spins, non-trivial sound speeds and chemical potentials.

\subsection{Universality from dimensional regulators}


To further confirm the one-loop universality of the imaginary part, we now turn to two other regularisation schemes that are known to preserve unitarity and analyticity. For simplicity, we only consider massless bulk fields for the verification here.
\paragraph{Via dimensional regularisation}

In dim reg, the one-loop wavefunction reads
\begin{align}
	\nonumber&\psi_n^{\rm 1L}(\{k\},\{s\},\{\mathbf{k}\})\\
    =&\int_{\mathbf{q}_1\in \mathbb{R}^d}\int_{-\infty}^0\left[\,\prod_{v=1}^V   \d\eta_v \, a^{d+1}(\eta_v)\,i\lambda_v\, D_{v}(\eta_v \partial_{\eta_v},-i\eta_v\{\mathbf{k}\})\right]\left[\,\prod_{e=1}^n K_e(\eta_e,k_e)\right] \left[\,\prod_{{\rm i}=1}^I G_{\rm i}(\eta_{{\rm i}},\eta_{{\rm i}}';q_{\rm i})\right]~,\label{oneloopPsiGeneraldimReg}
\end{align}
which after Wick rotation, becomes
\begin{align}
	\nonumber&\psi_n^{\rm 1L}(\{k\},\{s\},\{\mathbf{k}\})=\int_{\mathbf{q}_1\in \mathbb{R}^d}  e^{i\pi(d+1)V/2}\\
    &\times\int_0^{\infty}\left[\,\prod_{v=1}^V   \d\chi_v \, a^{d+1}(-\chi_v)\lambda_v\, D_{v}(\chi_v \partial_{\chi_v},\chi_v\{\mathbf{k}\})\right]\left[\,\prod_{e=1}^n K_e(i\chi_e,k_e)\right] \left[\,\prod_{{\rm i}=1}^I G_{\rm i}(i\chi_{{\rm i}},i\chi_{{\rm i}}';q_{\rm i})\right]~.\label{oneloopPsiGeneraldimRegWickRotated}
\end{align}
Since the bulk-boundary and bulk-bulk propagators behave as (see \eqref{argdimreg})
\begin{subequations}
    \begin{align}
        \arg [K_e(i\chi_e,k_e)] &= 0~,  \\
        \arg [G_{\rm i}(i\chi_{{\rm i}},i\chi_{{\rm i}}';q_{\rm i})] &=  -\frac{\pi (d-3)}{2}~,
    \end{align}
\end{subequations}
the regularised wavefunction thus behaves as
\begin{align}
    \nonumber\psi_n^{\rm 1L}&=e^{i\pi(d+1)V/2}\times 1\times e^{-i\pi (d-3)I/2}\times  k^{d-3}\left(\frac{f_0}{3-d}+\text{real and finite}\right)\\
    &=f_0\left(\frac{1}{\epsilon}-\log k+\text{finite \& real}\right)~,\label{oneloopPsiGeneraldimRegPhaseEstimate}
\end{align}
where we have applied the one-loop topology $1=I-V+1$. Here $f_0$ is expected to agree with that in the $\upeta$ regularisation thanks to the universality of the logarithmic divergences. The appearance of a scale-dependent term $\log k$ is expected as in Section \ref{dimRegSection}. This result being purely real agrees with the phase formula in \cite{Goodhew:2024eup}. Unlike in $\upeta$ regularisation, the counterterm contributes to the imaginary part. The power counting is basically the same as in the loop diagram, but with $0=I-V+1$,
\begin{align}
    \nonumber\psi_n^{\rm ct}&=e^{i\pi(d+1)V/2}\times 1\times e^{-i\pi (d-3)I/2}\times k^{d-3}\delta_{\rm ct} f_0\\
    \nonumber&=e^{-i\pi\epsilon/2} k^{-\epsilon} \delta_{\rm ct} f_0\\
    &=-f_0\left(\frac{1}{\epsilon}-\frac{i\pi}{2}-\log k\right)~,\label{coutnertermPsiGeneraldimRegPhaseEstimate}
\end{align}
where we have set $\delta_{\rm ct}=-1/\epsilon$ to remove the divergence. Combining \eqref{oneloopPsiGeneraldimRegPhaseEstimate} and \eqref{coutnertermPsiGeneraldimRegPhaseEstimate}, the scale-dependent terms cancel and we again arrive at the universal imaginary part after renormalisation,
\begin{align}
    \widehat\psi_n^{\rm 1L}=f_0\left(\frac{i\pi}{2}+\log\frac{\mu}{H}+\text{finite \& real}\right)~,
\end{align}
with $\mu>0$ being arbitrary. Interestingly, in dim reg, the imaginary part comes from the counterterm as opposed to the loop diagram.

\paragraph{Via mass-dimensional regularisation}

In mass-dimensional regularisation, we also extend the mass of the field such that the mode function is simple. The expression of the one-loop diagram after Wick rotation is the same as \eqref{oneloopPsiGeneraldimRegWickRotated}. The difference from dim reg lies in the phase information of the propagators (see e.g. \eqref{argumassdimregprop}),
\begin{subequations}
    \begin{align}
        \arg [K_e(i\chi_e,k_e)] &= -\frac{\pi (d-3)}{4}~,  \\
        \arg [G_{\rm i}(i\chi_{{\rm i}},i\chi_{{\rm i}}';q_{\rm i})] &=  -\frac{\pi (d-3)}{2}~.
    \end{align}
\end{subequations}
Thus the power counting for the regularised loop diagram reads
\begin{align}
    \nonumber\psi_n^{\rm 1L}&=e^{i\pi(d+1)V/2}\times e^{-i\pi (d-3)n/4}\times e^{-i\pi (d-3)I/2}\times (-\eta_0)^{n(3-d)/2 }k^{(n-2)(3-d)/2}\\
    \nonumber&\quad\times\left(\frac{f_0}{3-d}+\text{finite \& real}\right)\\
    &=f_0\left(\frac{1}{\epsilon}+\frac{i\pi n}{4}+\frac{n}{2}\log(-\eta_0)+\frac{n-2}{2}\log k+\text{finite \& real}\right)~,\label{oneloopPsiGeneralmassdimRegPhaseEstimate}
\end{align}
where the time-dependent terms appear due to the explicit $\eta_0$-dependence of the bulk-boundary propagators (see \eqref{massdimregprop}). The scale-dependent terms then arise because scale invariance in $d$ dimensions require $k$ and $\eta_0$ to appear together. Turning to the counterterm with $0=I-V+1$, we have
\begin{align}
    \nonumber\psi_n^{\rm ct}&=e^{i\pi(d+1)V/2}\times e^{-i\pi (d-3)n/4}\times e^{-i\pi (d-3)I/2}\times (-\eta_0)^{n(3-d)/2 }k^{(n-2)(3-d)/2}\delta_{\rm ct} f_0\\
    \nonumber&=e^{i\pi(n-2)\epsilon/4} (-\eta_0)^{n\epsilon/2} k^{(n-2)\epsilon/2}\delta_{\rm ct} f_0\\
    &=-f_0\left(\frac{1}{\epsilon}+\frac{i\pi(n-2)}{4}+\frac{n}{2}\log(-\eta_0)+\frac{n-2}{2}\log k\right)~.\label{coutnertermPsiGeneralmassdimRegPhaseEstimate}
\end{align}
With the notable exception of $n=2$, the counterterms contribute to the imaginary part. Combining \eqref{oneloopPsiGeneralmassdimRegPhaseEstimate} and \eqref{coutnertermPsiGeneralmassdimRegPhaseEstimate}, both the time-dependent and scale-dependent terms cancel and we again arrive at the universal imaginary part after renormalisation,
\begin{align}
    \widehat\psi_n^{\rm 1L}=f_0\left(\frac{i\pi}{2}+\log\frac{\mu}{H}+\text{finite \& real}\right)~,
\end{align}
with $\mu > 0$ being arbitrary. Note that in \mdreg, the imaginary part receives contribution from the both the loop contribution and the counterterm. Remarkably, their combination gives precisely the result $\Im\widehat\psi_n^{\rm 1L} = f_0\pi/2$.


\section{Conclusions and outlook} \label{ConclusionsSect}

Quantum field theory in cosmological spacetimes has a rich structure of UV divergences that is tightly constrained by fundamental principles of physics. In this work, focusing on wavefunction coefficients in de Sitter space, we have demonstrated how to perform consistent renormalisation for one-loop UV divergences that is compatible with unitarity and analyticity as encoded in the cosmological optical theorem. Taking a massless scalar EFT with shift-symmetric couplings as an example, we computed the two-point wavefunction coefficient to one-loop order using three variants of dimensional regularisation and found them to agree only after proper renormalisation. To further expose the necessary consistency requirements, we then introduced the general framework of $\upeta$ regularisation based on its recent Minkowski proposal. Within this framework, we found that the imaginary part of wavefunction coefficients can become arbitrary and unpredictable unless the regulator commutes with the discontinuity operation present in the cosmological optical theorem. This leads to our proposition of unitary and analytic $\upeta$ regulators from which one is granted with a unique prediction for the renormalised wavefunction coefficient. The implication of such unitary and analytic renormalisation is profound: we were thereby able to prove that in more general theories, the imaginary part of all $n$-point wavefunction coefficients is universally fixed in terms of the logarithmic running of the real part, indicating a deep connection between the loop-level emergence of imaginary part and the renormalisation group (RG) running of de Sitter EFTs.\\

This work opens up a variety of exciting avenues for future exploration:
\begin{itemize}
    \item \textbf{More on $\upeta$ regulators:} $\upeta$ regularisation proves to be useful in the classification of general UV divergences and plays a crucial role in demonstrating universality statements. However, the merit of $\upeta$ regularisation goes beyond formal aspects: it also greatly simplifies the practical computation of loops diagrams. For instance, one of its most satisfactory advantages is that it avoids the subtleties involving the change of spacetime dimensionality. Both the loops and the counterterms are constructed out of simple propagators and interaction vertices living in $3+1$ dimensions, enabling them to be exactly solved with a clear analytic structure. It is therefore tantalising to excavate more analytical results in realistic models utilizing the $\upeta$ regulators. Going beyond inflation, $\upeta$ regulators also straightforwardly generalise to general FLRW cosmologies, for example, in the computation of classical loops in the EFT of large scale structure.
    
    \item \textbf{Imaginary part and the RG running:} Our findings clearly point to a hidden relationship between the spontaneous emergence of the imaginary part and the RG running of the real part of the wavefunction coefficients. While it remains unclear how in practice one should interpret the RG running of de Sitter EFTs at the level of the observables (e.g. cosmological correlators), it is nevertheless interesting to follow the Minkowski analogy and search for a precise physical understanding for logarithmic UV divergences in de Sitter and why they are tied together to the imaginary part.

    \item \textbf{Parity anomaly vs trace anomaly:} One most important motivation for studying the imaginary part of wavefunction coefficient is the parity-odd correlators. At tree level, the no-go theorems forbidding the presence of parity-odd correlators may be understood as a sequence of classical symmetry requirements and $\ex{\phi^n}_{\rm PO}=0$ appears as a discrete Ward identity. The loop-level emergence of $\ex{\phi^n}_{\rm PO}\neq 0$ then suggests that some of the classical symmetries must be broken at quantum level, as required by the consistency of the loop regulators. Since in a broad class of models, only quantum loops are UV-divergent and hence relevant for regularisation, this phenomenon can be understood as a \textit{quantum anomaly}. In vague resemblance to this, in Minkowski quantum field theories, the emergence of RG running at loop levels is due to the regulators inevitably breaking (energy-) scale invariance\footnote{For an insightful take on this perspective, see Coleman's lecture notes \cite{Coleman:1985rnk}.}, a phenomenon widely known as the trace anomaly. It is therefore captivating to explore the potential connection between the parity anomaly and the trace anomaly in future works, especially in consideration of the previous point.

    \item \textbf{Extending the factorisation formulae:}  The correlator-to-correlator factorisation (CCF) formulae proposed in \cite{Stefanyszyn:2024msm,Stefanyszyn:2025yhq} are a direct consequence of the reality of tree-level wavefunction coefficients, and they serve to link different cosmological observables to each other. At one-loop order, the reality statement is anomalous, suggesting that the CCFs must be extended to include an anomalous contribution from the loop effects. It is interesting to investigate if such an anomalous term can be recast into cosmological observables and hence establishing an autonomous system of CCFs.

    \item \textbf{Loops of heavy fields and fermions?} Our universality relation applies to loops of light bosonic fields as their reality property is transparent under the Wick rotation. A natural question to ask is what happens when heavy fields are involved in the loop. It was known that heavy particles in de Sitter decay into themselves via the imaginary shift of their pole masses \cite{Marolf:2010zp,Krotov:2010ma,Jatkar:2011ju,Bros:2006gs}. How to incorporate this fact into our picture of unitary and analytic renormalisation is an interesting direction to explore. Similarly, one might expect fermion loops to enjoy different divergence structures due to different statistics, which is yet another fascinating route to pursue.
    
    \item \textbf{Higher loops?} So far we have restricted our attention to one loop results, which of course barely scratches the surface of renormalisation. As expected from the Minkowski experience, higher loop orders are much more complex and exhibit richer UV divergence structures. Although explicit calculations could soon become tedious and intractable, one might still learn from the general patterns that arise, for example, when $\upeta$ regulators are suitably implemented. An alternative regularisation scheme for higher loops involves going to Mellin space, where the divergences can be resolved by contour pinching techniques \cite{Premkumar:2021mlz}. It is thus interesting to consider the interplay of different schemes at higher loops.

    \item \textbf{How about classical loops?} In our setup, classical loops are convergent because of derivative couplings. However, this is by no means true in general. For non-shift-symmetric couplings, even the classical loops can be UV-divergent, meaning that to go from renormalised wavefunction coefficients to correlators, one needs extra counterterms to cancel the new divergences from contracting wavefunction coefficients (see also \cite{Huenupi:2024ztu} for the renormalisation of IR divergences in classical loops). Do we still have enough freedom to rearrange the counterterms in the Lagrangian to cancel the new divergences? If we cannot simultaneously renormalise the wavefunction coefficients (a.k.a. the boundary CFT correlators) and the correlators, which one then is preferred? These questions are necessary to address in the future.
    
    \item \textbf{Renormalisation in open EFTs:} Despite that we have rejected the possibility of \textit{unavoidable} unitarity violation in closed systems, non-unitary effects are nevertheless present had we begun with an open system \cite{Salcedo:2024smn,Salcedo:2024nex,Salcedo:2025ezu}. It is hence interesting to study the renormalisation of loop divergences in the density matrix/influence functional in open EFTs. One pressing question to ask is that if we apply $\upeta$ regularisation to an open EFT, what would be the consistency requirements now that unitarity is absent.
    
\end{itemize}


\paragraph{Acknowledgements} We would like to thank Ciaran McCulloch, Swagat S. Mishra, Antonio Padilla, Lucas Pinol, Zhehan Qin, Leonardo Senatore, David Stefanyszyn, Ayngaran Thavanesan, Aron Wall, Dong-Gang Wang, Yi Wang, and Yuhang Zhu for helpful discussions. This work has been supported by STFC consolidated grant ST/X001113/1, ST/T000694/1, ST/X000664/1 and EP/V048422/1. DJ is supported by the Simons Collaboration for Celestial Holography.

\newpage
\appendix

\section{Partial dimensional regularisation}\label{partialDimRegAppendix}

As pointed out in the main text, there are multiple ways to implement dimensional regularisation in de Sitter space. In this Appendix, we discuss the dimensional regularisation scheme introduced by Weinberg in \cite{Weinberg:2005vy}, which we refer to as \textit{partial dim reg}. In this approach, the momentum integrals are analytically continued to $d$ dimensions, while the mode functions are left unchanged. As a result, both the bulk-boundary and bulk-bulk propagators retain their three-dimensional forms. 
We will compute the one-loop contribution to $\psi_2$ using this scheme, examine the associated UV divergences, and identify the counterterms needed to cancel them. Finally, we will highlight some of the limitations and disadvantages of this regularisation method.


\subsection{Loop contribution}

In $3+1$ dimensions, the one-loop contribution to the two-point wavefunction coefficient arising from the Feynman diagram given in Figure \ref{figloop} is
\begin{align}
	\psi_2^{\rm 1L}
    =-\frac{\lambda^2}{2 H^2}\int\frac{\d^3 q_1}{(2\pi)^3}\int \frac{\d\eta_1}{\eta_1} \frac{\d\eta_2}{\eta_2}\partial_{\eta_1}K(\eta_1,k) \partial_{\eta_2}K(\eta_2,k) \partial_{\eta_1} \partial_{\eta_2} G(\eta_1,\eta_2; q_1)\, \partial_{\eta_1} \partial_{\eta_2} G(\eta_1,\eta_2; q_2)~,
\end{align}
where $\mathbf{q}_2\equiv \mathbf{q}_1-\mathbf{k}$ by momentum conservation and $K (\eta,k)$ and $G(\eta_1,\eta_2;q)$  denote the bulk-boundary propagator and the bulk-bulk propagator respectively. In $d=3$ spatial dimensions, the bulk-boundary and bulk-bulk propagators for a massless field take the following form
\begin{align}
	K(\eta,k)&=(1-ik\eta)e^{ik\eta}~,\label{MasslessBtobPropIn3dim}\\
	G(\eta_1,\eta_2;k)&=\frac{i H^2}{k^3}[k \eta_1 \cos(k\eta_1)-\sin (k\eta_1)] (1-i k\eta_2)e^{ik \eta_2}\theta(\eta_1-\eta_2)+(\eta_1\leftrightarrow\eta_2)~.\label{MasslessBtoBPropIn3dim}
\end{align}
Note that after Wick rotation i.e. $\eta\rightarrow e^{i\pi/2} \chi$, the bulk-boundary and bulk-bulk propagators become purely real, therefore
\begin{align}\label{argnaivedimreg}
	 \arg[K(e^{i\pi/2}\chi,k)] &= 0~,\\
     \arg[G(e^{i\pi/2}\chi_1,e^{i\pi/2}\chi_2;k)] &= 0~,
\end{align}
up to $\pm \pi$ phases and hence 
\begin{equation}
    \arg[\psi_2^{\rm 1L}] = 0~.
\end{equation}
Using \eqref{MasslessBtobPropIn3dim} and \eqref{MasslessBtoBPropIn3dim}, the 1-loop contribution reduces to
\begin{align}
		\psi_2^{\rm 1L}=-\frac{\lambda^2 H^2}{8}\int\frac{\d^3 q_1}{(2\pi)^3}k^4 q_1 q_2 & \int_{-\infty}^0 \d\eta_1 \d\eta_2 \eta_1^2 \eta_2^2 \Bigg[e^{i(k+q_1+q_2)(\eta_1+\eta_2)}\nonumber\\
        &+e^{i k (\eta_1+\eta_2)}e^{-i (q_1+q_2)(\eta_1-\eta_2)}\left(1-e^{2i q_1 \eta_1}-e^{2i q_2 \eta_1}\right)\theta(\eta_1-\eta_2) \nonumber\\
		&+e^{i k (\eta_1+\eta_2)}e^{-i (q_1+q_2)(\eta_2-\eta_1)}\left(1-e^{2i q_1 \eta_2}-e^{2i q_2 \eta_2}\right)\theta(\eta_2-\eta_1) \Bigg]~.
\end{align}\label{psi2originalIntegrand}
To perform the time-integrals, we Wick rotate\footnote{This contour rotation is justified because the Bunch-Davies initial condition ensures the fall-off behaviour of the integrand at $\Im \eta\to +\infty$.} $\eta_i$ to $e^{i\pi/2}\chi_i$ to obtain
\begin{align}
	\psi_2^{\rm 1L}=\frac{\lambda^2 H^2}{8}\int\frac{\d^3 q_1}{(2\pi)^3}k^4 q_1 q_2 &\int_{0}^\infty \d\chi_1 \d\chi_2 \chi_1^2 \chi_2^2 \Bigg[e^{-(k+q_1+q_2)(\chi_1+\chi_2)}\nonumber\\
    &+e^{-k (\chi_1+\chi_2)}e^{(q_1+q_2)(\chi_1-\chi_2)}\left(1-e^{-2 q_1\chi_1}-e^{-2 q_2 \chi_1}\right)\theta(\chi_2-\chi_1) \nonumber\\
	&+e^{-k (\chi_1+\chi_2)}e^{(q_1+q_2)(\chi_2-\chi_1)}\left(1-e^{-2 q_1\chi_2}-e^{-2 q_2 \chi_2}\right)\theta(\chi_1-\chi_2) \Bigg]~.
\end{align}
As expected, the integrand is purely real but divergent. The $\chi_i$ integrals can be easily done to obtain
\begin{align}
    \nonumber \psi_2^{\rm 1L}=\frac{\lambda^2 H^2}{16} k^3 \int\frac{\d^3 q_1}{(2\pi)^3}&\Bigg[q_1 q_2\frac{3(q_1+q_2)^2+9k(q_1+q_2)+8k^2}{k^4 (k+q_1+q_2)^3}\\
    &+\frac{k}{(k+q_1+q_2)^3} \left(\frac{\Poly_{4}}{(k+q_1)^5}+\frac{\Poly_{4}}{(k+q_2)^5}\right)
    +\frac{k\, \Poly_{2}}{(k+q_1+q_2)^6}\Bigg]~,\label{OriginalPsi2Integral}
\end{align}
where $\Poly_{n}(k,q_1,q_2)\sim q_1^n$ denotes a polynomial that is homogeneous in the three momenta of degree $n$. It is easy to check that the first term in the above expression diverges as $q_1^4$, which is expected since the interaction is a non-renormalisable operator of dimension 6. The other terms, however, are convergent and do give a real contribution to the result. Also note that the divergent piece purely comes from the Feynman part of the loop integral.

We can now regulate the loop integral using partial dim reg i.e. by replacing $\d^3 q_1 \to \d^d q_1$ with $d=3-\epsilon$. This gives
\begin{align}
	\psi_2^{\rm 1L}=\frac{\lambda_d^2 H^2}{16} k^3 \left[\int\frac{\d^d q_1}{(2\pi)^d}q_1 q_2\frac{3(q_1+q_2)^2+9k(q_1+q_2)+8k^2}{k^4 (k+q_1+q_2)^3}+\text{finite \& real const.}~\right].
\end{align}
Note that along with changing the dimension of the momentum integral, we also modified the coupling constant from $\lambda$ to $\lambda_d = \mu_0^{(3-d)/2} \lambda$, to keep the mass dimension of coupling $\lambda$ fixed. 
Using \eqref{genintans} with $n=2$ and 
\begin{equation}
    g(q_+,k) = \frac{3 q_+^2+9k q_++8k^2}{k^4 (k+q_+)^3}~,
\end{equation}
and expanding the result around $d=3$, we obtain
\begin{align}
	\psi_2^{\rm 1L}&=\frac{1}{15}\frac{\lambda^2 H^2}{(4\pi)^2}k^3\left[\frac{1}{\epsilon}-\log \left(\frac{k}{\mu_0}\right) +\text{finite \& real const.}~\right].\label{NDRResult}
\end{align}
We see that the one-loop contribution in the partial dim reg scheme gives a \textit{real and scale-dependent} wavefunction coefficient.


\subsection{Renormalisation}\label{rgndr}
In this subsection, we will renormalise the action given in \eqref{toy} to remove the divergent piece of $\psi_2^{\rm 1L}$. Note that the mode functions used in the partial dim reg are the three-dimensional mode functions, whereas the loop integral is performed in $d$ dimensions. This suggests that the free part of the Lagrangian (used to compute both bulk-boundary and bulk-bulk propagators) lives in $3+1$ dimensions, whereas the interaction terms live in $(d+1)$ dimensions (where $d=3-\epsilon$). Therefore, we can add counterterms that live either in $3+1$ dimensions or in $d+1$ dimensions. Below, we analyse both cases separately.
\paragraph{Scheme A : $3+1$ dimensional counterterms}

To renormalise $\psi_2^{\rm 1L}$, we add the following counterterm action
\begin{equation}
    S_{\rm ct} =  \int \d^d x \d\eta ~ \sqrt{-g_{3+1}}~\mathcal{L}_{\rm ct}
\end{equation}
where $\mathcal{L}_{\rm ct}$ is given in \eqref{ct1}. The wavefunction gets the following contribution from the counterterms
\begin{align}
	\psi_2^{\rm ct}=i\sum_{2l+p+q\geq 3}{c}_{l,p,q} k^{2l}\int_{-\infty}^{0}\d \eta [a(\eta)]^{4-2l}(-H\eta\partial_\eta)^{p} K(\eta,k)(-H\eta\partial_\eta)^{q} K(\eta,k)~,
\end{align}
where the bulk -boundary propagator $K(\eta,k)$ are given in \eqref{MasslessBtobPropIn3dim}. After Wick rotating $\eta$ to $i \chi$ and rescaling $\chi$ to $\tilde{\chi}/k$, we obtain
\begin{align}
	\psi_2^{\rm ct}= \sum_{2l+p+q\geq 3}{c}_{l,p,q} k^{3}\int_{0}^\infty \d \tilde\chi ~(-1)^l~[a(-\tilde\chi)]^{4-2l}(-H\tilde\chi\partial_{\tilde\chi})^{p} K\left(\frac{i\tilde\chi}{k},k\right)(-H\tilde\chi\partial_{\tilde\chi})^{q} K\left(\frac{i\tilde\chi}{k},k\right)~,
\end{align}
Since $K\left(\frac{i\tilde\chi}{k};k\right)$ is independent of $k$, the above integral 
\begin{equation}
    \tilde{I}_{l,p,q}(3) = \int_{0}^\infty \d \tilde \chi ~(-1)^l~[a(-\tilde\chi)]^{4-2l}(-H\tilde\chi\partial_{\tilde\chi})^{p} K\left(\frac{i\tilde\chi}{k},k\right)(-H\tilde\chi\partial_{\tilde\chi})^{q} K\left(\frac{i\tilde\chi}{k},k\right)
\end{equation}
is $k$-independent. To cancel the divergence in \eqref{NDRResult}, we choose 
\begin{equation}
 \sum_{2l+p+q\geq 3}{c}_{l,p,q} \tilde{I}_{l,p,q}(3)   = - \frac{1}{15}\frac{\lambda^2 H^2}{(4\pi)^2}\left( \frac{1}{\epsilon}+ \log\frac{\mu_0}{\mu} +\text{finite \& real const.\,}\right) ~
\end{equation}
where $\mu>0$ is an arbitrary constant and the finite \& real const. is chosen to match that in \eqref{NDRResult}. Adding the counterterm contribution to the one-loop contribution, we obtain the full renormalised result to be
\begin{equation}
 \widehat\psi_2^{\rm 1L} = - \frac{1}{15}\frac{\lambda^2 H^2}{(4\pi)^2} k^3\left( \log\frac{k}{\mu} \right)~.  \end{equation}
We find that renormalised result is \textit{scale dependent and real}.

\paragraph{Scheme B : $d+1$ dimensional counterterms}
As stated above, in principle, we can add $d+1$ dimensional counterterms to renormalise $\psi_2^{\rm 1L}$ given in \eqref{NDRResult}. Therefore, we add the following counterterm action,
\begin{equation}
    S_{\rm ct} =  \int \d^d x \d\eta ~ \sqrt{-g_{d+1}}~\mathcal{L}_{\rm ct}~,
\end{equation}
where again, $\mathcal{L}_{\rm ct}$ is given in 
\eqref{ct1}. The counterterm  contribution to $\psi_2$ is given by
\begin{align}
	\psi_2^{\rm ct}=i\sum_{2l+p+q\geq 3}{c}_{l,p,q} k^{2l}\int_{-\infty}^{0}\d \eta [a(\eta)]^{d+1-2l}(-H\eta\partial_\eta)^{p} K(\eta,k)(-H\eta\partial_\eta)^{q} K(\eta,k)~.
\end{align}
The bulk-boundary propagators appearing in the above expression are again $3+1$ dimensional propagators.\footnote{This is because the free action in partial dim reg lives in $3+1$ dimensions.} The difference between the above expression and the one obtained in the previous scheme is the $d$ dependent exponent of the scale factor. This factor arises because the action involves the determinant of the $d+1$ dimensional metric instead of $3+1$ dimensional metric. Again performing Wick rotation ($\eta = i \chi$) and rescaling $\chi = \tilde{\chi}/k$, we obtain
\begin{align}\label{ctndco}
    \nonumber\psi_2^{\rm ct}=i^{d+1} H^{3-d} k^d \Bigg[&\sum_{2l+p+q\geq 3}{c}_{l,p,q} \int_{0}^{\infty}\d \tilde\chi\\
    &(-1)^l~ H^{2l-4}\tilde\chi^{2l-d-1}(-H \tilde\chi\partial_{\tilde\chi})^{p} K\left( \frac{i\tilde\chi}{k},k\right)(-H \tilde\chi\partial_{\tilde\chi})^{q} K\left( \frac{i\tilde\chi}{k},k\right)\Bigg]~,
\end{align}
where the bulk-boundary propagators are the $3+1$ dimensional propagators given in \eqref{MasslessBtobPropIn3dim}. The term in the square brackets is purely real. To cancel out the UV divergence of the one-loop result given in \eqref{NDRResult}, we choose 
\begin{equation}
     \sum_{2l+p+q\geq 3}{c}_{l,p,q} \tilde{I}_{l,p,q}(d)   = - \frac{1}{15}\frac{\lambda^2 H^2}{(4\pi)^2}\left( \frac{1}{\epsilon}+ \log\frac{\mu_0}{\mu} +\text{finite \& real const.\,}\right) ~,
\end{equation}
where $\tilde{I}_{l,p,q}(d)$ is the integral in \eqref{ctndco} and $\mu$ is a finite real constant. Finally in $d = 3-\epsilon$ dimensions, $\psi_2^{\rm ct}$ is given by
\begin{equation}
    \psi_2^{\rm ct} =  -\frac{1}{15} \frac{\lambda^2 H^2}{(4\pi)^2} k^3 \left(\frac{1}{\epsilon}-\log \left(\frac{k}{H}\right) + \log \left(\frac{\mu_0}{\mu}\right) -\frac{i \pi}{2} +\text{finite \& real const.\,}\right) ~.
\end{equation}

After renormalisation, the full one-loop contribution is given by
\begin{equation}\label{ndresult}
\widehat{\psi}_2^{\rm 1L}= \psi_2^{\rm 1L} + \psi_2^{\rm ct} = \frac{1}{15} \frac{\lambda^2 H^2}{(4\pi)^2}k^3\left(   \frac{i \pi}{2} + \log\frac{\mu}{H}\right)~.
\end{equation}
In this scheme, the final result matches the one obtained via regularisation schemes discussed in the main text. 

\subsection{Properties and Issues}
Let us review some of the properties of the result obtained after partial dim reg.

\begin{itemize}
    \item \textbf{Diffeomorphism}: 
    Partial dim reg seems to be a straightforward generalisation of the flat space dim reg, but there are several issues with this approach. As reviewed at the beginning of this appendix, in this regularisation, $3+1$ dimensional mode functions are used. In flat space, the equation of motion and hence the mode functions ($e^{i p_\mu x^\mu}$) are independent of the dimension of spacetime, and hence, dim reg just changes the momentum integrals from $3+1$ dimensions to $d+1$ dimensions. In contrast, in curved spacetime and in particular in de Sitter spacetime, the equation of motion depends on the number of spacetime dimensions. But in partial dim reg, the $(3+1)$-dimensional mode functions are used. Therefore, in this regularisation scheme, the free action (which is used to compute mode functions) lives in $3+1$ dimensions, whereas the interaction terms live in $d+1$ dimensions. Therefore, the action (without adding counterterms)
    \begin{align}
        \nonumber S&=S_{\rm free}+ S_{\rm int}\\
        &=\int \d t \d^3x \sqrt{g_{3+1}}\,\mathcal{L}_{\rm free}+\int \d t \d^dx \sqrt{g_{3+1}}\, \mathcal{L}_{\rm int}
    \end{align}
    breaks diffeomorphism invariance. Even after adding the counterterms (both in Scheme A and Scheme B), the full action breaks diffeomorphism invariance. Hence, this is not a convenient regularisation scheme.

    \item \textbf{Scale invariance}:
    Both renormalisation schemes discussed above behave differently under the scaling symmetry.
    \begin{itemize}
        \item \textbf{Scheme A}: Under scaling $ k\rightarrow\alpha k$, the full renormalised result does not preserve scale invariance i.e.
        \begin{equation}
            \widehat\psi_2^{\rm 1L} (\alpha k) \neq \alpha^3 \widehat\psi_2^{\rm 1L} (k)~.
        \end{equation}
        This is due to the presence of $\log k$ piece in the full renormalised result.

        \item \textbf{Scheme B}: Even though the unrenormalised result contains $\log k$ piece, in this scheme, the counterterms cancel this contribution and hence restore the scaling symmetry of the full renormalised result. Therefore
        \begin{equation}
            \widehat\psi_2^{\rm 1L} (\alpha k) = \alpha^3 \widehat\psi_2^{\rm 1L} (k)~.
        \end{equation}
    \end{itemize}

    \item \textbf{Unitarity}:
    Let us check if the renormalised one-loop result obtained after partial dim reg obeys COT or not. Since the $\Disc$ of counterterms vanishes, the $\Disc$ of the full renormalised one-loop result is same in both the schemes and is given by
    \begin{equation}\label{discndr}
        \begin{split}
         i{\rm Disc} [i\widehat\psi_2^{\rm 1L}] &= i\Disc [i\psi_2^{\rm 1L}] \\
         &= \frac{1}{15}\frac{\lambda^2 H^2}{(4\pi)^2} k^3\left(\log k - (\log [e^{-i \pi} k])^*\right)\\
        &= \frac{k^3}{15}\frac{\lambda^2 H^2}{(4\pi)^2} \times (-i \pi)\,,
        \end{split}
    \end{equation}
    It matches the RHS of COT computed in  \eqref{cotright-hand side} and hence we conclude that $\widehat\psi_2^{\rm 1L}$ obeys the COT. 
\end{itemize}

\section{Symmetric two-body phase space}\label{2bdyPhaseSpaceAppendix}

When computing one-loop bubble diagrams in $d+1$ spacetime dimensions, we often encounter the momentum integral over the two-body phase space,
\begin{align}
	\mathcal{I}_d=\int \d^d q_1 \d^d q_2 \,\delta^d(\mathbf{q}_1+\mathbf{q}_2-\mathbf{k}) f(q_1,q_2,k)~.
\end{align}
In principle, one can directly integrate out one momentum using the $\delta$-function, leaving one layer of momentum integral albeit with an asymmetric integrand that is often complicated to analyse. Alternatively, one can find a much more elegant treatment based on the two-body phase space formula, which we will review in this appendix. 

We begin by changing variables to $\mathbf{Q}\equiv(\mathbf{q}_1+\mathbf{q}_2)/2$ and $\mathbf{P}\equiv \mathbf{q}_1-\mathbf{q}_2$, we have
\begin{align}
	\nonumber\mathcal{I}_d&=\int \d^d Q \d^d P \,\delta^d(2\mathbf{Q}-\mathbf{k}) f(|\mathbf{Q}+\mathbf{P}/2|,|\mathbf{Q}-\mathbf{P}/2|,k)\\
	&=\frac{1}{2^d}\int \d^d P \,f(|\mathbf{k}+\mathbf{P}|/2,|\mathbf{k}-\mathbf{P}|/2,k)~.
\end{align}
In polar coordinates, the angular integrals can be trivially finished to yield the area of a $(d-2)$-sphere,
\begin{align}
	\mathcal{I}_d=\frac{\Omega_{d-2}}{2^d}\int_0^\infty P^{d-1}\d P \int_{-1}^{1}\d x \,f(\sqrt{k^2+P^2+2k P x}/2,\sqrt{k^2+P^2-2k P x}/2,k)~.
\end{align}
Changing variables again to
\begin{align}
	q_1&=\sqrt{k^2+P^2+2k P x}/2~,\\
	q_2&=\sqrt{k^2+P^2-2k P x}/2~,
\end{align}
with a Jacobian,
\begin{align}
	\left|\frac{\partial(P ,\,x\,)}{\partial(q_1,q_2)}\right|=\frac{8 q_1 q_2}{k}~,\label{Pxq1q2Jacobian}
\end{align}
we arrive at
\begin{align}
	\mathcal{I}_d=\frac{\Omega_{d-2}}{2^{d-3}}\int_{q_1+q_2>k>0,\, q_1+k>q_2>0,\, q_2+k> q_1>0}\d q_1 \d q_2 \left[2(q_1^2+q_2^2)-k^2\right]^{\frac{d-3}{2}}\frac{q_1 q_2}{k}f(q_1,q_2,k)~.
\end{align}
In terms of $q_+\equiv q_1+ q_2$ and $q_-\equiv q_1-q_2$, we have the following expression for the $d$-dimensional symmetric two-body phase space integral:
\begin{align}\label{s2bdy}
	\mathcal{I}_d=\frac{\Omega_{d-2}}{2^d}\int_k^\infty \d q_+ \int_{-k}^k \d q_- \left[q_+^2 + q_-^2 -k^2\right]^{\frac{d-3}{2}} \frac{q_+^2-q_-^2}{k}f\left(\frac{q_+ +q_-}{2},\frac{q_+ -q_-}{2},k\right)~.
\end{align}
This form of the phase space integral is particularly convenient when the original integrand takes the form
\begin{align}
    f(q_1,q_2,k)=(q_1 q_2)^{n-1} g(q_+,k)~,
\end{align}
which is indeed the case in many examples. For such integrands, the $q_-$ integral can be directly finished to give
\begin{align}\label{genintans}
	\mathcal{I}_d=\frac{\Omega_{d-2}}{2^{d+2n-3}}\int_k^\infty \d q_+ q_+^{2 n} \left(q_+^2-k^2\right)^{\frac{d-3}{2}} F_1\left(\frac{1}{2};-n,\frac{3-d}{2};\frac{3}{2};\frac{k^2}{q_+^2},\frac{k^2}{k^2-q_+^2}\right)  g(q_+,k)~,
\end{align}
where $F_1$ denotes the Appell function of the first kind. For $n$ being an integer, the Appell $F_1$ function reduces to ordinary Hypergeometric functions. Due to the high transcendentality of the integrand, this last radial integral over $q_+$ is often not exactly computable. However, if we only care about the UV-divergent piece of the integral, we can perform a large-$q_+$ expansion for the integrand up to the desired order. After the expansion and truncation, the radial integral can be easily computed.

\section{Cosmological optical theorem at one loop revisited}\label{COTApplicationToOurModelSubAppendix}

In this appendix, we review the cosmological optical theorem (COT) and its application to our example. The COT is essentially a partial reflection of unitarity and analyticity in perturbative Feynman diagrams. Assuming these fundamental properties and a Bunch-Davies vacuum, the propagators satisfy
\begin{align}
    \Disc [K(\eta,k)]&=0~,\\
    \Disc_p[i G(\eta_1,\eta_2;p)]&=-i P(p)\Disc_p K(\eta_1,p) \times \Disc_p K(\eta_2,p)~,
\end{align}
while the interaction vertices also satisfy Hermitian analyticity,
\begin{align}
    \Disc V\left(\eta\partial_\eta,i\mathbf{k} \eta \right)=V\left(\eta\partial_\eta,i\mathbf{k} \eta \right)-\left[V\left(\eta\partial_\eta,i(-\mathbf{k}) \eta \right)\right]^*=0~.
\end{align}
Thus it was shown in \cite{Goodhew:2020hob} that the discontinuity of any contact tree diagram vanishes,
\begin{align}
    \Disc[i \psi_n^{\rm 0L,cont.}]=0~,\label{contactCOT}
\end{align}
and that the single-cut discontinuity of general tree-diagrams factorise into those of lower-point diagrams,
\begin{align}
    \Disc_s[i\psi_n^{\rm 0L}]=i P(s) \, \Disc_s [i\psi_{m}^{\rm 0L}]\times \Disc_s [i\psi_{n-m}^{\rm 0L}]~.
\end{align}
Later in \cite{Melville:2021lst}, these relations was generalised to cutting rules at arbitrary loop order,
\begin{align} 
   i \, \underset{\substack{ \text{internal} \\ \text{lines} } }{\Disc }  \left[ i \, \psi_n^{\rm (diagram)} \right]     &=  \sum_{\rm cuts}   \left[ \prod_{\substack{\rm cut \\ \rm momenta}} \int P \right]
   \prod_{\rm subdiagrams} (-i) \underset{\substack{ \text{internal } \& \\  \text{ cut lines} } }{\Disc } \left[ i \, \psi_{m}^{(\rm subdiagram)} \right]~.
\end{align}
Here the COT is to be understood diagram-by-diagram in practice.\\

To best illustrate how the COT works, let us study the COT corresponding to the one-loop wavefunction $\psi_2^{\rm 1L}$ in our example. We start with the propagator identity
\begin{align}
    \nonumber i\underset{q_1,q_2}{\Disc}[i\times i G(\eta_1,\eta_2;q_1) \,i G(\eta_1,\eta_2;q_2)]&=iP(q_2)\underset{q_1,q_2}{\Disc}\left[K(\eta_1,q_2)\, i G(\eta_1,\eta_2;q_1) \, K(\eta_2,q_2)\right]\\
    \nonumber&+iP(q_1)\underset{q_1,q_2}{\Disc}\left[K(\eta_1,q_1)\, i G(\eta_1,\eta_2;q_2) \, K(\eta_2,q_1)\right]\\
    & - P(q_1)P(q_2)\underset{q_1,q_2}{\Disc}\left[K(\eta_1,q_1) K(\eta_1,q_2)\right] \times\underset{q_1,q_2}{\Disc}\left[K(\eta_2,q_1) K(\eta_2,q_2)\right]~.\label{oneLoopBubbleCutIdentityDisc}
\end{align}
Dressing \eqref{oneLoopBubbleCutIdentityDisc} with bulk-boundary propagators, we obtain
\begin{align}
    \nonumber &i\underset{q_1,q_2}{\Disc}\left[i\frac{-\lambda^2}{2H^2}\int_{\mathbf{q}_1}\int\frac{\d\eta_1}{\eta_1}\frac{\d\eta_2}{\eta_2}\partial_{\eta_1} K(\eta_1,k) \partial_{\eta_2} K(\eta_2,k)\partial_{\eta_1}\partial_{\eta_2}G(\eta_1,\eta_2;q_1) \partial_{\eta_1}\partial_{\eta_2}G(\eta_1,\eta_2;q_2)\right]\\
    \nonumber&=-\frac{i}{2}\int_{\mathbf{q}_1}P(q_2)\underset{q_1,q_2}{\Disc}\left[i\frac{-\lambda^2}{H^2}\int\frac{\d\eta_1}{\eta_1}\frac{\d\eta_2}{\eta_2}\partial_{\eta_1} K(\eta_1,k) \partial_{\eta_1} K(\eta_1,q_2)  \partial_{\eta_1}\partial_{\eta_2}G(\eta_1,\eta_2;q_1) \partial_{\eta_2}K(\eta_2,q_2) \partial_{\eta_2}  K(\eta_2,k) \right]\\
    \nonumber&-\frac{i}{2}\int_{\mathbf{q}_2}P(q_1)\underset{q_1,q_2}{\Disc}\left[i\frac{-\lambda^2}{H^2}\int\frac{\d\eta_1}{\eta_1}\frac{\d\eta_2}{\eta_2}\partial_{\eta_1} K(\eta_1,k) \partial_{\eta_1} K(\eta_1,q_1)  \partial_{\eta_1}\partial_{\eta_2}G(\eta_1,\eta_2;q_2) \partial_{\eta_2}K(\eta_2,q_1) \partial_{\eta_2}  K(\eta_2,k) \right]\\
    \nonumber & - \frac{1}{2} \int_{\mathbf{q}_1} P(q_1)P(q_2)\int_{\mathbf{q}_1} \underset{q_1,q_2}{\Disc}\left[i\frac{i\lambda}{H}\int\frac{\d\eta_1}{-\eta_1}\partial_{\eta_1} K(\eta_1,k) \partial_{\eta_1}K(\eta_1,q_1) \partial_{\eta_1}K(\eta_1,q_2)\right] \\
    &\qquad\qquad\qquad\qquad\quad \times\underset{q_1,q_2}{\Disc}\left[i\frac{i\lambda}{H}\int \frac{\d\eta_2}{-\eta_2} \partial_{\eta_2}K(\eta_2,k) \partial_{\eta_2}K(\eta_2,q_1) \partial_{\eta_2} K(\eta_2,q_2)\right]~,
\end{align}
which reduces to\footnote{Notice that \eqref{oneloopBubbleCOT} differs from the COT in \cite{Melville:2021lst} by a factor of $1/(2!)$ since the symmetry factor is handled differently. In \cite{Melville:2021lst}, all symmetry factors are stripped out of wavefunction coefficients whereas they are included in ours. This leads to different numerical coefficients on the right-hand side of the COT.}
\begin{align}
	\nonumber i \Disc [i\widehat{\psi}_2^{\rm 1L}(k)]=&\int_{\mathbf{q}_1} P(q_1)\, (-i)\underset{q_1}{\Disc} \left[i\widehat\psi_4^{\rm 0L}(k,q_1,q_1,k)\right]\\
	&+\frac{1}{2}\int_{\mathbf{q}_1} P(q_1)P(q_2)\, i\,\underset{q_1,q_2}{\Disc} \left[i\widehat\psi_3^{\rm 0L}(k,q_1,q_2)\right]\times i\,\underset{q_1,q_2}{\Disc}  \left[i\widehat\psi_3^{\rm 0L}(q_1,q_2,k)\right]~.\label{oneloopBubbleCOT}
\end{align}
Here we have used the symmetry of the two-body phase space to reduce the two single-cut piece into one. We have also performed renormalisation and added the trivial counterterm contribution
\begin{align}
    i \Disc [i\psi_2^{\rm ct}(k)]=0~,
\end{align}
as all contact diagrams vanish due to \eqref{contactCOT}. Consequently, the wavefunction coefficients satisfy the COT both before and after renormalisation.\\

To evaluate the right-hand side of the COT \eqref{oneloopBubbleCOT}, we list the relevant tree diagrams,
\begin{align}\label{psi3psi4}
    \widehat\psi_3^{\rm 0L}(k,q_1,q_2)&=\frac{\lambda}{H}\frac{2k^2 q_1^2 q_2^2}{(k+q_1+q_2)^3}~,\nonumber\\
    \widehat\psi_4^{\rm 0L}(k,q_1,q_1,k)&=	-\frac{\lambda^2}{4} \frac{k^4 q_1^4}{(k+q_1+q_2)^3}\left[\frac{3}{(k+q_1)^2}+\frac{q_2}{(k+q_1)^3}+\frac{8 q_2}{(k+q_1+q_2)^3}\right]~.
\end{align}

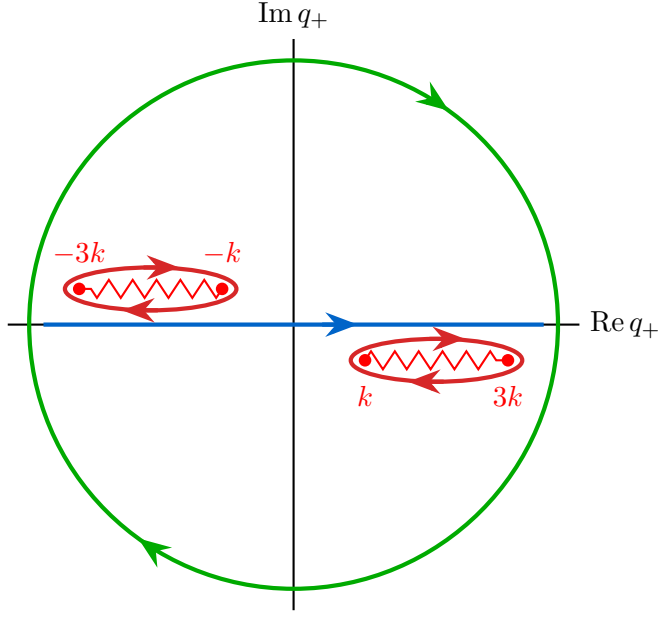
\begin{figure}
    \centering
  \begin{tikzpicture}[scale=0.95, >=stealth] 
  
    \draw[-, thick] (-4,0) -- (4,0) node[right] {$\Re q_+$}; 

    \draw[-, thick] (0,-4) -- (0,4) node[above] {$\Im q_+$}; 

    \draw[red, thick, 
      decoration={zigzag, amplitude=0.12cm, segment length=0.317cm}, 
      decorate] (1,-0.5) -- (3,-0.5);
    \fill[red] (1,-0.5) circle (2.5pt);
    \node[red] at (1,-1) {$k$};
    \fill[red] (3,-0.5) circle (2.5pt);
    \node[red] at (3,-1) {$3k$};

    \draw[red, thick, 
      decoration={zigzag, amplitude=0.12cm, segment length=0.317cm}, 
      decorate] (-1,0.5) -- (-3,0.5);
    \fill[red] (-1,0.5) circle (2.5pt);
    \node[red] at (-1,1) {$-k$};
    \fill[red] (-3,0.5) circle (2.5pt);
    \node[red] at (-3,1) {$-3k$};

    \draw[blue2, ultra thick] (-3.5,0) -- (3.5,0);

    \draw[blue2, -{Stealth[scale=1.2]}, ultra thick] (0.4,0) -- ++(0.5,0);

    \draw[red2, ultra thick] (2,-0.5) ellipse (1.2 and 0.3);
    \draw[red2, -{Stealth[scale=1.2]}, ultra thick] (1.9,-0.2) -- ++(0.5,0);
    \draw[red2, -{Stealth[scale=1.2]}, ultra thick] (2.1,-0.8) -- ++(-0.5,0);
    
    \draw[red2, ultra thick] (-2,0.5) ellipse (1.2 and 0.3);
    \draw[red2, -{Stealth[scale=1.2]}, ultra thick] (-1.9,0.2) -- ++(-0.5,0);
    \draw[red2, -{Stealth[scale=1.2]}, ultra thick] (-2.1,0.8) -- ++(0.5,0);

    \draw[green2, ultra thick, 
          postaction={
            decorate,
            decoration={
              markings,
              mark=at position 0.15 with {\arrow[scale=1.2,rotate=180]{Stealth}},
              mark=at position 0.65 with {\arrow[scale=1.2,rotate=180]{Stealth}}
            }
          }
         ] (0,0) circle (3.7cm); 
    
    \end{tikzpicture}
    \caption{We start from the \textcolor{blue2}{blue} contour in \eqref{InitialContourToTheRightCircle} and deform it to wind around the branch cut at $(k,3k)$. Then by the centrosymmetry of the integrand, we are allowed to extend this \textcolor{red2}{red} contour to include the \textcolor{red2}{red} loop around the other branch cut at $(-3k,-k)$. At last, we merge the two \textcolor{red2}{red} contours to form a large \textcolor{green2}{green} circle at infinity, which vanishes by the large-$q_+$ convergence of the integrand. Notice that $k=|k|-i\epsilon$ has an infinitesimal negative imaginary part which is exaggerated in this plot for better visual effect.}\label{singlecutIntegralContours}
\end{figure}

\paragraph{The single-cut piece.} Using the symmetric two-body phase space formula and integrating out $q_-$, the single-cut piece gives
\begin{align}
    \nonumber&\int_{\mathbf{q}_1} P(q_1)\, (-i)\underset{q_1}{\Disc} \left[i\widehat\psi_4^{\rm 0L}(k,q_1,q_1,k)\right]\\
    =&\frac{\lambda^2 H^2 k^3}{32\pi^2}\int_k^\infty \d q_+\left\{\left(q_+^2-3k^2\right)\left[-\frac{\log\left(\frac{q_+ - k}{q_+ -3k}\right)}{(q_+-k)^3}+\frac{\log\left(\frac{q_+ + k}{q_+ +3k}\right)}{(q_+ +k)^3}\right]-\frac{\Poly_{15}(q_+,k)}{15(q_+^2-k^2)^6 (q_+^2-9k^2)^2}\right\}~,\label{oneloopPsi2COTsingleCut1}
\end{align}
with
\begin{align}
    \nonumber\Poly_{15}(q_+,k)&=15 k q_+^{14} + 1235 k^3 q_+^{12} -9717 k^5 q_+^{10}-13809 k^7 q_+^8+118733
   k^9 q_+^6-109767 k^{11} q_+^4\\
   &\quad +21081 k^{13} q_+^2+8613 k^{15}~.
\end{align}
Apparently, the integral is ill-defined near the threshold $q_+=k$. The proposal in \cite{Melville:2021lst} is to extend the phase space to
\begin{align}
    \int_k^\infty\d q_+\to \int_0^\infty \d q_+~,\label{phaseSpaceExtension}
\end{align}
and go around the folded singularity $q_+=k$. We are allowed to do so because the extended phase space is symmetric under $k\leftrightarrow -k^*$ and so drops out from both sides of the COT. Doing so one realises that the integral \eqref{oneloopPsi2COTsingleCut1} is symmetric under $q_+\leftrightarrow -q_+$ and we are allowed to double it and uplift it to a contour integral
\begin{align}
    \int_0^\infty \d q_+=\frac{1}{2}\int_{-\infty}^\infty \d q_+~.
\end{align}
The second term in \eqref{oneloopPsi2COTsingleCut1} thus vanishes by the residue theorem,
\begin{align}
    \frac{1}{2}\int_{-\infty}^\infty \d q_+ \frac{\Poly_{15}(q_+,k)}{15(q_+^2-k^2)^6 (q_+^2-9k^2)^2}=-\pi i \left(\underset{q_+=k}{\Res}+\underset{q_+=3k}{\Res}\right)\frac{\Poly_{15}(q_+,k)}{15(q_+^2-k^2)^6 (q_+^2-9k^2)^2}=0~,
\end{align}
where we have deformed the contour to wind around the poles at $k=|k|-i\epsilon$ and $3k=3|k|-i\epsilon$. The first term in \eqref{oneloopPsi2COTsingleCut1} is slightly more subtle as it involves two branch cuts between $(-3k,-k)$ and $(k,3k)$. We proceed by first deforming the contour to wind around the branch cut at $(k,3k)$,
\begin{align}
    \frac{1}{2}\int_{-\infty}^\infty \d q_+ \left(q_+^2-3k^2\right)\left[-\frac{\log\left(\frac{q_+ - k}{q_+ -3k}\right)}{(q_+-k)^3}+\frac{\log\left(\frac{q_+ + k}{q_+ +3k}\right)}{(q_+ +k)^3}\right]=-\frac{1}{2}\oint_{(k,3k)} (\cdots)~.\label{InitialContourToTheRightCircle}
\end{align}
Due to the centrosymmetry of the integrand, the contour integral around the other branch cut at $(-3k,-k)$ is identical to that at $(k,3k)$. Therefore we can write
\begin{align}
    -\frac{1}{2}\oint_{(k,3k)} (\cdots)=-\frac{1}{4}\left(\oint_{(k,3k)}+\oint_{(-3k,-k)}\right)(\cdots)=\frac{1}{4} \oint_{\infty}(\cdots)=0~,
\end{align}
where in the last step we have merged the contours around the two branch cuts and blown it to a large circle at infinity. The integral then vanishes as the integrand is of order $(\cdots)=\mathcal{O}(q_+^{-2})$ at large $q_+$. The whole procedure is illustrated in Figure \ref{singlecutIntegralContours}. In summary, the single-cut piece turns out to vanish upon the extension of phase space \eqref{phaseSpaceExtension},
\begin{align}
    \int_{\mathbf{q}_1} P(q_1)\, (-i)\underset{q_1}{\Disc} \left[i\widehat\psi_4^{\rm 0L}(k,q_1,q_1,k)\right]=0~.
\end{align}

\paragraph{The double-cut piece.} The double-cut piece reads
\begin{align}
    \nonumber&\frac{1}{2}\int_{\mathbf{q}_1} P(q_1)P(q_2)\, i\,\underset{q_1,q_2}{\Disc} \left[i\widehat\psi_3^{\rm 0L}(k,q_1,q_2)\right]\times i\,\underset{q_1,q_2}{\Disc}  \left[i\widehat\psi_3^{\rm 0L}(q_1,q_2,k)\right]\\
    =&\frac{\lambda^2 H^2 k^4}{32\pi^2}\int_k^\infty \d q_+ \frac{q_+^2(q_+^2+3k^2)^2}{(q_+^2-k^2)^6}\left(q_+^4-\frac{2}{3}k^2 q_+^2+\frac{1}{5}k^4\right)~.
\end{align}
The folded singularity can again be avoided by extending the phase space and integrate from the origin. This integral can be easily evaluated by taking the residue at $k=|k|-i\epsilon$ to give
\begin{align}\label{cotright-hand side}
    \frac{1}{2}\int_{\mathbf{q}_1} P(q_1)P(q_2)\, i\,\underset{q_1,q_2}{\Disc} \left[i\widehat\psi_3^{\rm 0L}(k,q_1,q_2)\right]\times i\,\underset{q_1,q_2}{\Disc}  \left[i\widehat\psi_3^{\rm 0L}(q_1,q_2,k)\right]=\frac{1}{15}\frac{\lambda^2 H^2}{(4\pi)^2}k^3\times (-i\pi)~.
\end{align}


\section{The finite remainder in $\upeta$ regularisation}\label{FiniteRemainderAppendix}

In this appendix, we provide a recipe to compute the finite remainder of divergent integrals in $\upeta$ regularisation. We begin by a demonstration for the toy example \eqref{psi2OneLoopEtaRegAfterInsertion},
\begin{align}
    \psi_2^{\rm 1L}=\frac{\lambda^2 H^2}{16} \frac{\pi}{8(2\pi)^3 k}\int_{k}^\infty \frac{\d q_+}{q_+} \left(q_+^5 -\frac{2}{3} k^2 q_+^3+ \frac{1}{5}k^4q_+\right)\frac{3 q_+^2+9 k q_++8k^2}{(k+q_+)^3}\upeta\left(\frac{q_+/k}{\Lambda/H}\right)~.
\end{align}
This integral is quartically divergent in the absence of the regulator. Therefore, we can isolate the divergent piece by subtracting the leading polynomial in $q_+$,
\begin{align}
    \int_{k}^\infty \frac{\d q_+}{q_+} \Bigg[&3q_+^4-3k^2 q_+^2+\frac{64}{15}k^4\\
    \nonumber&+\left(q_+^5 -\frac{2}{3} k^2 q_+^3+ \frac{1}{5}k^4q_+\right)\frac{3 q_+^2+9 k q_++8k^2}{(k+q_+)^3}-\left(3q_+^4-3k^2 q_+^2+\frac{64}{15}k^4\right)\Bigg]\upeta\left(\frac{q_+/k}{\Lambda/H}\right)~.
\end{align}
The second line is now UV-finite and we can safely drop the regulator and finish the integral up to terms that are suppressed by $\Lambda$,
\begin{align}
    \nonumber\int_{H/\Lambda}^\infty \frac{\d x}{x} &\left[3x^4\left(\frac{\Lambda}{H}\right)^4-3x^2\left(\frac{\Lambda}{H}\right)^2+\frac{64}{15}\right]\upeta\left(x\right)\\
    &\quad -\frac{26+64\log 2}{15}+\mathcal{O}\left(\frac{H}{\Lambda}\right)~,
\end{align}
where we have also changed variables to $x\equiv (q_+/k)/(\Lambda/H)$ and pulled out overall factors of $k$. Using $\int_{H/\Lambda}^{\infty}=\int_{0}^{\infty}-\int_0^{H/\Lambda}$ and integrating by parts, we arrive at
\begin{align}
   \nonumber &3C_4[\upeta]\left(\frac{\Lambda}{H}\right)^4-3C_2[\upeta]\left(\frac{\Lambda}{H}\right)^2+ \frac{64}{15}\left[\upeta\left(\frac{H}{\Lambda}\right)\log\frac{\Lambda}{H}+\gamma[\upeta]+\int_{0}^{H/\Lambda} \d x \log x\,\upeta'(x)\right]\\
    & -\int_{0}^{H/\Lambda} \frac{\d x}{x} \left[3x^4\left(\frac{\Lambda}{H}\right)^4-3x^2\left(\frac{\Lambda}{H}\right)^2\right]\upeta\left(x\right)-\frac{26+64\log 2}{15}+\mathcal{O}\left(\frac{H}{\Lambda}\right)~,
\end{align}
where in the second line, we can approximate $\upeta(x)\approx 1$ and obtain another contribution to the finite term. Keeping terms up to $\Lambda^0$, we obtain
\begin{align}
    \nonumber\psi_2^{\rm 1L}&=\frac{\lambda^2 H^2}{16} \frac{\pi}{8(2\pi)^3}k^3\Bigg[3C_4[\upeta]\left(\frac{\Lambda}{H}\right)^4-3C_2[\upeta]\left(\frac{\Lambda}{H}\right)^2+ \frac{64}{15}\left(\log\frac{\Lambda}{H}+\gamma[\upeta]\right)\\
    & \qquad\qquad\qquad\qquad-\frac{59}{60}-\frac{64}{15}\log 2\Bigg]+\mathcal{O}\left(\frac{H}{\Lambda}\right)~,
\end{align}
in accordance with \eqref{psi2oneloopEtaRegResult}.

This method for extracting the finite remainder can be easily generalised to arbitrary one-loop UV divergences. Consider an integral over the radial loop momentum of the form \eqref{GeneralOneLoopIntegrals},
\begin{align}
    \psi_n^{\rm 1L}(\{k\},\{s\},\{\mathbf{k}\})&=\int_{q_+^{\rm min}}^{\infty} \frac{\d q_+}{q_+} f(\{k\},\{q\},\{\mathbf{k}\})\,\upeta\left(\frac{q_+/k_T}{\Lambda/H}\right)~.
\end{align}
We can likewise isolate the UV-divergent pieces as leading polynomials in the large-$q_+$ regime (see \eqref{generalOneloopLargeqExpansion} for the definition of $f_m$),
\begin{align}
    \nonumber \int_{q_+^{\rm min}}^{\infty} \frac{\d q_+}{q_+} \Bigg[&\sum_{m>0} \left(\frac{q_+}{k_T}\right)^m f_m+f_0\\
    &+f(\{k\},\{q\},\{\mathbf{k}\})-\left(\sum_{m>0} \left(\frac{q_+}{k_T}\right)^m f_m+f_0\right)\Bigg]\,\upeta\left(\frac{q_+/k_T}{\Lambda/H}\right)~.
\end{align}
Here the second line contributes as a part of the finite remainder upon dropping the regulator, whereas the first line can be rewritten as
\begin{align}
    \nonumber&\sum_{m>0}f_m C_m[\upeta]\left(\frac{\Lambda}{H}\right)^m+f_0\left[\log\left(\frac{\Lambda}{H}\frac{k_T}{q_+^{\rm min}}\right)\upeta\left(\frac{q_+^{\rm min}/k_T}{\Lambda/H}\right)+\gamma[\upeta]+\int_0^{\frac{q_+^{\rm min}/k_T}{\Lambda/H}}\d x \log x \,\upeta'(x)\right]\\
    &-\sum_{m>0}f_m\int_0^{q_+^{\rm min}/k_T}\frac{\d y}{y}\, y^m \,\upeta\left(\frac{Hy}{\Lambda}\right)\\
    =&\sum_{m>0}f_m C_m[\upeta]\left(\frac{\Lambda}{H}\right)^m +f_0\left[\log\frac{\Lambda}{H}-\log\frac{q_+^{\rm min}}{k_T}+\gamma[\upeta]\right]-\sum_{m>0}\frac{f_m}{m}\left(\frac{q_+^{\rm min}}{k_T}\right)^m+\mathcal{O}\left(\frac{H}{\Lambda}\right)~.
\end{align}
Combining everything, we arrive at
\begin{align}
    \psi_n^{\rm 1L}&=\sum_{m>0}f_m C_m[\upeta]\left(\frac{\Lambda}{H}\right)^m +f_0\left[\log\frac{\Lambda}{H}+\gamma[\upeta]+g\right]+\mathcal{O}\left(\frac{H}{\Lambda}\right)~,\label{generalpsin1LForm}
\end{align}
with the full finite remainder $g=g(\{k\},\{s\},\{\mathbf{k}\})$ given by
\begin{align}
     g=-\log\frac{q_+^{\rm min}}{k_T}+f_0^{-1}\left\{-\sum_{m>0}\frac{f_m}{m}\left(\frac{q_+^{\rm min}}{k_T}\right)^m+\int_{q_+^{\rm min}}^{\infty} \frac{\d q_+}{q_+}\left[f-\left(\sum_{m>0} \left(\frac{q_+}{k_T}\right)^m f_m+f_0\right)\right]\right\}~.
\end{align}
Crucially, the finite remainder does not depend on the choice of regulator function but only on the loop integrand $f$. All functional dependence on the regulator $\upeta$ is clearly exposed as its Mellin transforms in \eqref{generalpsin1LForm}, a pleasant merit of the $\upeta$ regularisation scheme.

\bibliographystyle{JHEP}
\bibliography{2ptRefs}

\end{document}